# Knowledge Graph Building Blocks: An easy-to-use Framework for developing FAIREr Knowledge Graphs

Vogt, Lars[1] orcid.org/0000-0002-8280-0487;

Konrad, Marcel[1] orcid.org/0000-0002-2452-3143

Prinz, Manuel[1] orcid.org/0000-0003-2151-4556

[1] *TIB Leibniz Information Centre for Science and Technology, Welfengarten 1B, 30167 Hanover, Germany*

Corresponding Author: lars.m.vogt@gmail.com

# Abstract

Knowledge graphs and ontologies provide promising technical solutions for implementing the FAIR Principles for Findable, Accessible, Interoperable, and Reusable data and metadata. However, they also come with their own challenges. Nine such challenges are discussed and associated with the criterion of cognitive interoperability and specific FAIREr principles (FAIR + Explorability raised) that they fail to meet. We introduce an easy-to-use, open source knowledge graph framework that is based on knowledge graph building blocks (KGBBs). KGBBs are small information modules for knowledge-processing, each based on a specific type of semantic unit. By interrelating several KGBBs, one can specify a KGBB-driven FAIREr knowledge graph. Besides implementing semantic units, the KGBB Framework clearly distinguishes and decouples an internal in-memory data model from data storage, data display, and data access/export models. We argue that this decoupling is essential for solving many problems of knowledge management systems. We discuss the architecture of the KGBB Framework as we envision it, comprising (i) an openly accessible KGBB-Repository for different types of KGBBs, (ii) a KGBB-Engine for managing and operating FAIREr knowledge graphs (including automatic provenance tracking, editing changelog, and versioning of semantic units); (iii) a repository for KGBB-Functions; (iv) a low-code KGBB-Editor with which domain experts can create new KGBBs and specify their own FAIREr knowledge graph without having to think about semantic modelling. We conclude with discussing the nine challenges and how the KGBB Framework provides solutions for the issues they raise. While most of what we discuss here is entirely conceptual, we can point to two prototypes that demonstrate the principle feasibility of using semantic units and KGBBs to manage and structure knowledge graphs.

# Introduction

According to the **FAIR Guiding Principles** (1), data and metadata must be Findable, Accessible, Interoperable, and Reusable for machines and humans alike. In recent years, the FAIR Guiding Principles have increasingly attracted the attention of research, industry, and developers of knowledge management tools (1–6). Stakeholders within science and research policy have also become aware of FAIR. FAIR is also relevant for the economy—in 2018, the cost-benefit analysis for FAIR research data, commissioned by the European Union (EU), estimated that not having FAIR research data costs the EU economy at least 10.2 billion Euro each year, and when considering the positive impact of FAIR data in terms of data quality and machine-readability they estimated another 16 billion Euro on top (7). The High Level Expert Group of the [European Open Science Cloud](#) (EOSC) therefore recommended building an Internet of FAIR Data and Services ([IFDS](#)) (8), through which all relevant data-rich institutions, research projects, and citizen-science projects can make their data and metadata accessible in line with the FAIR Guiding Principles, while retaining full control over their data's ethical, privacy, or legal aspects (following Barend Mons' *data visiting as opposed to data sharing* (9)). This requires providing rich machine-actionable data and metadata, organizing them into **FAIR Digital Objects** (10,11), each of which can be individually referenced by its own Unique Persistent and Resolvable Identifier (UPRI), and developing adequate concepts and tools for human-readable interface outputs and search capabilities. While efforts to build the IFDS are already under way (see [GO FAIR Initiative](#) and [EOSC](#)), the situation at many data-rich institutions and companies regarding FAIRness of their data and metadata is still far from ideal.

With their transparent semantics, highly structured syntax, and standardized formats, **knowledge graphs** in combination with **ontologies** provide a promising technical solution for implementing the FAIR Guiding Principles (12,13). In a knowledge graph, instances, classes, and their relationships can all be represented as resources with their own UPRIs, representing relations between entities following the triple syntax of *Subject-Predicate-Object*. Each particular relation is thus modelled as a structured set of three distinct data points. In relational databases, in contrast, with their tabular format, the relationships between entities are modeled between the columns of data tables and not between individual data points. The consequence is that knowledge graphs outperform relational databases for complex queries over densely connected data, which is often the case with research data and their associated metadata. Therefore, knowledge graphs are particularly suitable in the context of FAIR research and development and for all tasks that require optimal and detailed data retrieval.

However, using a knowledge graph for documenting data and metadata does not guarantee their FAIRness, which would require following additional guidelines such as consistently applying the same semantic data models for modelling the same type of data and metadata statements to provide schematic interoperability (14), and organizing data and metadata into FAIR Digital Objects (10,11). Moreover, as a fairly new concept and technology, knowledge graphs also bring their own specific technical, conceptual, and societal challenges. Just looking at the somewhat fuzzy concept of *knowledge graph* (12) and the lack of commonly accepted standards illustrates this clearly, covering such different technical and conceptual incarnations like property graphs such as [Neo4J](#) and approaches based on the Resource Description Framework (RDF), the use of RDF-stores, and, with the Web Ontology Language (OWL), also applications of Description Logics. Below, we discuss nine user stories that exemplify some of these technical, conceptual, and societal challenges that users and developers of knowledge graph applications face.

We discuss these user stories because we have the impression that in the context of FAIR and GoingFAIR strategies, the focus has been mainly on achieving machine-actionability of data and metadata, with their human-actionability and the usability of corresponding knowledge management frameworks (e.g., RDF) having been lost sight of in the process. Each of the nine user stories reflects an aspect of that lack in human-actionability of data, metadata, and knowledge management frameworks. To counter this, we suggested (14) to extend the **EOSC Interoperability Framework (EOSC IF)** (10) to include, in addition to the four interoperability layers of **1) technical interoperability** (i.e., information technology systems must work with other information technology systems in implementation or access without any restrictions or with controlled access), **2) semantic interoperability** (i.e., contextual semantics related to common semantic resources), **3) organizational interoperability** (i.e., contextual processes related to common process resources), and **4) legal interoperability** (i.e., contextual licenses related to common license resources), also the layer of **5) cognitive interoperability**.

We understand **cognitive interoperability** as

> *"a characteristic of an information technology system to communicate its information with human users in the most efficient ways, providing them with tools and functions that intuitively support them in getting an overview of the data, find the data they are interested in, and allow them to explore other data points from a given data point in semantically meaningful and intuitive ways, thereby taking into account the general cognitive conditions of human beings—not only in terms of how humans prefer to interact with technology (human-computer interaction) but also in terms of how they interact with information (human-information interaction). In the context of knowledge graphs, the tools should make the user aware of their contents, help to understand their meaning, support communicating their contents, provide means that increase the trustworthiness of their contents, support their integration in other workflows and software tools, and make transparent what actions can be taken. Additionally, cognitive interoperability of an information technology system can also be understood as the system's characteristic to be easily implemented and employed by developers and easily managed by operators."* (14)

Cognitive interoperability thus focuses on the usability of data structures and knowledge management systems for human users and developers—an aspect that has been somewhat overlooked so far, especially in the context of knowledge graphs and Semantic technologies. As a guideline for improving cognitive interoperability, we suggested extending the FAIR Guiding Principles to include the **Principle of human Explorability**, resulting in the **FAIREr (Explorability raised) Guiding Principles** (14) (for a list of the Principles, see Box 1).

---

**Box 1 | The FAIREr Guiding Principles (14). They include the FAIR Guiding Principles (1) (in regular font) and proposed additions to them, including the human Explorability principle (in bold font).**

**To be Findable:**
F1.  (meta)data are assigned a globally unique and persistent identifier
F2.  data are described with rich metadata (defined by R1 below)
F3.  metadata clearly and explicitly include the identifier of the data it describes
F4.  (meta)data are registered or indexed in a searchable resource
**F5.1   terms with the same extension must be mapped to each other**
**F5.2   terms ideally include multilingual labels and specify all relevant synonyms**
**F6.1   the same (meta)data schema is used for the same type of statement, and the schema is referenced with its**

|   |   |
|---|---|
|   | identifier in the statement's metadata |
| F6.2 | (meta)data schemata for the same type of statement must be aligned and mapped to each other |

| **To be Accessible:** |   |
|---|---|
| A1. | (meta)data are retrievable by their identifier using a standardized communications protocol |
| A1.1 | the protocol is open, free, and universally implementable |
| A1.2 | the protocol allows for an authentication and authorization procedure, where necessary |
| **A1.3** | **the protocol is compliant with existing data protection regulations (e.g., [General Data Protection Regulation](#), GDPR)** |
| A2. | metadata are accessible, even when the data are no longer available |

| **To be Interoperable:** |   |
|---|---|
| I1. | (meta)data use a formal, accessible, shared, and broadly applicable language for knowledge representation |
| I2. | (meta)data use vocabularies that follow FAIR principles |
| I3. | (meta)data include qualified references to other (meta)data |
| **I4.** | **(meta)data specify the logical framework that has been used for their modelling (e.g., OWL and thus description logic or common logic interchange framework and thus first-order-logic)** |
| **I5.** | **(meta)data use a formalism to clearly distinguish between lexical, assertional, contingent, prototypical, and universal statements** |
| I6.1-2 | see F5.1-2 |
| I7.1-2 | see F6.1-2 |

| **To be Reusable:** |   |
|---|---|
| R1. | (meta)data are richly described with a plurality of accurate and relevant attributes |
| R1.1 | (meta)data are released with a clear and accessible data usage license |
| R1.2 | (meta)data are associated with detailed provenance |
| R1.3 | (meta)data meet domain-relevant community standards |

| **To be Explorable:** |   |
|---|---|
| **E1.** | **(meta)data are structured into semantically meaningful subsets, each represented by its own globally unique and persistent identifier that enables its referencing and its identification and that instantiates a semantically defined corresponding class** |
| **E1.1** | **each binary and n-ary proposition contained in the (meta)data is structured into its own statement subset** |
| **E1.2** | **each granularity tree contained in the (meta)data is structured into its own granularity tree subset** |
| **E1.3** | **each frame of reference contained in the (meta)data is structured into its own context subset** |
| **E2.** | **(meta)data subsets can be (recursively) combined and encapsulate complexity to manageable units** |
| **E2.1** | **some subsets are organized into different levels of representational granularity** |
| **E3.** | **human-readable display of (meta)data is decoupled from machine-actionable (meta)data storage, reducing the complexity of the displayed (meta)data to what is relevant to a human reader** |
| **E3.1** | **a user interface provides a form-based textual and a mind-map like graphical option for accessing and exploring (meta)data** |
| **E3.2** | **a user interface enables zooming in and out of (meta)data, leveraging different levels of representational granularity, granularity trees, and frames of reference, with statement subsets having the actual statement as their label** |
| **E3.3** | **a user interface supports making statements about statements** |
| **E3.4** | **a user interface enables querying the (meta)data without knowledge of (graph) query languages** |

After discussing the nine user stories, we briefly summarize our concept of semantic units and how they can be applied to organize and structure a knowledge graph into semantically meaningful subgraphs. Next, we introduce the concept of a Knowledge Graph Building Block (KGBB) as a modular management unit for types of semantic units. We discuss different types of KGBBs and some of their functionalities and features, and how KGBBs can be used to specify KGBB-driven FAIREr knowledge graph applications. We continue with describing the different components of the KGBB Framework, including besides the various KGBBs also a KGBB-Engine that, based on the information from KGBBs, operates FAIREr knowledge graph applications and provides automatic provenance tracking, editing changelog, and versioning. After briefly introducing our idea for KGBB-Functions that can add rule-based functionalities to the knowledge graph for purposes of data analyses and visualization, we discuss the KGBB-Editor for defining new KGBBs and combining existing KGBBs to specify a KGBB-driven knowledge graph application in the form of a KGBB specification graph. Finally, we discuss the impact and potential challenges of the KGBB Framework regarding the development and usability of FAIREr knowledge graph applications, including its potential to provide solutions for the problems exemplified in the user stories.

We want to point out that most of what we present here is the result of conceptual work and still needs development, testing, and implementation. So far, only two applications have been developed using the KGBB Framework, one of which is a small prototype for a crowdsourced scholarly knowledge graph that implements basic ideas from KGBBs and semantic units and thus functions as a very basic proof-of-concept for some of their core components (https://github.com/LarsVogt/Knowledge-Graph-Building-Blocks) (16). The other prototype has been developed in a master thesis project and already allows for a generic specification for certain KGBB types, making it possible to define different kinds of basic knowledge graph applications. It also supports storing the knowledge graph in multiple databases, namely RDF and Neo4j (https://gitlab.com/TIBHannover/orkg/semantic-building-blocks).

---

**Box 2 | Conventions**

In this paper, we refer to FAIR knowledge graphs as machine-actionable semantic graphs for documenting, organizing, and representing lexical, assertional (e.g., empirical data), universal, and contingent statements and thus a mixture of ABox and TBox expressions (thereby contrasting knowledge graphs with ontologies, with the latter containing mainly universal statements and thus TBox expressions and lexical statements). We want to point out that we discuss KGBBs against the background of RDF-based triple stores, OWL and Description Logics as a formal framework for inferencing, and labeled property graphs as an alternative to triple stores, because these are the main technologies and logical frameworks used in knowledge graphs that are supported by a broad community of users and developers and for which accepted standards exist. We are aware of the fact that alternative technologies and frameworks exist that support an n-tuples syntax and more advanced logics (e.g., First Order Logic) (15,16), but supporting tools and applications are missing or are not widely used to turn them into well-supported, scalable, and easily usable knowledge graph applications.

Throughout this text we use regular underlined to indicate ontology classes, *italicsUnderlined* when referring to properties (i.e., relations in Neo4j), and use ID numbers to specify each. ID numbers are composed of the ontology prefix followed by a colon and a number, e.g., *isAbout* (IAO:0000136). If the term is not yet covered in any ontology, we indicate it with \*, e.g., the class \*metric measurement statement unit\*. We use 'regular underlined' to indicate instances of classes, with the label referring to the class label and the ID number to the class. Moreover, when we use the term *resource*, we understand it to be something that is uniquely designated (e.g., a Uniform Resource Identifier, URI) and about which you want to say something. It thus stands for something and represents something you want to talk about. In RDF, the *Subject* and the *Predicate* in a triple statement are always resources, whereas the *Object* can be either a resource or a literal. Resources can be either properties, instances, or classes, with properties taking the *Predicate* position in a triple and with instances referring to individuals (=particulars) and classes to universals.

For reasons of clarity, in the text and in all figures we represent resources not with their UPRIs but with human-readable labels, with the implicit assumption that every property, every instance, and every class has its own UPRI.

---

# User story 1: Tom needs his data and metadata to be FAIR but does not want to look at large, complex graphs

Tom is a biologist and wants to store the data and metadata of his genome-phenotype experiment for later analysis and documentation. He found various causal relationships between particular genes and phenotype characters and wants to document them in a FAIR and thus machine-actionable knowledge graph together with the relevant data and metadata and then integrate them with other comparable data, resulting in a very large and highly complex graph. He wants to explore this graph, but is overwhelmed by its complexity and the fact that it contains many nodes that seem to be irrelevant to him.

Tom, like all human beings, is an expert in efficiently communicating information, leaving out background knowledge, and using somewhat fuzzy statements that refer to general figures of thought and that use metaphors and metonymies, because he knows that other humans will nevertheless be able to understand him and infer missing information from the context. For a machine, on the other hand, all relevant information must be explicitly stated. Tom is thus confronted with the dilemma that arises from the conflict between machine-actionability and human-actionability of data and metadata: **the more he pushes data representations toward machine-actionability, the more complex and therefore the less human-actionable they become** (14) (see Fig. 1, middle). This represents an impedance mismatch and Tom is frustrated by it.

If Tom wanted to store data and metadata in a knowledge graph in a machine-actionable format and in the same time represent them in an easily comprehensible human-readable way in the user interface (UI), he would have to increase the **cognitive interoperability of the graph** by developing new approaches and tools for exploring and navigating it, zooming in and out across different levels of representational granularity, thereby reducing the complexity of the graph to only those bits of information that are momentarily relevant to him. Ideally, he would **decouple data storage in the graph from data presentation in the UI [E3.;E3.1]**, so that information that is only necessary for machines but irrelevant to humans is only accessed by machines but not shown in the UI (see Fig. 1, bottom).

Since Tom's project is part of a larger program, Tom even wants to distinguish **different defined views on his data**, depending on which device is being used for their presentation (e.g., browser versus smartphone), which access rights a user has, or which level of detail is needed. Tools for describing graph patterns that enforce a standardized way of modelling and representing data of the same type, such as the Shapes Constraint Language [SHACL](#) and Data Shapes [DASH](#) (17), Shape Expressions [ShEx](#) (18,19), or the Reasonable Ontology Templates [OTTR](#) (20,21), provide solutions to this problem only to a certain extent. Tom needs more comprehensive solutions.

**Observation:**

*ObjectX weighs 5 kilograms, with a 95% confidence interval of 4.54 to 5.55*

**Machine-Actionable RDF Graph:**

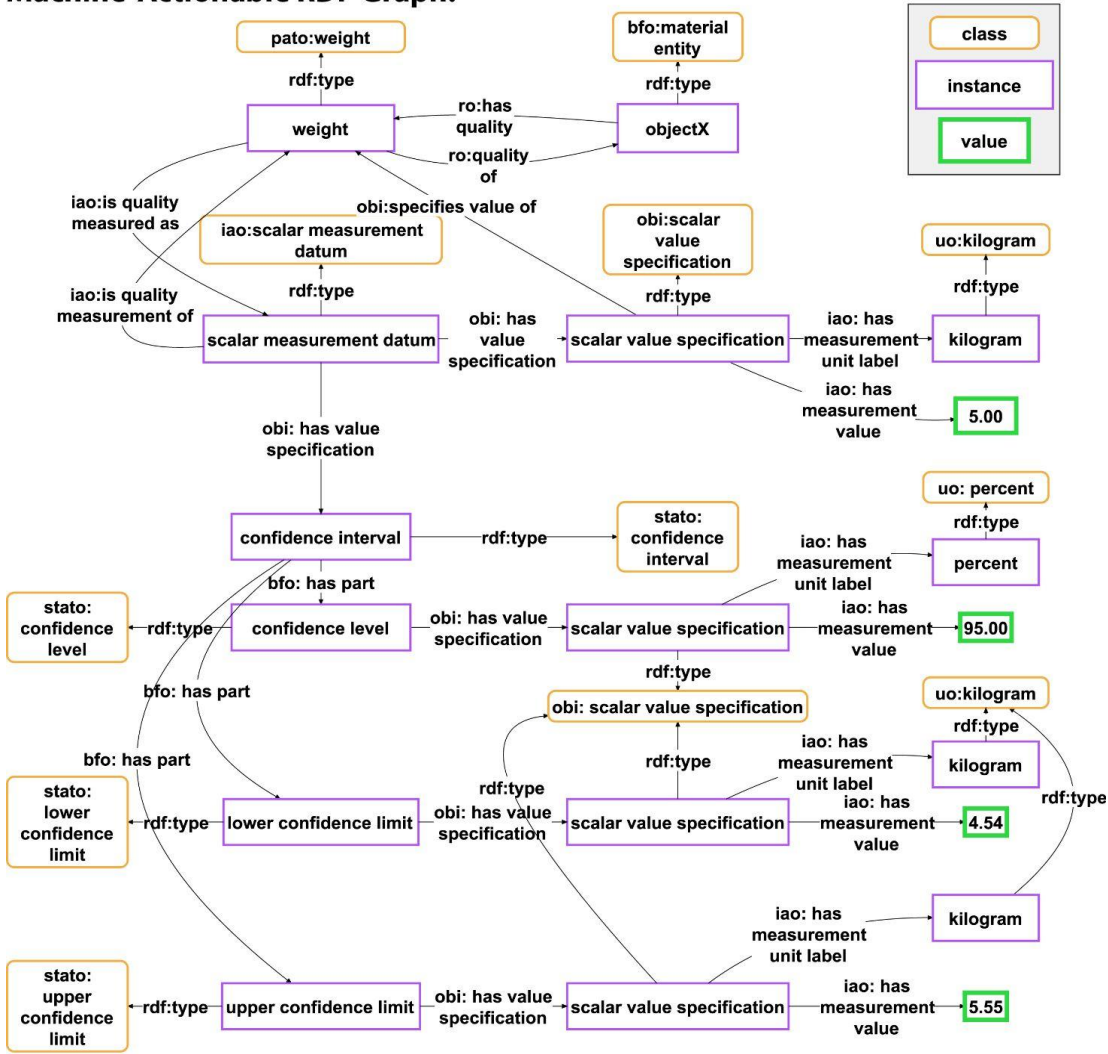

**Human-Actionable Mind-Map like Graph:**

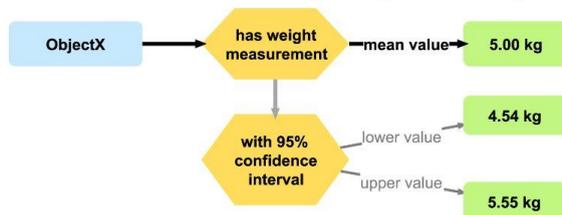

**Figure 1: Comparison of a human-readable statement with its machine-actionable representation and its human-actionable representation. Top**: A human-readable statement about the observation that ObjectX weighs 5 kilograms, with a 95% confidence interval of 4.54 to 5.55. **Middle**: A machine-actionable representation of the same statement as an ABox semantic graph, using RDF syntax and following the general pattern for measurement data from the Ontology for Biomedical Investigations (OBI) (22) of the Open Biological and Biomedical Ontology (OBO) Foundry. **Bottom**: A human-actionable representation of the same statement as a mind-map like graph, reducing the complexity of the RDF graph to the information that is actually relevant to a human reader. [Figure taken from (14)]

## User story 2: Catherine and Felix want to add FAIR data to a crowdsourced knowledge graph with a broad scope

Catherine and Felix are domain experts and colleagues. They both work on describing the anatomy of evolutionary closely related organisms. They have published several papers in their field and now want to add their published anatomy descriptions to a crowdsourced scholarly knowledge graph that allows its users to add the contents and not only the bibliographic metadata of scholarly publications to its graph. The scope of this knowledge graph is very broad, as it aims to cover scholarly knowledge from all possible research domains. Unfortunately, after comparing the graphs they created, they realize that they modelled their data differently, leaving the machine no chance to integrate and compare their anatomy descriptions. In other words, their data are not interoperable. After all the work they put into adding the descriptions, Catherine and Felix are frustrated by this result. They will not continue using this application.

Isabelle is the developer of the application. Because its scope is very broad, she allows users to freely add their data, coining new properties and classes on-the-fly. She basically follows the Wikipedia method of a community-driven crowdsourced approach. She suggested Catherine and Felix to use her knowledge graph, and now she is frustrated by their feedback. Isabelle understands the criticism, but does not know how to improve her application.

RDF/OWL as well as labeled property graphs are both highly expressive. Any given statement can be modelled in many different ways. As a consequence, if users can freely add contents without any restrictions, the data will likely neither be interoperable, nor easily findable and thus only minimally machine-actionable and not truly FAIR.

If appropriate graph patterns would be openly available that knowledge graph developers could reuse, shape constraining languages and frameworks such as SHACL, ShEx, or OTTR could be used for implementing them. Their usage would guarantee that all data and metadata added through them are interoperable (i.e., schematic interoperability). Moreover, if each such pattern would have its own UPRI, data statements can reference them in their metadata. With this information, one can identify potentially interoperable data by their commonly shared graph-pattern identifiers. Practically, this implies that all data statements in a knowledge graph must reference their associated graph pattern (e.g., in the form of a shape specification) and, ideally, also allow identifying the subgraphs that correspond with a specific graph pattern (i.e., all instantiations of a given pattern). When these criteria were met, empirical data and their metadata documented in a knowledge graph would truly comply with the FAIR Guiding Principles (23) **[F6.1/I7.1]**.

## User story 3: Sara and Bob do not know graph query languages and struggle learning them

Sara is a software developer who has been hired for a research project to develop a database for storing and documenting the project's data and metadata. She has experience with relational databases, but not with graph-based databases and their **graph query languages**, such as SPARQL or Cypher. Their complexity has kept her from learning them, in part also because she can already implement most of the things required for the project with her current relational database expertise. Sara will argue that the missing parts could only be implemented with considerable technical and financial effort.

Bob is a domain expert in engineering, and he also has no experience in graph query languages. When Bob heard about a new database for engineering-related research data and realized that querying its contents requires the use of SPARQL or Cypher, he became frustrated. He is very busy with his research and does not want to invest additional time in learning a rather complex query language. When asked, he would always prefer a relational database with its typical search functionality over a knowledge graph with a SPARQL or Cypher query interface.

The described scenario reflects our personal experience that most users and software developers have no experience with graph-based databases and do not know graph query languages and their advantages, and therefore do not see the need to learn them. And even those who are familiar with them report that writing more complex queries can be challenging. **Apparently, having to write SPARQL or Cypher queries represents an entry barrier for interacting with a knowledge graph and hinders their broader usage** (24). Openly available and reusable query patterns that link to specific graph patterns would contribute to a solution to this problem, or UIs that allow writing queries using forms or natural language and that automatically translate the input into SPARQL or Cypher queries **[E3.4]**.

## User story 4: Karl wants to enable users to intuitively make statements about statements in a knowledge graph

Karl is a software developer and wants to develop a knowledge graph application for documenting measurement data from a research project and for identifying possible relationships between empirical data and various causal hypotheses. Part of his job is to enable users to make statements about statements in the knowledge graph and organize this information in a transparent and user-friendly way in the application and its UI. Whereas RDF reification (25) and RDF-star (26,27) are feasible for referring to individual triples, making statements about larger subgraphs such as a measurement datum with a 95% confidence interval that is modelled with more than 20 triples becomes inefficient and complicated for querying (see Fig. 1, middle). Using Named Graphs instead, seems to be a better solution (25) but needs clearly stated criteria that specify which triples belong to which Named Graph and thus criteria for systematically structuring a data graph into separate subgraphs. Karl is not satisfied with either solution.

Automatically structuring and organizing the data graph into different **semantically meaningful and identifiable subgraphs** at different levels of representational granularity would provide an efficient way of structuring a knowledge graph to allow users to intuitively make statements about statements [**E1.1-3;E2.;E3.3**].

## User story 5: Dan and Anna describe the same type of measurement data but use different community-specific data models

Dan is a botanist and Anna an ecologist. Dan describes the anatomy of plant organisms, including various quality measurements. He wants to document the descriptions in a knowledge graph following the model of the Open Biological and Biomedical Ontologies (OBO) Foundry (Fig. 2, top). Anna describes plant-insect relationships. In this context, specific plant qualities are relevant, so she also describes them. Without knowledge of Dan's descriptions, she documents her descriptions in the same knowledge graph. However, since she is an ecologist, she follows the model of the Extensible Observation Ontology (OBOE) (Fig. 2, bottom) which deviates considerably from the OBO model. After adding her descriptions, she realizes that she could use some more plant measurement data and wants to use the data from Dan. However, she soon becomes frustrated, because their data are not easily made interoperable and Anna has no programming skills to write a script that transfers Dan's data to the OBOE model.

Standards are often developed by communities. While it may be straightforward to develop graph patterns that comply with an established standard that does not change over time, it becomes quite challenging to comply with different and frequently changing standards across several communities. Figure 2 shows two significantly different graph patterns for modelling a weight measurement. If both are used in the same knowledge graph, respective data are not directly interoperable.

**Decoupling data access from data storage** by aligning and mapping the data from a defined storage model to different defined access models, thus defining **data and metadata schema crosswalks**, would solve this problem. If new access models could be added to the knowledge graph at any time, its data and metadata could be adapted to newly proposed standards and formats, and thus to an ever-evolving research landscape. **[F6.2/I7.2]**

### OBO Foundry Model:

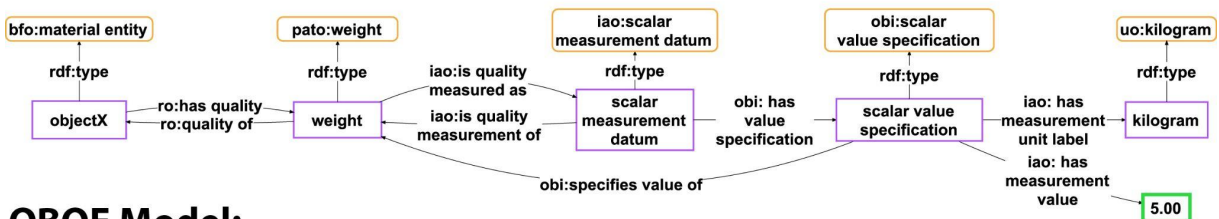

### OBOE Model:

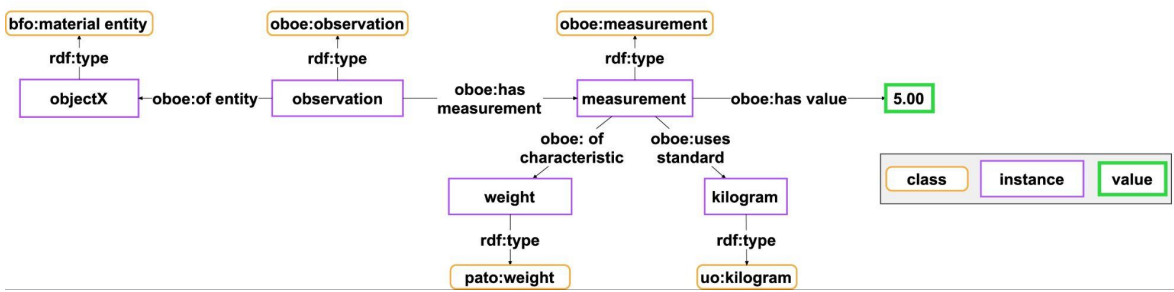

**Figure 2: Comparison of two graph patterns for modelling the same statement**. Frequently, different communities have agreed on different models for representing the same type of information. For example, the statement *"ObjectX has a weight of 5 kilograms"* can be modeled according to the two graph patterns shown here. **Top**: The pattern of the Ontology for Biomedical Investigations (OBI) (22) of the Open Biological and Biomedical Ontology (OBO) Foundry, which is frequently used in the biomedical domain. **Bottom**: The pattern of the Extensible Observation Ontology (OBOE), which is frequently used in the ecology community.

# User story 6: John wants to change the database technology of the backend of his biodiversity knowledge graph

John does research on biodiversity loss. He is part of a project in which a specific habitat is monitored using, among others, various sensors for recording local weather data. Each sensor provides this data related to a specific location every ten minutes. John is frustrated, because he started building his biodiversity knowledge graph in the last five years using a specific database technology, and while in the first years, everything was looking good, with increasing amounts of monitoring stations, he realizes that the database he is using does not scale well—querying the spatio-temporal data is taking increasingly longer each year.

John does some research and realizes that storage and query technologies have significantly improved over the last years. He would like to change the database for his knowledge graph, but this is not that easily done and John does not have the required resources for that right now.

Virtual knowledge graph systems such as [Ontop](Ontop) that implement the ontology-based data access (OBDA) paradigm have been suggested for translating SPARQL queries expressed over the knowledge graph into SQL queries executed by relational data sources (28–31), thus enabling federated queries over RDF triple stores and relational databases. By **decoupling the storage model from the storage technology** and, via an appropriate API, by allowing the data graph to be stored in (i) various graph databases such as RDF-stores, [Wikibase](Wikibase), or property graph databases like [Neo4j](Neo4j), (ii) other NoSQL databases like key-value databases, or (iii) in the form of a semantic table that is semantically isomorph to the graph and stored in a relational database, the OBDA approach could be extended to include even more storage technologies. This would enable storing spatio-structural data as semantic tables in a relational database. Moreover, if an RDF or a property graph representation of the data is required, the table could be exported any time and stored in respective alternative technologies. Knowledge management frameworks that allowed decoupling storage model from storage technology would increase their **cognitive interoperability for developers of knowledge graph applications**.

## User story 7: Mila wants to import semantically unstructured (legacy) data from a tabular format into a FAIR knowledge graph

Mila is the director of a natural history museum that houses a large collection of specimens. The museum is also hosting various collection-related datasets from past projects, each with its own particular format and data scheme. Mila wants her museum to participate in an upcoming initiative that aims at building a large national knowledge graph about collection-related data. This knowledge graph requires data to be FAIR. Initially, Mila was enthusiastic to contribute the museum's data. It did not take long, however, that she realized that transferring the data to the knowledge graph and thereby FAIRifying it requires converting it to the knowledge graph's data schema. This is a very time-consuming task and requires experience in Semantics and ontologies, which the museum does not have.  This frustrates Mila, and she backs off.

Many valuable research datasets are semantically unstructured and stored on local machines using tabular formats (e.g., CSV, Excel sheets). As a consequence, these legacy datasets are not FAIR and thus cannot be easily found, accessed, integrated with other existing data, and reused for research. Converting them into a FAIR format can be very challenging and time-consuming. It usually involves either sophisticated semi-automatic NLP approaches that utilize Deep Learning algorithms or a lot of manual work. Knowledge management frameworks that support the FAIRification of legacy data increase the **cognitive interoperability of their knowledge graph applications**.

# User story 8: Nick has to distinguish universal, prototypical, and assertional statements in his knowledge graph

Nick is a data scientist working for a project that collects data from clinical studies on the potential impact of certain drugs on the fetal development of human gross anatomy during pregnancy. The data will be documented in a FAIR knowledge graph using terms from a human anatomy ontology, and cover anatomy descriptions of individual fetuses based on ultrasound recordings, protocols about the drugs that have been taken during pregnancy, and representations of statistically averaged fetal anatomy development in the form of prototypical anatomy descriptions. Because human fetal anatomy shows considerable disparity (i.e., extent of morphospace across individual fetal anatomies), a certain degree of variation across the anatomy descriptions is expected.

Nick's responsibility is to develop adequate graph patterns for modelling this data. The data include ontology terms with their class axioms, which will be used for reasoning over the data. Since class axioms specify what is necessarily true for every instance of the class, they are **universal statements**, which in OWL form a TBox (Terminology Box). The anatomy descriptions of individual fetuses, on the other hand, consist of statements that are only true for a particular individual and are thus **assertional statements** (see also *instance anatomy* (32)), which in OWL form an ABox (Assertion Box). Finally, the **prototypical** anatomy descriptions include statements that are only true for most, but not necessarily for all instances of a class, such as that a hand typically has a thumb as its part, but a hand without a thumb would still be a hand. They are **contingent (prototypical) statements** (see also *canonical anatomy* (32)).

Nick would prefer using a labeled property graph over an RDF tuple store, because the former is intuitively usable and allows for adding property-value pairs to each node and relation in the graph, which supports making statements about statements (see *user story 4*). Unfortunately, however, property graphs lack a unified set of standards for formal semantics and reasoning and thus do not allow for formally differentiating between assertional, prototypical, and universal statements (33,34). Solutions for representing universal statements in a property graph have been suggested. They involve annotating edges with logical properties such as existential restriction axioms, resulting in an object-property mapping (33,34). A standardized, W3C recommended, formalism for mapping OWL into a property graph, however, is still missing and with it also reasoning over the data. So Nick abandons his first idea to use property graphs and takes a look at RDF tuple stores and OWL as an alternative technical solution.

Because OWL is based on Description Logics, it distinguishes assertional from universal statements, but it does not include a formalism for distinguishing prototypical statements. Moreover, OWL-based representations of universal statements often lead to unnecessarily complex graphs that are not intuitively comprehensible for a human reader and also involve blank nodes, which makes querying over them difficult (34). Furthermore, universal statements in the form of TBox expressions are documented as axioms of ontology classes in an OWL file and thus lie outside the domain of discourse of a knowledge graph, so users cannot make statements about them within the knowledge graph. Nick is frustrated since no technology meets all of his requirements. Nick needs a framework for knowledge graph applications that facilitates the formal distinction between assertional, prototypical, and universal statements **[I5.]**.

## User story 9: Paula wants to build a FAIR knowledge graph, but has neither experience in Semantics nor the required resources

Paula is a biologist who works in the field of biodiversity research. She usually gets funding for smaller, specialized projects that do not come with enough resources to employ an expert in Semantics for semantically modelling the project's data and documenting it in a FAIR knowledge graph. This frustrates Paula, because she believes in sharing and reusing data and metadata and wants her data and metadata to be FAIR so that they contribute to larger biodiversity datasets. Her institution has an IT department who could support her in setting up existing software solutions, but they have no experience in semantic data modelling.

What Paula needs is a system that allows her to document her data and metadata in a FAIR knowledge graph without her having to model this data semantically in OWL or in a labelled property graph. All that Paula needed to bring to the table would be her domain knowledge. Unfortunately, knowledge management frameworks based on knowledge graph technologies typically **lack the cognitive interoperability required for their widespread adoption**.

## Organizing a knowledge graph into different types of semantic units

In (23), we propose to structure a knowledge graph **into identifiable sets of triples, i.e., subgraphs that are semantically meaningful to a human reader and thus represent units of representation (short: semantic units)** (23) **[E1.]**. Each semantic unit is represented in the knowledge graph with its own UPRI **[F1.;F4;E1.]** that identifies the semantic unit as a resource and at the same time its associated subgraph, i.e., the semantic unit's data graph (Fig. 3). Consequently, referring to the UPRI of a semantic unit is equivalent to referring to the contents of its data graph.

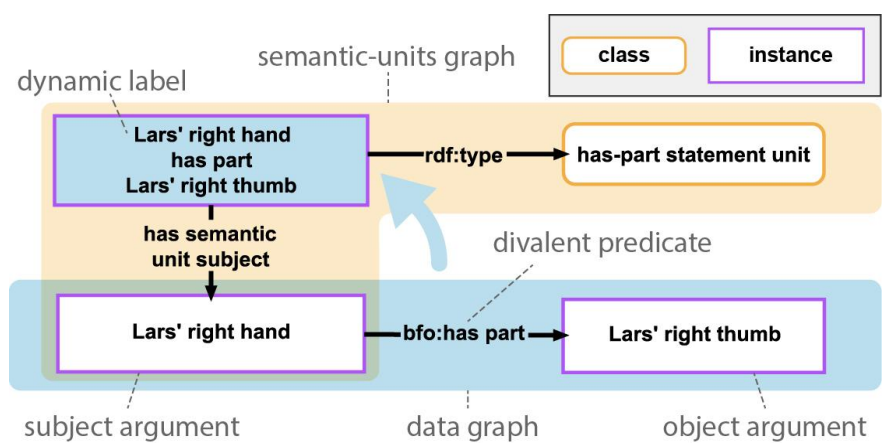

**Figure 3: Example of a statement unit** that models a proposition based on a has-part relation between two instances. The data graph expressing the statement is shown in the blue box at the bottom, with 'Lars' right hand' being the subject argument and 'Lars' right thumb' the object argument of the statement. The semantic-units graph is shown in the peach-colored box and contains the triples representing the semantic unit. It specifies that the resource representing the statement unit (blue box with border, here shown with its dynamic label), which represents the has-part statement in the data graph (indicated by the blue arrow), is an instance of *has-part-of_statement unit* and that 'Lars' right hand' is the subject of this semantic unit. The UPRI of *'has-part-of_statement unit'* is also the UPRI of the semantic unit's data graph (the subgraph in the blue box without border). [Figure taken from (23)]

Semantic units introduce a new type of **representational entity** (35) in addition to instances and classes. The instances and classes with their relations constitute what we call the **data graph layer** of a knowledge graph, whereas the semantic units with their relations add a **semantic-units graph layer** to a knowledge graph (Fig. 3). Each semantic unit is represented in the semantic-units graph layer by its UPRI, whereas its corresponding subgraph, i.e., its data graph, is part of the data graph layer of the knowledge graph.

A semantic unit resource instantiates a corresponding **semantic unit class [F1;F4;E1.]**. The class includes a human-readable description of the type of information its instances and their associated data graphs cover. A semantic unit can be organized like a nanopublication (36–38) and, when properly implemented, also functions as a FAIR Digital Object (10,11) and thus meets the requirements suggested by the *European Commission Expert Group on FAIR Data* for a realization of FAIR.

Applied to a knowledge graph, different types of semantic units add an organizational framework to the graph by structuring its data graph into several layers of partially overlapping, partially enclosed subgraphs, each of which constitutes the data graph of a particular semantic unit that, in turn, is represented in the semantic-units graph layer of the knowledge graph by its own node. **A semantic unit is thus a resource representing a higher level of abstraction than the low-level abstraction of individual triples.** We distinguish statement units, compound units, and question units as basic categories of semantic units (14,23).

## Statement unit

When organizing and structuring a knowledge graph with the goal to substantially increase the human-actionability of its contents, the focus must lie on identifying **propositions (i.e., statements)** and on making them referenceable, since propositions are the basic units of information that humans want to communicate and document. Even datasets consisting of tables with values implicitly represent collections of propositions that can be derived from contextual information associated with the table or with its rows and columns.

Therefore, we conceive **statement units** to be the most basic category of semantic unit, on which all other categories build on. A statement unit is a **unit of information (i.e., subgraph) that represents the smallest, independent proposition that is semantically meaningful for a human reader [E1.1]** (23). Statement units **mathematically partition the data graph layer** of a knowledge graph so that each of its triples belongs to the data graph of exactly one statement unit.

Statement units add another and coarser-grained **level of representational granularity** to the semantic-units graph layer of a knowledge graph, on top of the fine-grained level of individual triples **[E2.1]**.

Propositions are communicated via statements, and according to the **predicate-argument-structure** from linguistics, the main verb of a statement, together with its auxiliaries, represents the statement's predicate. A predicate has a **valence** that determines the number and types of **arguments** it requires to complete its meaning. **Adjuncts** can additionally relate to the predicate, but they are not necessary for completing the predicate's meaning. Adjuncts provide optional information, such as a time specification in a has-part statement. Therefore, each statement possesses a subject phrase as one of its arguments, and may possess one or more object phrases as further arguments and as additional adjuncts, depending on its underlying predicate.

We can say that each statement unit documents a particular proposition by relating a resource that is the subject argument of the predicate of the proposition to some literal or to some other resource, which is one of its object arguments or adjuncts (see Fig. 3).

Different types of statement units can be distinguished based on the type of predicate they model (e.g., *has-first-name*, *has-part*, *derives-from*, *has-value*, *gives-to, travels-by-from-to-on-the* statement unit class). The predicate of a statement, thereby, is not necessarily always binary as in the statement "*subject* has part *object*", but **n-ary [E1.1]** and thus cannot be modelled using a single triple such as in "*subject* gives *object_A* to *object_B*"; "*subject* travels by *object_A* from *object_B* to *object_C* on the *object_D*".

In addition and orthogonal to classifying statement units based on their underlying predicate, they can also be classified based on the category of statement into lexical, assertional, contingent (prototypical), and universal statement units:

1. **Lexical statement units** are statements about linguistic items such as terms, and comprise information such as the label of a given resource or its synonyms. Ontologies model lexical statements using annotation properties.

In the case the statement is not a lexical statement, all other categories of statements are distinguished by the type of resource their subject has. By introducing some-instance and every-instance resources as new types of representational entities in addition to named individuals (i.e., instances), classes, relations (i.e., RDF properties), and semantic units, statement units that are not lexical statement units can be classified as follows:

2. Any statement unit that has a **named-individual resource** as its subject is an **assertional statement unit**. The class-affiliation of named-individual resources is modelled via the property *type* (RDF:type).
3. Any statement unit that has a **some-instance resource** as its subject is a **contingent statement unit**. A some-instance resource (23) refers to some unspecified instance of a given class and its class-affiliation is modelled via the property **someInstanceOf**. When a some-instance resource is used in a statement unit, it is meant as '$\exists\ i \in C$' (with i = instance and C = class), reading 'there exists an instance *i* of class *C* for which holds'. Contingent statements can be used to state what prototypically but not necessarily holds for instances of a given class. Respective **prototypical statement units** form a subclass of contingent statement unit.
4. Any statement unit that has a **class** or an **every-instance resource** as its subject is a **universal statement unit**. An every-instance resource (23) refers to every instance of a given class, and its class-affiliation is modelled via the property **everyInstanceOf**. When an every-instance resource is used in a statement unit, it is meant as '$\forall\ i \in C$' (with i = instance and C = class), reading 'for every instance *i* of class *C* holds'.

Every given statement unit instantiates at least one predicate-based and one statement-category-based statement unit class. A weight measurement of a particular object, for example, when modeled in a knowledge graph, would be modelled using multiple triples, each of which would belong to the data graph of the same statement unit that would instantiate *weight measurement statement unit* as well as *assertional statement unit*, with its subject being a named individual that represents some material entity (BFO:0000040).

Semantic units also provide **formal semantics** for distinguishing assertion, contingent, prototypical, and universal statements **[I5.]** (see *user story 8*). Moreover, semantic units provide a formalism for representing universal statements without having to use blank nodes because they can be modelled and documented directly in a knowledge graph utilizing some-instance and every-instance resources. Consequently, and contrary to OWL axiomatic expressions, **universal statements can become part of the domain of discourse of a knowledge graph and are not restricted to class-axioms of ontologies**. Semantic units also provide formalisms for expressing **negation and cardinality restrictions as instance graphs** (i.e., ABox), without the need to include any blank nodes and with a **substantially simplified representation when compared to OWL axiomatic expressions**, especially when using dynamic labels or dynamic mind-map patterns (compare C with D and E in Fig. 4). These 'features' of semantic units substantially increase the overall expressivity of knowledge graphs and their cognitive interoperability.

### A) Observation
*This head has no antenna* (negated parthood assertion)

### B) Conventional OWL model

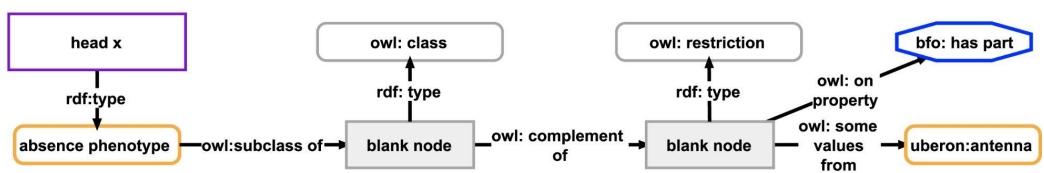

### C) Assertional statement units model

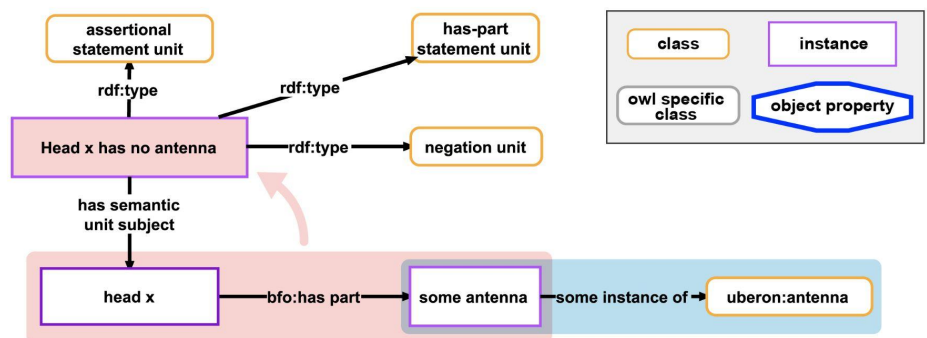

### D) Negation represented with dynamic label
Head x has no antenna

### F) Negation represented with dynamic mind-map pattern

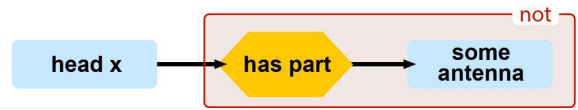

**Figure 4: Relation between an absence observation, its OWL-based modelling, its semantic units based modelling, and two alternative ways to represent the statement in a human-actionable way in the user interface (UI). A)** A human-readable statement about the observation that a given head has no antenna. **B)** The translation of the assertion from A) into an OWL expression mapped to RDF. Note how *absence phenotype* is defined as a set of relations of subclass and complement restrictions involving two blank nodes. **C)** The same statement can be modelled using two semantic units. One of them is modelling the has-part relation and negates it (red box with purple borders and its dynamic label, with its data graph in the red box without borders). It is therefore an instance of *has-part statement unit* as well as *assertional statement unit* and *negation unit*. The other semantic unit is an instance of *some-instance identification unit* and relates 'some antenna' to antenna (UBERON:0000972) via the property *some instance of*. Its data graph is shown in the blue box. Together, they model the observation from A). **E), F)** Utilizing the dynamic display patterns of the statement unit, the graph can be displayed in the UI of a knowledge graph application in a human-readable form, either as text through a

dynamic label E) or as a mind-map through a dynamic mind-map pattern F). *For reason of clarity of representation, the relation between 'head x' and head (UBERON:0000033) is not shown in B) and C).* [Figure taken from (14)]

By assigning to each statement unit class a specific graph model that specifies how statement units of that type must be modelled, for instance in the form of a SHACL shape specification (17) that ideally possesses its own identifier and includes constraints for its slots, semantic units also support **schematic interoperability** (14) **[F6.1;I7.1;I1.;I2.]**. By including a specification of the logical framework that has been applied (e.g., Description Logic or first-order-logic), **logical interoperability** (14) can be achieved for all statement units that are based on the same logical framework **[I4.]**.

Based on defined slots in a SHACL shape, one can map labels of specific resources to strings with corresponding variables to create **dynamic labels** for representing semantic units in visualizations of the knowledge graph in a UI **[E3.1;E3.2]** (see Fig. 4D). A dynamic label is generated by parsing the label from the subject resource, the labels from the various object resources, and the object literals of a given statement unit instance, and by creating from them a human-readable statement following a pre-defined template. For example, the template '*PERSON_subject* travels by *TRANSPORTATION_objectA* from *DEPARTURE_LOCATION_objectB* to *DESTINATION_LOCATION_objectC* on the *DATETIME_objectD*' with the input '*Anna | train | Berlin | Rome | 5th of August 2019*' would return the dynamic label '*Anna* travels by *train* from *Berlin* to *Rome* on the *5th of August 2019*'.

In the same way, **dynamic mind-map patterns** can be specified for the graphical representation of statement units in the UI of a knowledge graph **[E3.1]** (see Fig. 4E). Like dynamic labels, dynamic mind-map patterns parse the labels and literals from the subject and objects of statement units and create mind-map like graphical representations of the statements. As such, these representations do not necessarily have to follow the RDF syntax of *Subject-Predicate-Object* and n-ary predicates can be represented directly in the graph and do not have to be translated into sets of binary triple statements. Moreover, since dynamic mind-map patterns serve to increase the human-actionability of statement units, they do not have to show all information in the data graph of a statement unit, but can restrict themselves to only what is actually relevant to a human reader **[E3.]**. They can ignore all information that is additionally necessary for achieving the machine-actionability of the statement unit.

## Compound unit

A compound unit is a semantic unit that organizes statement units and other compound units into larger but still **semantically meaningful** data graphs (Fig. 5) **[I3.;E1.;E2.]**. Technically, a compound unit is thus a container of semantic units. It does not possess its own data graph—merging the data graphs of its associated semantic units constitutes the data graph of the compound unit. Analog to statement units, each compound unit is a semantic unit that is represented in the graph by its own node, possesses its own UPRI, and instantiates a corresponding compound unit ontology class **[E1.]**. Several subcategories of compound units can be distinguished (23):

1. An **item unit** is a compound unit that associates all statement units that share the same subject resource, which is then also the subject resource of the item unit **[E2.;E3.2]**. For example, all statement units with a resource representing a particular user in their subject position would be associated to a profile item unit of that user, including statement units about the username, their email address, homepage, etc. Item units represent semantic units at the **next coarser level of representational granularity, above the level of statement units**. Information contained in an item unit is well suited to be presented on a dedicated UI page or as a knowledge panel, since it includes all statements made with the respective resource in the subject position. Depending on the type of the subject resource, one can distinguish **instance item units** (i.e., named-individual resource) and **class item units** (i.e., class, some-instance, or every-instance resource).

2. An **item group unit** is a compound unit that comprises at least two item units that are semantically related to each other via at least one statement unit that shares its subject as the subject of one and one of its objects as the subject of another item unit **[E2.;E3.2]**. We refer to statement units that link an item unit to another semantic unit in this way as **item links**. All item units that are linked to each other via chains of such item links form an item group unit. The information of an item group unit is well suited to be presented as a collection of interlinked UI pages, with each page showing the contents of an item unit. For instance, when describing the parts of a material entity or the mixing steps of a recipe, we can describe each part or step in a corresponding item unit that is presented in the UI as a separate page, with one of its statement units linking to the item unit that describes one of its parts or sub-steps respectively, resulting in a tree of interlinked pages that can be browsed and explored by a user. Because item units can be linked to each other in both directions, one cannot unambiguously identify a specific subject of an item group unit. An item group unit is an **instance item group unit** if the subjects of all its associated semantic units are named-individual resources, and a **class item group unit** if they are all some-instance, every-instance, or class resources. Item group units represent semantic units at the **next coarser level of representational granularity above the level of item units**. With statement units, item units, and item group units, a knowledge graph can be organized into **five levels of representation granularity**, ranging from triples, statements, items, item groups, to the knowledge graph as a whole (Fig. 6) (14,23).

3. A **granularity tree unit** is a compound unit that comprises two or more statement units that in combination represent a **granularity tree** (39–41) **[E1.2;E3.2]**. Any type of statement unit that is based on a **partial order relation** such as parthood, class-subclass subsumption, or derives-from, can give rise to a corresponding **granularity perspective** (42–44), of which a particular granularity tree is an instance. The statement units associated with a granularity tree unit are linked to each other following a tree-hierarchy, where the object of one statement unit is the subject of another. Due to the resulting nested structure and its implicit directionality from root to leaves, one can specify the subject of a granularity tree unit to be the subject of the statement units that share their objects with the subjects, but not their subject with the objects of other statement units associated with the same granularity tree unit. Granularity tree units organize a knowledge graph somewhat orthogonally to the representational granularity mentioned above. An example of a granularity tree is the partonomy of the assembly parts of a car or that of the anatomical parts of a multicellular organism, but also the taxonomy of class terms in an ontology.

4. A **granular item group unit** is a compound unit that takes a granularity tree unit and includes the corresponding item unit for each resource in the tree **[E2.;E3.2]**. In other words, it comprises all item units whose subject resources are part of the same granularity tree unit. The subject of this granularity tree unit is then also the subject of the granular item group unit. If the data in a knowledge graph are highly connected, item group units tend to become very large. Granular item group units can provide additional structure to such large item group units that can be utilized for graph exploration (14).

5. A **context unit** is a compound unit that includes all semantic units for which merging their data graphs forms a connected graph, with every resource and triple being connected to every other resource via a series of triples that also belong to the data graph of the context unit, except for triples that have *isAbout* (IAO:0000136) as their property and which thus belong to an **is-about statement unit**. Is-about statement units relate an information artifact to an entity that the artifact contains information about. They are thus indices that point at positions in the graph where changes of the **reference frame** occur **[E1.3; E3.2]**. Thus, every is-about statement unit demarcates the border between two different frames of reference in a graph and with it two different context units. Similar to granular item group units, context units additionally structure item group units, and knowledge graphs in general, by organizing them into different context units and thus subgraphs that belong to different frames of reference. A special case of context unit is the set of lexical statement units in a graph. Since they comprise statements that are about linguistic items, they share the same **terminological frame of reference** and thus form a **terminological context unit**.

6. A **dataset unit** is a compound unit that is defined as an ordered collection of particular semantic units **[E2.;E3.2]**. Dataset units can be homogeneous, comprising only semantic units of the same type, or compilations of different types of statement and compound units. The UPRI of a dataset unit can be used for identifying and referencing any arbitrary collection of semantic units that is not covered by the other semantic unit categories. For instance, for referencing any dataset that has been imported to a knowledge graph or for referring to all semantic units contributed by a specific institute or created by a specific user.

7. **List units** are specific ordered and unordered lists of individual resources, where membership to the list is specified via a membership statement unit.

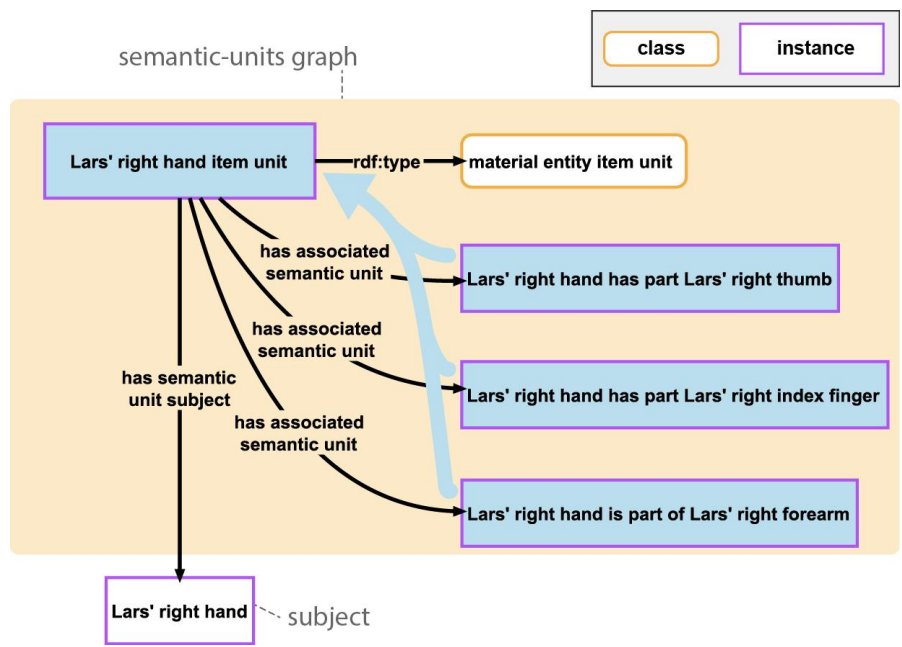

**Figure 5: Example of a compound unit** that comprises several statement units. Compound units possess only indirectly a data graph, through merging the data graphs of their associated statement units. The compound unit resource (here, *'Lars' right hand item unit'*), however, stands for this merged data graphs (indicated by the blue arrow). Compound units possess a semantic-units graph (shown in the peach-colored box), which documents the semantic units that are associated with it. [Figure taken from (23)]

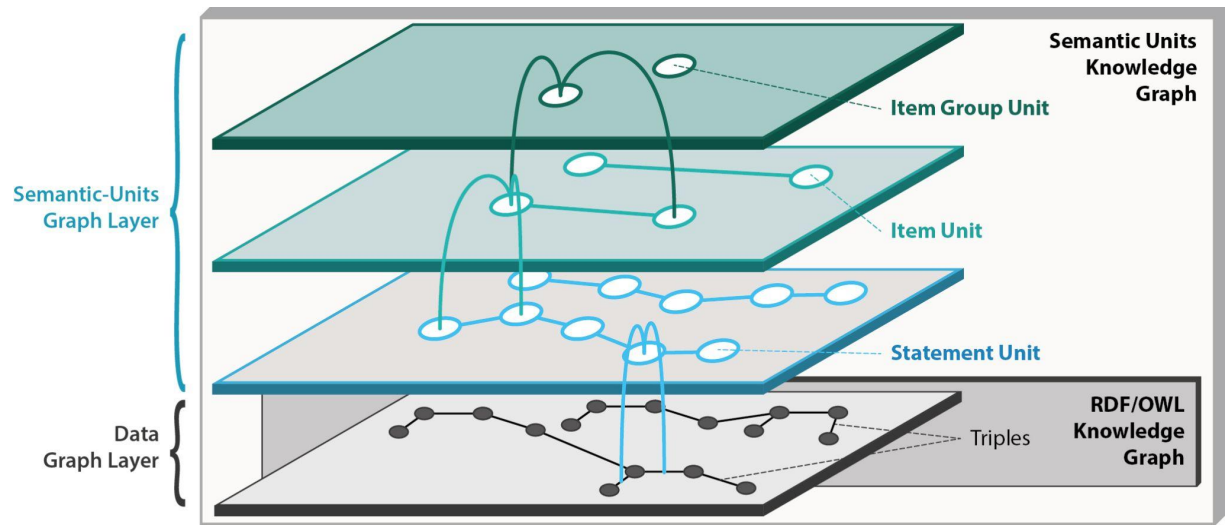

**Figure 6: Five levels of representational granularity.** The introduction of semantic units to a knowledge graph adds a semantic-units graph layer to its data graph layer, which adds a level of statement units, a level of item units, and a level of item group units to the level of triples and the level of the graph as a whole, resulting in five levels of representational granularity. [Figure taken from (23)]

## Question unit

In a **question unit** (14), an existing semantic unit class is taken as the **source** and is "rephrased" as a question that can be documented in the graph similar to any other semantic unit. This results in searches becoming objects in the knowledge graph. By using some-instance, every-instance, or class resources like variables for resource objects and subjects, and datatype specifications with (underspecified) values as variables for literal objects, and by interpreting them as wild-cards in the data graph of any question unit, respective queries will return matching subgraphs. A statement question unit thus either defines an instance of an assertional statement unit using named-individual resources and datatype specifications with fully specified values, which asks for a Boolean answer in return, or it defines a statement unit class using a combination of named-individual, some-instance, every-instance, and class resources and datatype specification with underspecified values, which asks for a list of statement units that instantiate the defined class in return. If we understood a statement unit as a string with variables (e.g., *Anna* travels by *train* from *Berlin* to *Rome* on the *5th of August 2019*) that is modeled in the knowledge graph using a specific graph pattern that is documented in the statement unit class as an n-ary predicate, we could use this same string for asking the knowledge graph about respective statements, either as a Boolean *true/false* question (Did *Anna* travel by *Train* from *Berlin* to *Rome* on the *5th of August 2019*?) or by replacing one of the variables with a some-instance, every-instance, or class resource for resource-subjects and objects or a datatype specification with an underspecified value for literal-objects. This way, one can underspecify a given subject- or object-position in a statement unit by, for instance, referring to *someCity* (ENVO:00000856) instead of *Berlin* (geonames:2950159), which then functions as an underspecified subject or object in the data graph of any question unit, and ask a question that assumes a list of single resources as the answer ('Did *somePerson* travel by *Train* from *Berlin* to *Rome* on the *5th of August 2019*?'), or by replacing more variables and assuming a list of statements as the answer (Did *somePerson* travel by *someTransportation* from *Berlin* to *Rome* on the *Date[year2019]* ?). Moreover, by reusing object resources from one statement question unit as the subject resource of another one, more complex question units can be built. Based on the SHACL shape specification of the statement unit classes involved, a query-builder can derive corresponding graph queries that could be answered by the knowledge graph. Question units together with a query-builder and corresponding UIs would not only provide an intuitively usable framework for specifying SPARQL or Cypher queries without having to know graph query languages (see *user story 3*) **[E3.4]**, but it could also support ontology development by documenting specific competency questions in the graph (45) in the form of question units.

## Representing different frames of reference and distinguishing the ontological, diagnostic, and discursive layer in knowledge graphs

Communication usually involves statements about different layers of information, i.e., the ontological, diagnostic, and discursive layers (cf. (23)). The **ontological layer** comprises statements about some reality and covers domain knowledge and empirical observations such as "*The melting point of lead is at 327.5 °C*". This ontological information provides answers to questions of the type *"What is it?"* and thus carries semantic meaning and increases the receiver's inferential lexical competence (46,47). The **diagnostic layer** provides information on how to identify an instance of a given concept or class. This often involves the specification of certain methods and a description of the expected results, such as "*Sodium sulfide kits indicate the presence of lead by turning black or grey*". The diagnostic information provides answers to *"How does it look?"* or *"How can I recognize/identify it?"* questions and thus carries information about the epistemological appearance of entities that helps to unambiguously reference them and increases the receiver's referential lexical competence (46,47). The **discursive layer** provides discursive contextual information, because we sometimes make statements about the statements made by others, such as "*author A asserts that the melting point of lead is at 327.5 °C*". Moreover, when we communicate, we frequently change the **frame of reference**. For instance, when we talk about a research activity that has a dataset as its output, which in turn contains facts about some entity such as a description of the anatomy of a multicellular organism, we change from the research-activity reference frame when describing the research activity to the research-subject reference frame when describing the multicellular organism. Because all these different types of statements can be modeled and classified using corresponding types of semantic units, representing all three layers within a knowledge graph together with possible interrelations and changes of reference frame is straightforward in a framework that uses semantic units (23).

## Comparing semantic units and ontologies

What ontologies are, in relation to kind terms and terminology, are semantic units in relation to statements and collections of statements. You always need more than a single term to communicate **meaning**—you need several terms put into context and relate them via predicates. In other words, it takes **statements**. Therefore, if we wanted to communicate meaning in a FAIR way, we would need both: terms and statements.

With their class definitions taking the form of OWL axioms, RDF/OWL-based ontologies are basically collections of interrelated terms, with each term functioning like a placeholder for one or more universal statements–the set of OWL axioms of that term. However, these class axioms only model what necessarily applies to every instance of the class and does not cover all statements that can be made for a given set of instances of that class.

We see a similarity between the structure of sentences defined by a language's syntax and grammar and the structure of data and metadata schemata (Fig. 7). The **positions** of terms in a natural language sentence take specific grammatical and **semantic roles** that significantly contribute to the statement's meaning (48–50). The same applies to the positions of values or nodes in a tabular data schema or a knowledge graph, respectively. Therefore, we can understand data schemata to be attempts of translating the structure of natural language sentences into a machine-actionable data structure. Semantic units provide a formal representation and formal semantics for FAIR statements and collections thereof by providing such a structure, whereas ontologies provide formal semantics for FAIR terms.

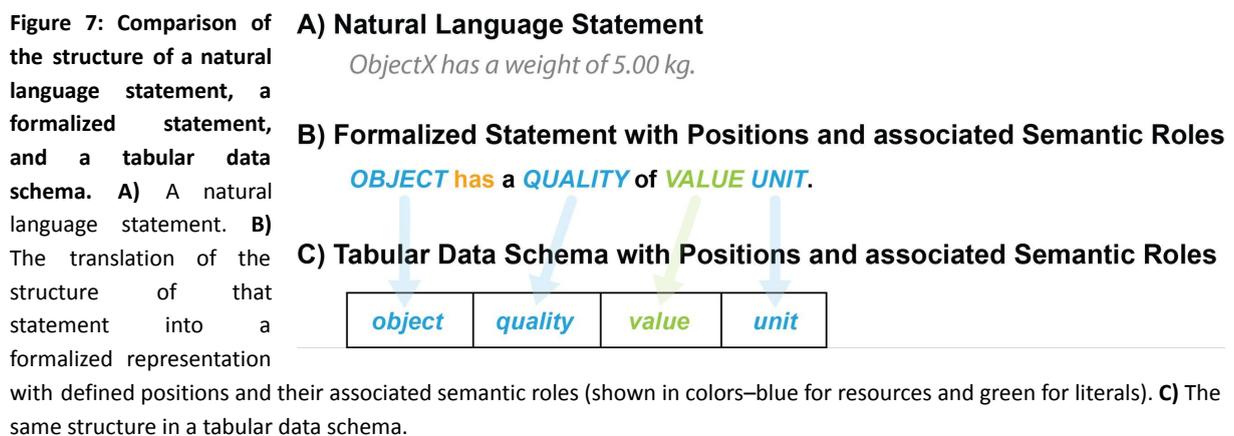

**Figure 7: Comparison of the structure of a natural language statement, a formalized statement, and a tabular data schema. A)** A natural language statement. **B)** The translation of the structure of that statement into a formalized representation with defined positions and their associated semantic roles (shown in colors–blue for resources and green for literals). **C)** The same structure in a tabular data schema.

The structure of an RDF triple cannot provide the analog to a sentence structure, because its predicate does not necessarily map to the predicates of natural language statements. As we have discussed above, many predicates are n-ary, whereas RDF/OWL properties—at least technically—are necessarily binary. Therefore, ontology properties do not map in a one-to-one relation to natural language predicates, and we often have to model natural language statements using several triples. Consequently, if we want to create FAIR data and metadata, and we realize that data and metadata always takes the form of statements, we not only have to use controlled vocabularies for the terms we need to make the statements, but also controlled statement types and associate with them respective graph patterns with the corresponding positions and semantic roles. Semantic units provide such controlled statement types in the form of semantic unit classes, with associated predicates and shape specifications, the latter of which specify how to model the statements in RDF.

Establishing interoperability across terms (i.e., **terminological interoperability** (14)) requires ontology terms that share the same meaning and referent to be mapped across different ontologies. Mappings between ontology terms are **homogeneous definition mappings** (51), where there is only one vocabulary element on the left side, which needs to be mapped, and several others on the right side of the definition, which do not have to be mapped[1].

Establishing interoperability across statements (i.e., **schematic interoperability** (14)), on the other hand, requires the alignment of different positions and their associated semantic roles across different sentence structures of the same type of statement. In other words, the subject, predicate, and object slots of different data and metadata schemata that model the same type of statement must be aligned, and their terms be mapped across controlled vocabularies. This requires the schemata to be formally specified in the first place, e.g., in the form of graph patterns specified as SHACL shapes. Then, shapes that share the same statement-type referent would have to be aligned and mapped. These are **ontology pattern alignments** (i.e., TBox alignments) (51) or **ABox alignments**, where several vocabulary elements must first be aligned and then mapped in **schema crosswalks**.

---

[1] However, in order to establish full terminological interoperability between two ontology terms that share the same meaning and referent(s), the terms used in their ontological definitions (i.e., their class axioms) must also be mapped if they share the same meaning and referen(s), because the meaning of a term is communicated via its definition, which takes the form of a statement.

So we can summarize: Since terms and statement structures are required for reliably communicating meaning, we need controlled vocabularies (i.e., **ontologies**) for **FAIR terms** and **schema crosswalks** for **FAIR statements** that provide alignments of term-positions across schemata to achieve semantic interoperability of data and metadata. The latter can be provided with the help of semantic units.

## Semantic units and FAIREr knowledge graphs

The classification of different types of semantic units with their human-readable content descriptions, together with additional information from their subject and, in case of compound units, their associated semantic units, provide information that can be utilized to assist users in exploring the graph and identifying the portions of the graph that interest them.

The modular and recursive organization of semantic units also supports making **statements about statements [E3.3]**. Because each semantic unit is represented in the semantic-units graph layer with its own node and own UPRI, you can make assertions about them using the unit's UPRI. Such an assertion can either be a statement about the contents of a particular semantic unit, such as specifying its certainty or confidence level or indicating its time-validity, or a statement that relates two semantic units to each other, such as specifying that the empirical observation expressed in one semantic unit contradicts the causal hypothesis expressed in another semantic unit. By structuring a knowledge graph into identifiable and semantically meaningful units, UIs can be developed that allow in very intuitive ways to make statements about statements. The development of such UIs and their accompanying tools and services thereby benefits from semantic units being implemented following a clear and straightforward schema.

Semantic units can be used for graph-alignment, subgraph-matching, knowledge graph profiling, and for managing access restrictions to sensitive data **[A1.3]**. Based on semantic units, graph patterns for modelling absence statements, negations, and cardinality restrictions can be specified for the use within instance-based graphs (ABox expressions). Moreover, organizing a FAIR knowledge graph into semantic units supports the separation of ontological, diagnostic, and discursive information. Finally, structuring and organizing a FAIR knowledge graph into semantic units contributes in general to **increasing the cognitive interoperability** of its data and metadata, turning the graph into a **FAIREr knowledge graph** (FAIR + <u>E</u>xplorability <u>r</u>aised), i.e., a FAIR knowledge graph that can be **intuitively explored** by human users, by reducing the graph's complexity to what interests them at the moment. If you want to read more about the concept of semantic units and the FAIREr Guiding Principles, we ask you to refer to (14,23).

# The Knowledge Graph Building Block (KGBB) Framework

Partitioning the data graph layer of a knowledge graph into statement units, structuring it further with compound units, and documenting questions (i.e., queries) as question units, provides the foundation for our concept of a knowledge graph that is driven by Knowledge Graph Building Blocks. **A Knowledge Graph Building Block (KGBB) is a small information module for knowledge-processing that is associated with a particular type of semantic unit** (see also *knowledge graph cells* (52) for a first discussion of a similar idea). Based on the distinction of statement units, compound units, and question units, we distinguish statement KGBBs, compound KGBBs, and question KGBBs as top-categories of KGBBs.

A KGBB only provides information, including the storage model and display patterns for its associated semantic unit type. To bring this information into action, a **Knowledge Graph Building Block Application Engine (short: KGBB-Engine)** is required that processes the information and applies it to a corresponding application logic. The KGBB-Engine uses the information from various KGBBs and communicates through corresponding APIs with the persistence-layer and the presentation-layer of a knowledge graph application. The KGBB-Engine also includes a **KGBB-Query-Builder** that translates form-based input from users that is based on and structured by the storage models provided by the KGBBs into SPARQL, Cypher, or MySQL queries **[A1.;A1.1]**, so that users of KGBB-driven knowledge graphs do not have to know (graph) query languages for writing queries **[E3.4]**. Additionally, the KGBB-Engine enables **versioning** of semantic units, **automatic provenance tracking [F1.-F4.;F6.1/I7.1]**, and ideally also provides a **detailed editing history for each semantic unit**.

Each KGBB provides information required by the KGBB-Engine for managing data and metadata in the knowledge graph that belong to its corresponding type of semantic unit. Possible interactions between all the KGBBs of a knowledge graph are specified in the KGBB specification graph. The Engine manages the knowledge graph application, and the specification graph defines its **possible data- and proposition-space** by limiting it to combinations of statements and contexts that meet the scope of the knowledge graph. To do so, a KGBB and its possible interactions with other KGBBs must be specified for each type of semantic unit used in the knowledge graph.

In addition to a collection of relevant KGBBs and the KGBB-Engine, a **KGBB-Editor** is required that enables domain experts who do not necessarily have a background in semantics, programming, or graph query languages to describe new KGBBs, with their **display patterns** and their connections and potential interactions to other KGBBs, alongside with their corresponding semantic unit classes. The KGBB-Editor provides a visual interface that can also be used to specify additional **access and import models** for each KGBB at any point in time, allowing for adapting to newly emerging data formats and data standards.

Ultimately, the idea for KGBBs is that each KGBB interacts with the KGBB-Engine to function independently of other KGBBs and that KGBBs can be stored and made openly available in a GitHub **KGBB-Repository**, so that they can be reused by others **[F6.1/I7.1]**. Each KGBB must be documented as an ontology class **[F1.;I1.;I2.]** that inherits all of its properties to its subclasses. An existing KGBB can be loaded to the KGBB-Editor, together with its corresponding semantic unit class, to add further constraints to it in the form of additional properties, and then, together with the new semantic unit class, saved as a subclass of the loaded KGBB and the loaded semantic unit class, respectively. Together, all KGBBs then form a taxonomy of KGBBs with their corresponding taxonomy of semantic unit classes. Based on this taxonomy, one can then use the KGBB-Editor to describe a particular

FAIREr knowledge graph application by taking instances of these KGBB classes and relate them to each other, specifying their potential interactions in a **KGBB specification graph**.

A **KGBB-Frontend** provides a basic UI so that users can access the contents of a KGBB-driven FAIREr knowledge graph, search for specific information, browse and explore its contents in textual as well as in mind-map like graphical displays, thereby benefitting from the modular organization of the graph achieved via semantic units. The KGBB-Frontend also provides UI tools that support exploring the graph by enabling zooming, filtering, and providing contextual information, thus supporting different **visual information seeking strategies**, by utilizing the structure and organization of the graph into statement, item, item group, granularity tree, granular group, question, and context units, thereby increasing the **cognitive interoperability** of its data and metadata and meeting the criteria of the **human Explorability Principles of FAIREr [E1.-E3.]** (for a detailed discussion, see (14)).

By following a **hexagonal architecture** approach (53,54), **KGBB-Function** modules can be developed that add further data analysis, processing, and visualization functions to a specific KGBB or to a KGBB-driven knowledge graph application as a whole.

The KGBBs, KGBB specification graphs, the KGBB-Engine, KGBB-Frontend and KGBB-Editor, the KGBB-Function modules, and the KGBB-Repository provide the different components of what we call the **KGBB Framework**. The KGBB Framework allows domain experts to create their own FAIREr knowledge graph applications in a straight-forward way by reusing and combining available KGBBs and creating new KGBBs as they need. In the following, we provide a detailed description of that framework.

## The general semantic-units graph storage model

To meet the criteria of the FAIREr Guiding Principles that relate to schematic interoperability **[F6.1-2;I7.1-2]** (14), all data and metadata in a knowledge graph that belong to the same semantic unit must be modelled in the same way using the same schema. In other words, all data and metadata of the same type that are added to the knowledge graph through user input or data import and all automatic provenance-tracking procedures should apply the same storage model, respectively. The overall storage model for a given semantic unit must thus cover both its semantic-units graph storage model part and its data graph part (cf. Fig. 3). These two parts of the overall semantic unit storage model can be specified in the respective KGBB independently of each other, each in its own YAML file, following the notation of LinkML, which can be translated into for instance SHACL shapes.

Each YAML file specifies slots/attributes with constraints that can be utilized by the frontend and the KGBB-Engine for validation and input-control purposes. Constraints include specifications of datatypes as ranges such as '*xsd:float*', max-min, or pattern constraints, and class restrictions for resources.

Since each type of semantic unit has its own associated KGBB, the taxonomy of class-subclass relations between different types of semantic units maps to a corresponding taxonomy of KGBBs. The storage model specifications of each KGBB class is inherited along these class-subclass paths to its KGBB subclasses. This allows for efficient storage-model-management, where changes in the storage model at a parent KGBB will be automatically passed on to all its child KGBBs.

The class *KGBB* is at the root of the taxonomy of KGBBs. A **general semantic-units graph storage model [I1.;I2.;R1.]** is specified for this KGBB that covers the semantic-units graph part (cf. peach-colored box in Fig. 3) that belongs to a semantic unit and inherits it to all other KGBB classes.

This storage model covers mainly provenance metadata relating to the semantic unit resource itself—it does not cover the data graph of a semantic unit, because only statement units require a storage model for their data graphs. Compound units have no constraints that add to their data graph, since their data graph results from merging the data graphs of their associated statement units. Consequently, statement KGBBs require, in addition to a storage model for their semantic-units graph part, also a storage model for their data graph part (cf. blue-colored box without borders in Fig. 3), which we discuss further below.

Despite inheriting the storage models from their parent KGBBs, any statement or compound KGBB can complement the model by adding further constraints that narrow down what is a valid subject and what are valid objects to meet the requirements that are specific to the particular type of semantic unit they manage. These added constraints are then inherited down to their child KGBBs, respectively.

The specification of the storage model for the semantic-units graph for *KGBB* includes the following slots (*ranges*) (Fig. 8):

1. **Label** (*string*): a human-readable label that indicates what type of information this semantic unit carries, such as 'weight measurement'. This label is specified via the property *label* (RDFS:label).
2. **Type** (*semantic unit class UPRI*): when creating a new instance of a specific semantic unit class, the KGBB-Engine automatically classifies it in reference to that class, the UPRI of which is specified by its managing KGBB. If the KGBB manages a statement unit, the KGBB-Engine will classify it as a type of predicate-based statement unit class **[F1.;A1.;E1.]** and additionally as either lexical or, depending on its type of subject-resource, as assertional, contingent (or even prototypical), or universal statement unit **[I5.]**. Thus, every given statement unit is always classified as an instance of at least two statement unit classes. These instantiation relations are specified via the property *type* (RDF:type).
3. **HasSemanticUnitSubject** (*UPRI of subject resource*), optional: specifies the resource that takes the subject position in the proposition modeled in this semantic unit. The property *\*hasSemanticUnitSubject\** specifies the UPRI of that subject resource. Every statement unit, and many but not every compound unit, have a subject.
4. **KGBB_URI** (*KGBB UPRI*): each semantic unit indicates which KGBB is managing its data via the property *\*KGBB_URI\** and the corresponding UPRI as its value.
5. **Semantic unit provenance [F2.;F3;F4;A1.;R1.]**
   a. **Creator** (*user UPRI*): information on who created this semantic unit using the property *\*creator\**.
   b. **CreationDate** (*dataTime*): information on when this semantic unit was created using the property *\*creationDate\**.
   c. **CreatedWithApplication** (*UPRI of knowledge graph application*): identifies the application with which the data has been created, using the property *\*createdWithApplication\**.
   d. **ImportedFrom** (*UPRI of imported dataset*), optional: identifies the dataset from which the content of this semantic unit has been imported from, using the property *\*importedFrom\**.
   e. **ImportDate** (*dataTim*), optional: information on when the dataset has been imported, using the property *\*curationDate\**.

       f. **Curator** (*user UPRI/software UPRI*), optional: information on who curated this semantic unit, using the property **curator*. The curator may be a user with a specific curator role or a software agent.

       g. **CurationDate** (*dataTim*), optional: information on when this semantic unit has been curated, using the property **curationDate*.

       h. **DeletedBy** (*user UPRI*), optional: information on who deleted this semantic unit, using the property **deletedBy*. This is only added when the semantic unit is deleted. The semantic unit resource itself is not deleted and tracks this information, thus keeping the metadata of the semantic unit still accessible **[A2.]**.

       i. **DeletionDate** (*dataTime*), optional: information on when this semantic unit was deleted using the property **deletionDate*. This is only added when the semantic unit is deleted.

6. **Data production process metadata [F2.;F3;F4;A1.;R1.]**

       a. **DataProductionMetadata** (*UPRI of semantic unit describing the data production process*), optional: in some cases, the semantic unit contains information that represents the output of a specific process (e.g., a measurement process, using a specific method and specific instruments), the metadata of which is critical for evaluating the data graph of the semantic unit. Since descriptions of processes typically comprise several statement units and thus represent item or even item group units in their own right, such metadata must be documented with their own semantic units, to which the semantic unit containing the data can refer using the property **dataProductionMetadata* **[F6.1/I7.1]**.

7. **VersionID** (version UPRI), optional: information that indicates whether the semantic unit belongs to a specific documented version of itself or of some other semantic unit to which it is associated. This is tracked via the property **versionID*. Since a given semantic unit can be associated with more than one other semantic unit, this slot can be realized multiple times.

8. **DatasetUnitID** (dataset unit UPRI), optional: information that indicates whether the semantic unit belongs to a specific documented dataset unit. This is tracked via the property **datasetUnitID*. Since a given semantic unit can be associated with more than one dataset unit, this slot can be realized multiple times.

9. **Editable** (*Boolean*): information that indicates whether the semantic unit should be locked or be open for editing. If no editing is allowed, the value is '*false*'. The Boolean property **editable* indicates whether the semantic unit is open for being edited.

In addition to the metadata that is directly stored with each semantic unit, the KGBB-Engine dynamically fetches the list of contributors from the list of creators of any object-position instance (for object-position instances, see further below) of the semantic unit and of all of its associated semantic units. It also fetches the last-update metadata from the latest creation-date of the same list of object-position instances and associated semantic units as well as their legal statement metadata specifying their copyright licenses, any access rights restrictions, and information about the logical framework they used. The engine makes this information available to the presentation-layer as '*contributors:*', '*last updated on:*', '*copyright license:*'[2], '*access restriction:*'[3], and '*logical framework:*'[4].

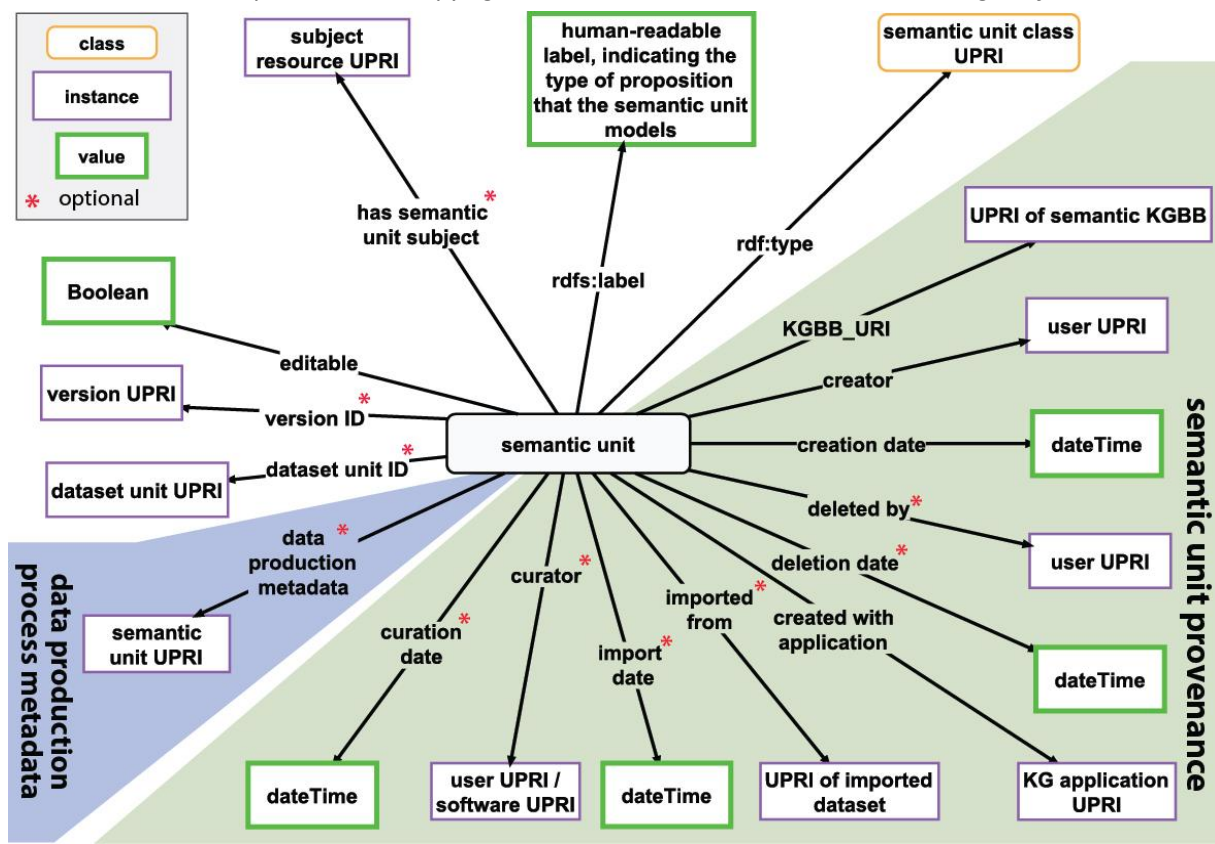

**Figure 8: The general semantic-units graph storage model specified for \*KGBB\*.** The graph shows the information tracked within the semantic-units graph of every semantic unit. Different types of metadata tracked with the semantic unit are shown with their own background color. Some relations are optional, depending on context, and are indicated with a red asterisk, others can be realized multiple times. Some compound units have no subject resource and the respective relation is omitted. This model is inherited from \*KGBB\* down to all other KGBB classes, but can be complemented with additional constraints to meet the requirements of any specific KGBB class.

## Statement KGBB

Statement units represent the most fundamental type of semantic unit and instantiate corresponding statement unit classes. Each statement unit class that is defined in reference to the type of predicate that underlies its propositions (i.e., predicate-based criterion) must have its own statement KGBB class specified **[F1;F4;E1.]**. A statement KGBB class provides the following information:

1) The specification of a **data model** for storing the subgraphs belonging to a specific type of statement unit in a machine-actionable, standardized, and FAIRer way. The model is based on the **general semantic-units graph storage model** described above, which can be extended with additional specifications required by the corresponding type of statement unit, and is complemented by a **storage model for the data graph** that is specific for each statement KGBB. The models and extensions are specified in LinkML notation and provide input-constraints that can be utilized by the frontend for input-control and by the KGBB-Engine for input-validation **[F6.1/I7.1;I1.;I2.;R1.;E1.1]**.

---

[2] Using as its license the most restrictive copyright license from all of its associated semantic units.

[3] This information is used to indicate from which type of semantic unit information is present but cannot be accessed due to specific access restrictions, such as because it is person-related information **[A1.3]** or because it contains geo-references for occurrence data about an endangered species.

[4] This specifies the logical framework that can be applied to the data, e.g., description logics with OWL or first-order-logic with common logic interchange format (CLIF) **[I4.]**.

2) One or more **display templates** (i.e., data views) that are used for identifying data points from the machine-actionable data graph that are of interest for a human reader. Display templates are used for translating the information in the data graph of a statement unit into a human-readable textual statement or mind-map-like graphical representation that can be presented in a user interface (UI) in the form of a dynamic label or a dynamic mind-map pattern, respectively, therewith **decoupling data display from data storage [E3.;E3.1]**.

3) Optionally, one or more **access templates**. With an access template, the KGBB-Engine can automatically transform data that complies with the storage pattern into data representations that comply with a defined format and pattern for accessing the data graph of a statement unit following different graph patterns and for exporting data and metadata in various formats following different established standards, including, e.g., RDF/OWL, CSV, JSON, and Python. Consequently, **data storage is not only decoupled from data access, but the data's and metadata's fitness to be readily usable by other applications is increased and thus the barrier for software developers to utilize knowledge graphs reduced [F6.2/I7.2]**.

4) Optionally, one or more **import templates** for supporting the import of legacy data into a knowledge graph.

5) In addition to the predicate-based type specification, each statement unit must also be specified in terms of whether it is a lexical, assertional, contingent, prototypical, or universal statement and the statement KGBB together with the KGBB-Engine must provide the corresponding formal semantics for this type of statement **[I5.]**.

**Statement KGBB storage models**

**Statement-unit-specific extension to the general semantic-units graph storage model**
The data graphs of two statement units can be connected with each other when one unit shares one of its object resources with the subject resource of the other statement unit (see also *subject-object connection* further below), for example when forming item links that link two item units in an item group unit. Therefore, and because only statement units possess their own data graph and thus require additional metadata, the general semantic-units graph storage model inherited from *KGBB* needs to be extended to include the following additional slots (ranges):

10. **ObjectDescribedBySemanticUnit** (*UPRI of semantic unit*), optional: if the statement unit connects semantic unit *A* with semantic unit *B* by sharing its subject resource with *A* while its object resource is the subject of *B*, then the statement unit relates to semantic unit *B* via the property *objectDescribedBySemanticUnit*.

11. **BasedOnGraphPattern** (*data graph storage model UPRI*): for each statement unit, information must be provided about the underlying data schema or graph pattern that has been used for modelling the data graph of this statement unit. The property *basedOnGraphPattern* links to the UPRI of the data graph storage model of the statement unit. This information is important for evaluating which parts of a knowledge graph are comparable and interoperable, and thus contributes to the data's FAIRness (see *schematic interoperability* (14)) **[F2.;F3;F4;F6.1/I7.1;A1.;I1.;I2.;E1.1]**.

12. **HasConstraintNode** (*UPRI of constraint node*)[5]: information that enables identifying the constraints that applied to a specific input event and thus a specific object-position instance. The information is relevant for editing purposes and is context-sensitive. This relation is specified via the property *hasConstraintNode*. The constraint node resource, in turn, links (i) to information about the constraint itself via the property *hasConstraint* and a LinkML constraint expression as its value, and (ii) to information about the object-position to which the constraint applies via the property *appliesToObjectPosition* and the UPRI of the corresponding object-position class as its value.

13. **Legal statements' metadata**
    a. **License** (*copyright license UPRI*): information about the copyright license for the contents of the statement unit's data graph using the property *license* **[R1.1]**.
    b. **AccessRestrictedTo** (*user-group or user-role UPRI*), optional: in case the knowledge graph application distinguishes different user-groups or roles, a specification of who is allowed to access the statement unit's data graph can be specified using the property *accessRestrictedTo* **[A1.3]**.

14. **LogicalFramework** (*UPRI of a resource representing a logical framework*): information about the logical framework (e.g., description logics, first-order-logic) under which this statement has been modelled via the property *logicalFramework* **[I4.]**.

15. **Integrity and truthfullness**
    a. **ConfidenceLevel** (*UPRI of a resource representing a particular confidence level*), optional: information about the level of confidence or degree of certainty of the statement via the property *hasConfidenceLevel*. Specifying the level of confidence of a statement is very important, especially in the scholarly context (55,56), and lack of it can cause problems such as citation distortion (57).
    b. **Validity period**[6]
        i. **ValidityStartDate** (*dataTime*), optional: information on the starting date for which the statement is valid using the property *validityStartDate*.
        ii. **ValidityEndDate** (*dataTime*), optional: information on the ending date for which the statement is valid using the property *validityEndDate*.

16. **Reference** (*UPRI of a resource representing a particular reference*), optional, multiple: information about a particular publication that is the source of the evidence for the statement using the property *reference*.

In addition to the metadata that is directly stored with each statement unit, the KGBB-Engine dynamically fetches the imported-from provenance from all object-position instances (see [General object-related data graph storage models](#)) that are part of the current version of the statement unit and makes the information available to the presentation-layer as '*contains data imported from:*'. As described above, the KGBB-Engine does the same with the last-update and contributor provenance for the statement unit.

---

[5] The reason, why this information is required, is discussed later in this paper.

[6] In many cases, particular statements are only valid, i.e., true, for a specific time period. For example, the statement '*George W. Bush* role *U.S. President*' has the validity start date of January 20th, 2001, and the validity end date of January 20th, 2009.

Moreover, we extend the concept of a statement unit from (23) to include a specification of some logical properties of the statement's predicate, i.e., whether its predicate is **transitive**, **symmetric**, or **asymmetric**. This information is stored with the statement unit class alongside a **human-readable definition of the meaning of the statement's predicate**. The specification of the logical properties is not only required for identifying predicates that model partial order relations and thus types of statement units that potentially give rise to granularity trees, but can also be leveraged for rule-based functions and thus KGBB-Functions.

**A generic data graph storage model for statement KGBBs**

Every triple in the data graph layer of a knowledge graph is part of the data graph of exactly one statement unit. Consequently, statement units contain the actual data of the knowledge graph, including all input from users and data imports. By understanding statement units as the most basic type of semantic units, with all other semantic units referring to them, the KGBB Framework also takes into account that propositions are the main units of communication for humans, and thus organizes knowledge graphs in a way that is compatible with human cognition. Therefore, and to guarantee the FAIRness of all data and metadata in a knowledge graph, every statement KGBB class must provide the specification of a **data graph storage model [F6.1/I7.1]** for its corresponding type of statement unit. Information from these two parts—from the semantic-units graph model and the data graph model of a statement unit—is then used by the KGBB-Engine as input for generic query patterns for CRUD operations, i.e., for creating (=writing), reading (= searching), updating, and deleting the corresponding type of statement units in the knowledge graph using the KGBB-Query-Builder. The data graph model not only defines the graph pattern, but also any input constraints that can be used for input-control. Each statement KGBB provides for its associated statement unit class its own storage model that can be referenced unambiguously as metadata for each of its statement instances **[F6.1/I7.1]**.

The data graph storage model of each statement KGBB provides the model for storing a specific type of statement. In other words, we have to find a good way to model propositions. When we started thinking about such models, we decided to deliberately follow a minimalistic approach, preferring **lean models** over complex models, with the goal to **reduce the overall modelling complexity**. We also wanted to be **as close as possible to the structure of natural language statements** to also address the need to make data and metadata FAIREr by increasing their **cognitive interoperability** (14) and thus **reduce the overall modelling burden**.

The predicate-argument-structure of statements in natural language is well known and intuitive to most users. We wanted to harness this knowledge in the context of semantic modelling so that specifying new statement KGBBs with their corresponding statement unit classes and associated storage models becomes a task that is straight forward and does not require any background in semantics because **the semantic modelling step can be automated**. This could only be achieved, if we came up with a **very generic structure for the model**. That structure must be applicable to every type of statement, independent of its **n-aryness**. To be lean, it should only store what is necessary to recover the proposition's meaning, which should be the same as storing only the information that a user has to provide to create a new statement. For example, instead of creating the entire subgraph shown in Figure 2, top, for a weight measurement statement, it should be sufficient to only store the resources for the measured object and quality together with the value and unit, with the main focus on **always being able to reconstruct the original user input or data import for a given statement** **unit** so that it can be used and mapped to various access formats and syntaxes. Another criterion that the model had to meet was that it must facilitate seamlessly deriving queries from it.

In other words, the goal was to develop an overall generic structure from which the data graph storage models for all possible statement KGBB classes could be derived, so that the **KGBB-Editor can automatically specify the data graph storage model for a newly defined statement KGBB class based on some general information provided by a domain expert** who is not necessarily experienced in semantics.

In the search for such a generic structure, we realized that when one compares the propositions of different types of statement units on a very abstract level, two independent general characteristics can be distinguished, which can be used as general parameters for defining a generic modelling structure:

**1) Syntactic positions, semantic roles, thematic labels—number of arguments and adjuncts.** Every statement unit models a smallest, independent proposition that is semantically meaningful for a human reader (23). Different types of propositions can be differentiated based on their underlying predicates (i.e., relations), resulting in the already mentioned predicate-based classification of statement units. As described above, following the **predicate-argument-structure**, each predicate has a **valence** that determines the number and types of **arguments** that it requires to complete its meaning. For instance, a has-part statement requires two arguments—the subject representing the entity that possesses some part and an object representing the part—resulting in the proposition 'SUBJECT *has part* OBJECT'. Without the specification of a subject and an object, the has-part predicate cannot complete its meaning. **Adjuncts** can additionally relate to the predicate, however, they are not necessary for completing the predicate's meaning but provide optional information, such as a timestamp specification in the has-part statement, e.g., 'SUBJECT *has part* OBJECT *at* TIMESTAMP'. Subject and object phrases are the most frequently occurring arguments and adjuncts. So we can say that each statement unit documents a particular proposition by relating a resource that is the subject argument of the predicate of the proposition to one or more literals or resources which are the object arguments and adjuncts of the predicate. The subject argument of a proposition of a statement unit is what we call the **subject** of the statement unit and the object argument(s) and adjunct(s) its **required and optional object(s)**. Every statement unit has one such subject and one or more objects.

According to this model, we can distinguish different types of predicates (and thus different types of relations on which each statement unit is based) by the overall number of subjects and objects they relate within a given proposition. A statement like '*Sarah* met *Bob*' is a proposition with a **binary relation**, where we refer to '*Sarah*' as the statement's subject and '*Bob*' as its object. If we add a date to the proposition, such as in '*Sarah* met *Bob* on *4th of July 2021*', it becomes a **ternary relation** with two objects[7]. When also including the location, it even becomes a **quaternary relation**, as in '*Sarah* met *Bob* on *4th of July 2021* in *New York City*'. This is in principle open-ended, although it is limited to which dimensionality human readers are still capable of comprehending n-ary relations[8]. Independent of that limitation, based on the number of subjects and objects their underlying predicates put into relation, propositions can be differentiated into binary propositions, ternary propositions, quaternary propositions, etc. Moreover, **based on the distinction of arguments and**

---

[7] Many properties of the Basic Formal Ontology 2.0 are actually ternary relations, as they are time-dependent (58). For instance, "*subject* located in *object_A* at *t*".

[8] Humans can hold only 5-9 items in memory (59).

**adjuncts, we can distinguish objects that are necessary and thus required for completing the meaning of the statement's predicate from objects that are optional**.

Each argument and adjunct of a predicate can be understood to have a specific **position** with a particular **semantic role** in the syntax of a statement. This relates to Kipper et al.'s (48) verb lexicon [VerbNet](), where they extend Levin verb classes (49) to include abstract representations of syntactic frames for each class, with explicit correspondences between **syntactic positions** (i.e., positions in a syntax tree) and the **semantic roles** (i.e., thematic roles sensu (48)) these positions express (Fig. 9). Each verb class lists semantic roles that the predicate-argument-structure of the class's instances allow and basic syntactic frames that the instances share.

The list of arguments of a predicate-argument-structure can be described by a list of **thematic labels** taken from a set of pre-defined possible labels (e.g., Agent, Patient, Theme, etc.), and the syntactic frames are represented by an ordered sequence of such thematic labels. The thematic labels function like descriptors of semantic roles that are mapped onto positions in a given syntactic frame (48).

In [PropBank](), Palmer et al. (50) define semantic roles on the level of individual verbs by numbering the verb's arguments, with the first argument generally taking the semantic role of an **agent** and the second argument typically but not necessarily that of a **patient** or a **theme**. For higher numbered arguments, no consistently defined roles are specified. In our approach, we define what Palmer et al. call a **frameset**, which is a specified set of semantic roles, and we define a single storage pattern that provides the associated syntactic frame, with their first argument mapping to our subject-position.

Considering that every datum is a structured representation of a proposition, a data schema is nothing else but a formalized representation of a frameset with, for instance, a table defining syntactic positions with associated semantic roles for a given type of proposition. In many cases, it represents a translation of the human-actionable syntactic structure of a natural language statement into a machine-actionable tabular structure (see also Fig. 7). Here, we need a generic storage model that supports reconstructing the actual proposition underlying a datum and that at the same time documents it using a formalized structure to guarantee human- and machine-actionability of the data. Abstracting statements to their syntactic positions and associated semantic roles seems to provide a promising solution for meeting both requirements.

Since we do not need to be able to represent the full expressivity of natural language statements, we can restrict ourselves to statements with a rather simple structure of subject, transitive verb or predicate, and a number of objects. We are not considering passive forms, no tenses (the latter can be dealt with by specifying the validity period on the statement unit level), and we do not need to distinguish different possible syntactic alternations in which a verb or predicate can express its arguments. **Our storage model is thus similar to a highly simplified frameset that specifies a subject-position and a number of required and optional object-positions, each with its associated semantic role in the form of a thematic label and a corresponding constraint-specification**. Its structure is an abstraction of the structure of a syntax tree and thus a tree of phrases that is used to translate a web of ideas in the mind of a sender into a string of words that can be understood by a receiver to translate it back to the web of ideas (60). In other words, it is based on a structure that is interoperable with the cognitive conditions of humans.

### A) Natural Language Statement

*Anna travels by train from Berlin to Rome on the 5th of August 2019.*

### B) Syntax Tree with Positions and associated Semantic Roles

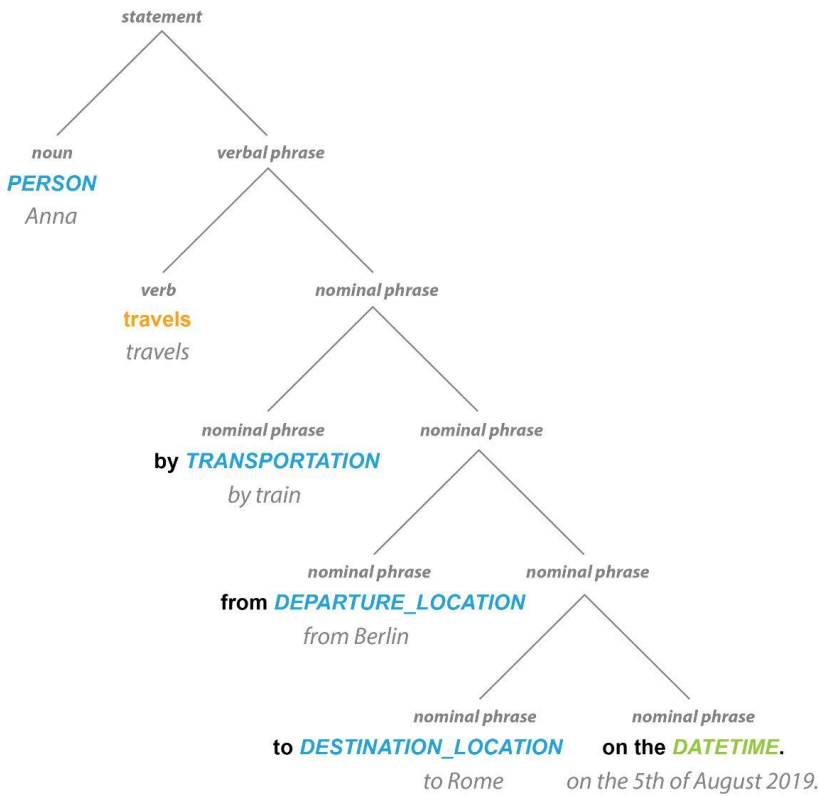

### C) Formalized Statement with Positions and associated Semantic Roles

**PERSON** *travels* by **TRANSPORTATION** from **DEPARTURE_LOCATION** to **DESTINATION_LOCATION** on the **DATETIME**.

### D) Storage Model for a *travels* Statement with Positions and associated Semantic Roles

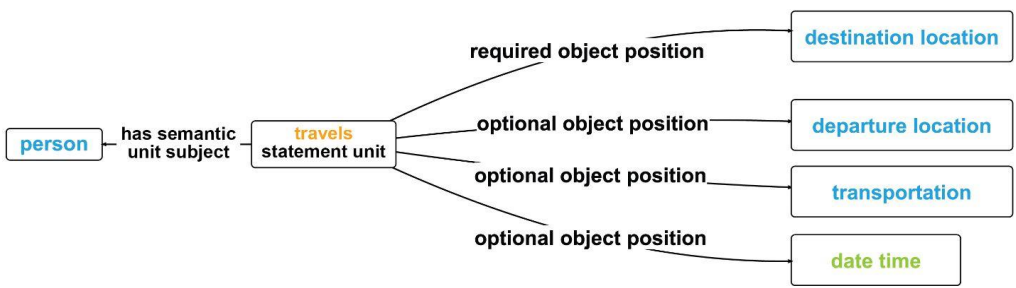

**Figure 9: From the structure of a natural language statement to the structure of a formalized storage model of this statement in a knowledge graph. A)** A natural language statement with the verb *travels*. **B)** The corresponding syntax tree with the syntactic positions and their associated semantic roles in color. **C)** The translation back into a formalized natural language statement with the same syntactic positions and semantic roles. **D)** The translation of the formalized natural language statement into a graph pattern, covering the same syntactic positions and semantic roles.

**2) Object-type and object-position class.** Having another look at the example from above, we see that when we model the statement '*Sarah* met *Bob* on *4th of July 2021*' in a knowledge graph, the objects '*Bob*' and '*4th of July 2021*' would be modelled differently. Whereas '*Bob*' is likely modelled as a resource that instantiates the class person (foaf:Person), '*4th of July 2021*' is likely modelled as a literal associated with the datatype xsd:date. Therefore, in addition to the above-mentioned distinction of arguments (i.e., subject and required objects) and adjuncts (i.e., optional objects), each with their associated semantic roles and thematic labels, one can distinguish objects by their type into resources and literals and specify **class/datatype constraints based on their associated semantic roles**. Resources, in turn, can be the URIs of named-individual, some-instance, every-instance, class, or property resources. We refer to them from here onwards as **resource-objects** and literal-based objects as **literal-objects**. Propositions can be differentiated based on the number of resource- and the number of literal-objects they comprise.

In a statement KGBB, each distinguished object-position is defined in its own **object-position class**. Instances of such a class then take the role of the corresponding object-position in a particular statement unit, through which data input is documented by linking to the object-position instance. Object-positions are only meaningful within the context of a given type of statement and thus a given statement KGBB and its associated statement unit class. The introduction of object-position classes meets specific requirements of the versioning and history-tracking approach discussed below (see [Versioning and tracking editing history](#)) and also documents any input-restrictions that apply to its corresponding type of input.

The number of object-position classes that a given storage model distinguishes depends on the n-aryness of the underlying type of proposition and is thus KGBB-specific. Every object-position instance of a statement unit is linked to the statement's subject resource via a specific property, the type of which depends on whether its linked object is necessary and thus required or optional, indicated by the properties *_requiredObjectPosition_* and *_optionalObjectPosition_*, respectively. This is also documented at the class level: Each statement unit class as well as each statement KGBB class links to its corresponding required and optional object-position class(es) via the annotation properties *_hasRequiredObjectPositionClass_* and *_hasOptionalObjectPositionClass_*, respectively.

These two criteria, i.e., (i) the syntactic positions with associated semantic roles, specified as a subject-position and a number of required and optional object-positions and (ii) the corresponding object-position classes with their respective object-types—either resource or literal—set the dimensions and boundaries for our **approach for a generic data graph storage model for statement KGBBs**[9]: subject-position, predicate, required object-positions, and optional object-positions and their corresponding object-position classes with semantic role specifications represent all the elements needed for modelling any type of statement in a graph.

What is still missing for a data graph storage model, however, is a specification of a storage model for the information that is linked to an object-position instance. The data graph storage model thus must consist of two parts: (1) the **subject-related data graph storage model**, that links the subject resource to the various object-position instances and which differs depending on the type of proposition that underlies the statement unit, and (2) the general **object-related data graph storage model** that applies to each object-position instance in the data graph of a statement unit.

The overall data graph storage model is specified for *statement KGBB*, which inherits it to all other statement KGBB classes. The object-related part covers mainly provenance metadata relating to each object-position instance. It complements the subject-related part to constitute the data graph storage model which, in turn, combines with the general semantic-units graph storage model inherited from *KGBB* to form the overall storage model of a statement KGBB. This overall storage model is inherited by *statement KGBB* to all specific types of statement KGBBs along class-subclass paths.

---

[9] Remember: Only statement units have a data graph and thus require a data graph storage model.

**Subject-related data graph storage model**

Depending on the number of argument objects and adjunct objects of the underlying predicate and thus the n-aryness of the statement's relation, a particular statement KGBB can have multiple object-positions of each type specified—one for each argument and one for each adjunct object-position. The subject-related model distinguishes these two types of object-position: one for specifying an argument object and another one for specifying an adjunct object (see also Fig. 11):

1. **RequiredObjectPosition** (*UPRI of the instance of the object-position class specific for this type of argument*): specifies the UPRI of an instance of the corresponding object-position class for an input required for completing the meaning of the predicate of the statement and thus for an argument. It is documented using the property **requiredObjectPosition*.
2. **OptionalObjectPosition** (*UPRI of the instance of the object-position class specific for this type of adjunct*), optional: specifies the UPRI of an instance of the corresponding object-position class for an optional input and thus an adjunct. It is documented using the property **optionalObjectPosition*.

In the LinkML specification of the storage model, the slot belonging to a required object-position is declared to be '*required: true*', indicating that this slot must have a value.

For example, in a time-stamped-has-part statement KGBB, the subject-related data graph storage model has three object-positions that link to the subject resource. Two optional positions for adjuncts that specify the time interval and thus for literal-objects, whereas the required object position specifies the subject's part and is therefore a resource-object. Consequently, the subject-related data graph storage model of this time-stamped-has-part statement KGBB would consist of one required and two optional object-positions.

**Object-related data graph storage models**

As already mentioned, the number of object-positions being distinguished in a storage model is KGBB-specific, and each object-position resource in a corresponding statement unit is documented as an instance of a specific object-position class. A KGBB class relates to its object-position classes via the annotation property **hasObjectPositionClass*. Each object-position resource thus instantiates its respective object-position class and tracks the actual input for its object-position, alongside with metadata about the particular input event. Therefore, one can think of **an object-position instance of a particular statement unit to document a particular input event belonging to a specific object-position** (e.g., the part-specification-position in a has-part statement).

Whereas the input itself depends on the particular statement KGBB, and any constraints for it must be specified at the KGBB class level, the specification of how to store the metadata is general and applies to any statement unit and therefore is specified for *statement KGBB*, which inherits it to all other statement KGBB classes. The specification of the general object-related data graph storage model includes the following slots (ranges) for each object-position, independent of whether it is a required or an optional object (Fig. 10):

1. **ObjectPositionType** (*UPRI of the object-position class*): information that enables identifying to which object-position this instance belongs. This way, the application can, for instance, distinguish a mean-value object-position from an upper- and a lower-confidence-interval-value object-position within the graph of a measurement-with-confidence-interval statement unit. This relation is specified via the property *type* (RDF:type).

2. **InputTypeLabel** (*string*): This provides information for a human reader to identify the object-position's input type. It takes the function of a thematic label and is tracked via the property \**inputTypeLabel*\* with a human-readable label as its value (e.g., 'weight-unit' or 'weight-value'). Alternatively, this information could also be provided by the corresponding object-position class, from which it could be fetched.

3. **Input**; the actual input provided through user input or data import, depending on whether the object is a resource-object or a literal-object. By linking the input to the object-position instance instead of using that instance as the actual input, resource-objects and literal-objects can be treated in the same way (except for the difference of their respectively linked input), and metadata can be tracked for each input event. The specification of any input-constraints, however, must be specified in the respective object-position class.
    a. **ResourceURI** (*resource UPRI*), applies only to object-positions with resource-objects: The property \**resourceURI*\* with the UPRI of the input-resource as its value documents the object-resource, which represents either a named-individual, a some-instance, an every-instance, a class, or a property resource **[F1.;A1.]**. This applies only
    b. **Literal** (*literal*), applies only to object-positions with literal-objects: The value of the property \**literal*\* documents the object-literal.

4. **LogicalProperty** (*UPRI specifying a logical property*), optional: information on whether a particular logical property (i.e., transitivity, symmetry, asymmetry) of the predicate underlying the statement applies to this object via the property \**logicalProperty*\*.

5. **CurrentVersion** (*Boolean*): information that indicates whether the object-position instance contains input belonging to the live-version of this statement unit or input that already has been updated and thus is outdated. This information is required for tracking the editing history of semantic units, which is discussed in detail further below. The Boolean property \**currentVersion*\* indicates whether the object-position instance belongs to the unit's live version.

6. **Input-provenance [F2.;F3;F4;A1.;R1.]**
    a. **Creator** (*user UPRI*): information on who added this input to the data graph using the property \**creator*\*.
    b. **CreationDate** (*dataTime*): information on when this input has been added to the data graph using the property \**creationDate*\*.
    c. **CreatedWithApplication** (*UPRI of knowledge graph application*): information on with which application the input has been added to the data graph using the property \**createdWithApplication*\*.
    d. **ImportedFrom** (*UPRI of external dataset*), optional: information on from which external dataset this input has been imported using the property \**importedFrom*\*.

7. **VersionID** (version UPRI), optional: information that indicates whether the object-position instance belongs to a specific documented version of this statement unit or of some semantic unit to which this statement unit is associated. This is tracked via the property \**versionID*\*. Since a given input can belong to more than one version of the statement unit and since a statement unit can be associated with more than one other semantic unit, this slot can be realized multiple times.

8. **DatasetUnitID** (dataset unit UPRI), optional: information that indicates whether the statement unit with this specific object-position instance belongs to a specific documented dataset unit. This is tracked via the property *datasetUnitID*. Since a given statement unit can be associated with several dataset units, this slot can be realized multiple times.

In the LinkML specification of the storage model, the slot belonging to a required object-position is declared to be '*required: true*' and slots that can be instantiated multiple times to be '*multivalued: true*', whereas any value constraints for a slot can be declared to have a '*range:*', with either a datatype specification as a value or a URI of a class, and to have a '*minimum value:*', '*maximum value:*', and '*pattern:*' for any literal values. Constraints for literal- and resource-objects for specific object-positions can be specified in LinkML using these declarations.

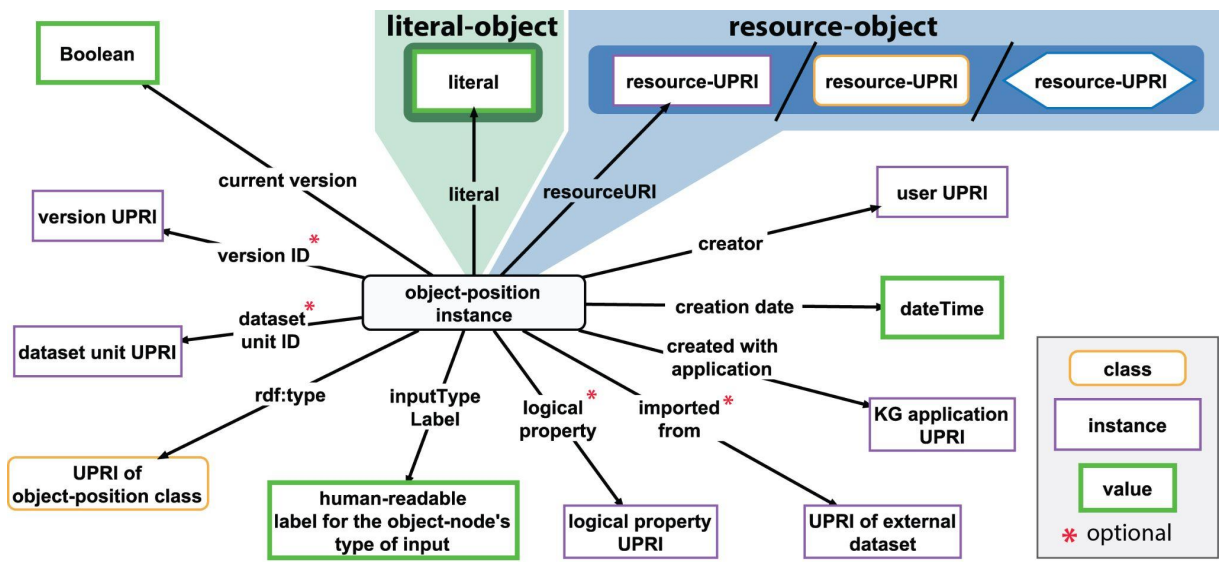

**Figure 10: The general object-related data graph storage model specified for *statement KGBB*.** The graph shows the information tracked with each object-node. Depending on the type of object-node, parts of the graph are not realized (blue background not realized for a literal-object, green background not for a resource-object). Some relations are optional, depending on whether the statement unit is part of a versioned semantic unit or dataset unit, or whether the input results from the import of an external dataset, or logical properties apply (red asterisks). Some relations can be realized multiple times. This model is inherited from *statement KGBB* down to all other statement KGBB classes, but can be complemented with additional constraints to meet the requirements of any specific statement KGBB class. The objects of the proposition underlying a respective statement unit are indicated by a green box with rounded corners in the case of a literal-object and a blue box with rounded corners in the case of a resource-object.

When a new statement unit is created, the KGBB-Engine will automatically classify it depending on the type of resource in its subject-position. If this resource is a **named-individual resource**, the statement will be classified as an **assertional statement unit** and the KGBB-Engine will allow only named-individual resources for any of the statement's resource object-positions. If the subject resource is a **class** or an **every-instance resource**, the KGBB-Engine will classify the statement unit as a **universal statement unit** and will allow only some-instance or class resources for any of its resource object-positions. For **some-instance resources** in the subject-position, it is a bit more complicated. Here, the user must decide whether the statement unit should be classified as a **contingent statement unit or a prototypical statement unit**, which has logical implications regarding reasoning (23). In either case, however, the KGBB-Engine will allow only some-instance resources for any of the statement's resource object-positions.

**The overall statement KGBB storage model—combining the parts**

To summarize the above, the storage model for any given statement KGBB class is the combination of three parts (see Fig. 11):

1. The **general semantic-units graph storage model**, extended by nine additional slots as described above,
2. the **subject-related part of the data graph storage model**, and
3. the **object-related part of the data graph storage model**.

Each of these parts can be extended by adding input constraints and specifying object-positions along the class-subclass path from *KGBB* to any specific statement KGBB class. The constraints narrow down what subjects and objects are considered to be valid for the respective statement KGBB class.

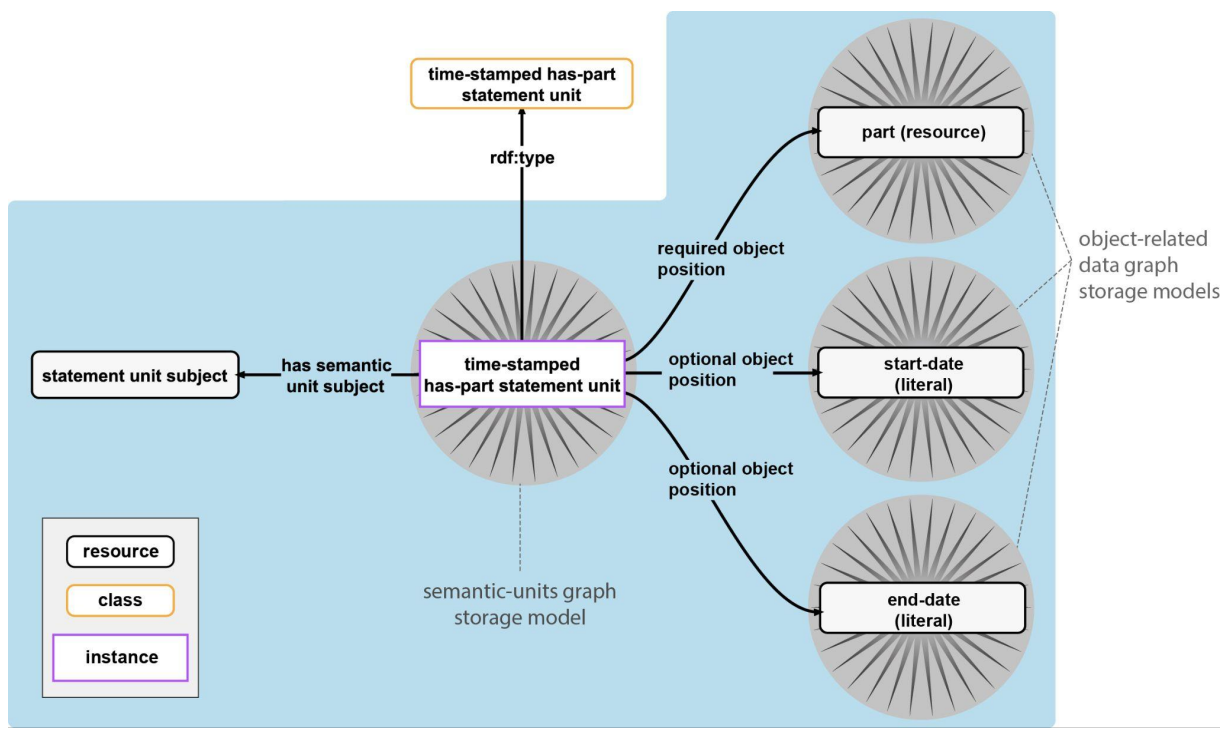

**Figure 11: Example of the storage model for a quaternary time-stamped has-part statement KGBB class.** The time-stamped has-part proposition has one required resource-object for specifying the part and two optional literal-objects for specifying a time interval. The storage model of the corresponding time-stamped has-part statement KGBB class consists of: 1) The data graph that models the relation between the subject and the object-nodes of the particular has-part proposition that the statement KGBB models, here shown in the blue box. The graph follows the generic storage pattern for quaternary propositions, with a single required resource object-position and two optional literal object-positions—that this data graph models a has-part relation can be derived from the fact that the statement unit instantiates the *time-stamped has-part statement unit* class. 2) The representation of the data graph in the semantic-units graph in the form of a single resource, i.e., the statement unit resource, which in turn instantiates the class *time-stamped has-part statement unit* and links to the subject resource of the data graph. 3) The object-related information linked to each object-node. The four gray circles with outward rays indicate where this graph is complemented and which storage model parts provide the required graph patterns.

With its slots for the subject and its slots for the various object-positions and their linked object inputs, the storage model necessarily refers to a set of variables that the KGBB-Engine must specify when creating a new statement unit. They must be specified based on input provided by users, input automatically depending on context, or input via data import, before the KGBB-Engine can specify an adequate create query. The input, in turn, can be validated against the constraints specified for each object-position and any imported data can be validated against the overall storage model. Figure 11 indicates how the different parts of the storage model are combined to form the overall storage model of a statement KGBB class.

**Access templates**

Each storage model specified for a statement KGBB class (or any other type of KGBB class) can have one or more access models associated with it that the KGBB-Engine can use to automatically convert data and metadata from the unit's data graph or semantic-units graph to a representation of the same data and metadata in a pattern and format that complies with the access model.

Since the storage model considers each object and thus each input individually and assigns its own object-position to it, it allows the systematic identification and reconstruction of the original user input or data import. The storage model for the data graph and with it all relevant information, therefore, can be aligned with any of the access models and its subject resource and the different object-positions can be mapped to the aligned positions of the access models, resulting in **schema crosswalks [F6.2/I7.2]** (see Fig. 12). All resources in the access schema that cannot be directly aligned to the subject or any of the storage model's object-positions are typically added in the access models for providing the machine more contextual information for properly processing and analyzing the data, to enable reasoning over it, to improve querying, or to meet established community standards.

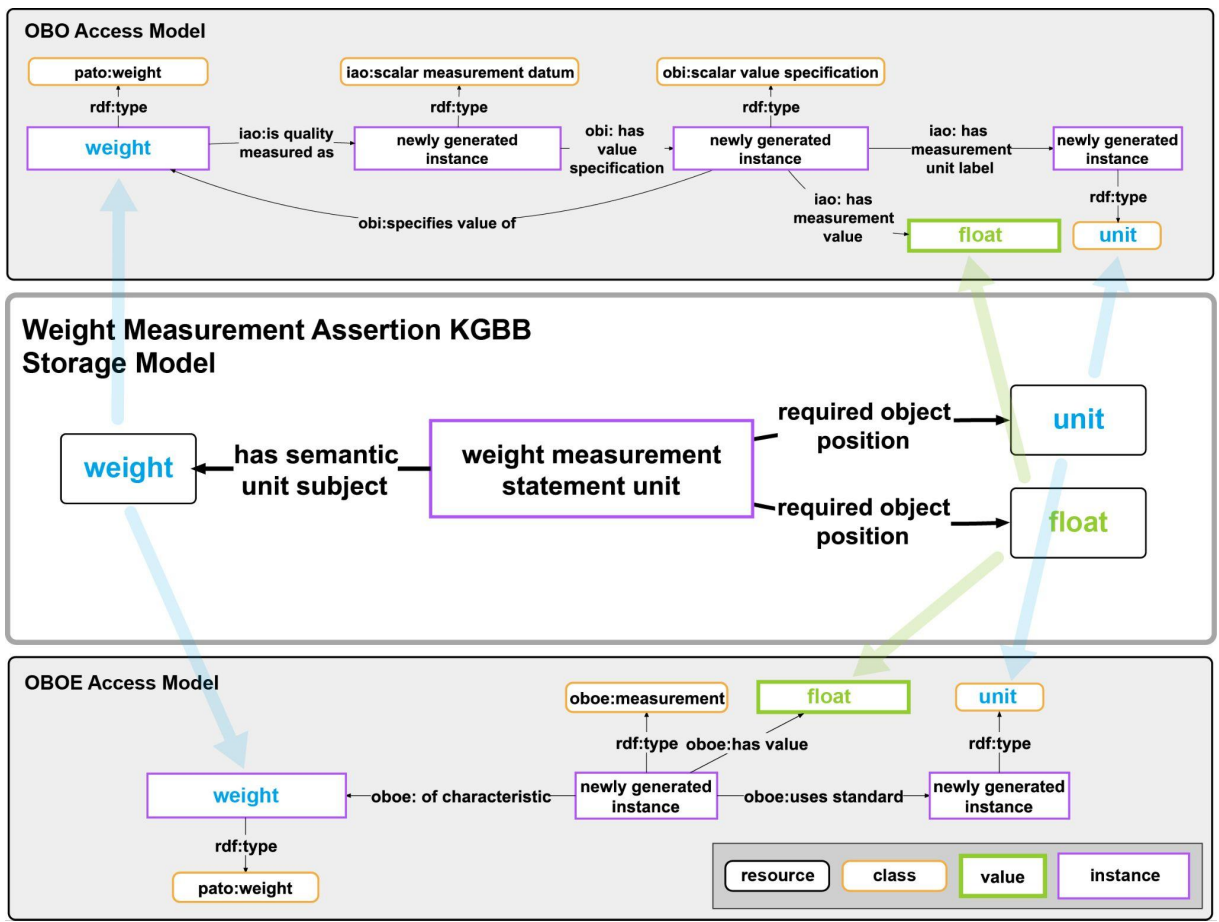

**Figure 12: Mapping of the subject-related data graph storage model of a statement KGBB class onto two of its access models**. **Middle:** The subject-related storage model of a weight measurement statement KGBB class. **Top** & **Bottom:** Two different access data models. Notice that the variables of the storage model are reused in the access model and thus can be

easily mapped from storage to access model (see arrows). The access models introduce additional variables for resources that have to be newly created for each data export ('newly generated instance' boxes).

All provenance and metadata associated with a particular semantic unit or one of its object-position instances can be mapped to any provenance and metadata schema or model if a corresponding access model has been specified, using, e.g., PROV-O, PAV ontology (61), Dublin Core Metadata Initiative, or DataCite Schema, resulting in **metadata schema crosswalks [F6.2/I7.2]**.

In addition to allowing accessing the data and metadata according to different graph models, the KGBB-Engine must make the data and metadata also accessible in different formats such as RDF/OWL, GraphQL, Python or Java data classes, JSON, and CSV. Since all these formats must provide data-slots for a given statement unit class that map to its positions and their associated semantic roles, the mapping to non-graph-based formats should be analog to that of graph-based formats.

Using the KGBB-Editor, new access models and formats can be specified and added to KGBB classes whenever needed, thus allowing for adapting to newly emerging data and metadata formats and standards[10]. This takes into account the observation that FAIRness is not sufficient as an indicator for a high data and metadata quality—usage of data often depends on its **fitness-of-use**, that is, data must be available in adequate formats complying with established standards and protocols that allow their direct usage as for instance when a specific analysis software requires data in a specific format.

Access models can be inherited across the taxonomy of KGBB classes, just like storage models. Moreover, a given access model can be associated with one or more **access model families**. By referring to a specific access model family, a user can access the data managed by different KGBBs, with each KGBB applying the access model that belongs to a specific family of access models. This way, a user can access the data in the graph using a consistent and interoperable set of models (e.g., OBO/OBI-compliant data models).

Each access model should provide, where possible, information about relevant literature references, curators, the logical framework that can be applied to it **[I4.]**, a UPRI for identifying the graph pattern **[F6.1/I7.1;A1.;I1.]**, besides the typical provenance data **[F1.;F2.;F3;F4;A1.;I1.;I2.]**.

Statement KGBBs and their corresponding statement unit classes can be understood as a formalized approach for modelling n-ary predicates. Each statement unit class represents an n-ary predicate that is modelled as an ontology class instead of an ontology property. As a consequence, reasoning over property axioms such as transitivity or domain and range specifications is not straightforward within the KGBB Framework, and tools established for OWL-based frameworks cannot be readily reused in the KGBB Framework. It is therefore important for the KGBB-Framework to provide **interoperability with OWL and description logics** by providing, per default, for every statement KGBB an **OWL-based access template**. When specifying a new statement KGBB, the KGBB-Editor uses the information provided for the newly specified statement unit class and for the storage model of the newly created statement KGBB to automatically specify an OWL-based access template for that statement KGBB.

Taking the traveling statement unit with a destination location as a required object and all other objects as optional objects as an example, we can define the following properties for translating a traveling statement unit into an OWL-based statement, each of which is annotated to be a member property of the *travelingStatementUnit* class (Fig. 13):

---

[10] In the case of new formats, this might involve having to specify new adaptors to the API of the KGBB-Engine and making adjustments to the KGBB-Editor.

1. the object property *travelsTo* as a subproperty of *requiredObjectPosition*, with the domain and range specification of 'PERSON' and 'LOCATION', respectively;
2. the object property *travelsFrom* as a subproperty of *optionalObjectPosition*, with the domain and range specification of 'PERSON' and 'LOCATION', respectively;
3. the object property *travelsWith* as a subproperty of *optionalObjectPosition*, with the domain and range specification of 'PERSON' and 'TRANSPORTATION', respectively; and
4. the data property *travelsOn* as a subproperty of *optionalObjectPosition*, with the domain specification of 'PERSON' and the range specification of 'datatype:DATETIME'.

The labels of the properties can be automatically obtained from the textual display template (see discussion below), the domain and range specifications from the input-restrictions of the storage model and any logical property axioms such as transitivity from the respective specifications on the object-position classes (see [KGBB-Editor](KGBB-Editor)). This approach would allow a generic translation from the storage pattern of any statement KGBB to an OWL-based access template.

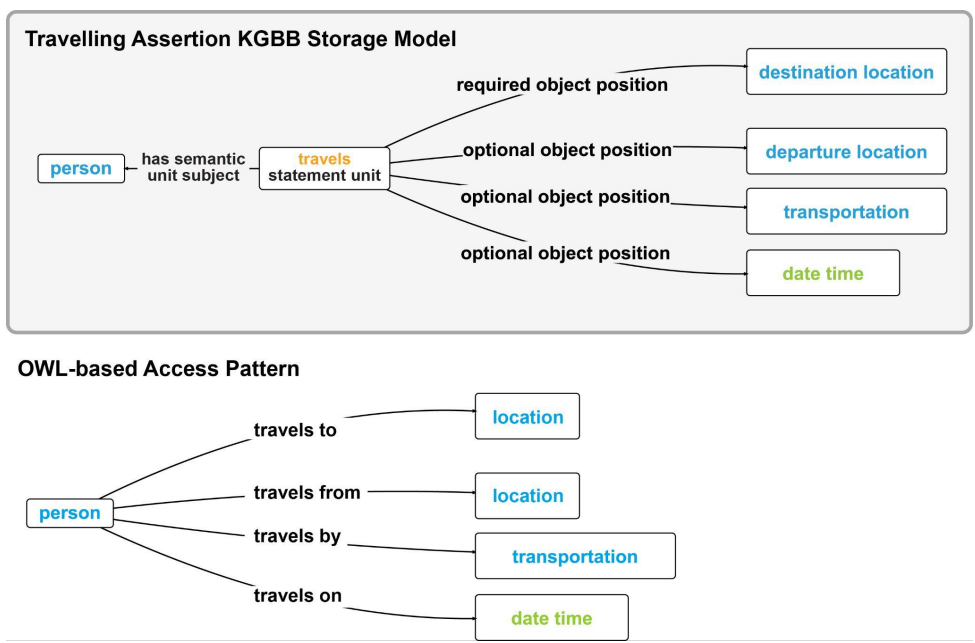

**Figure 13: comparison of the storage model of a statement KGBB and its OWL-based access model. Top:** The subject-related storage model of a travels statement unit, provided by its statement KGBB class. **Bottom:** The transformation into an OWL-based access model, where the restrictions on the subject and the various object positions of the KGBB storage model are translated into domain and range specifications in the OWL-based model. Here, logical axioms such as transitivity can also be assigned to each of the properties if required.

**Importing and exporting data and metadata**

Import templates can be specified in various formats in the same way as the subject and the different object-positions and their associated semantic roles can be mapped onto various access models and formats. This allows a systematic import of data from different formats. The import still requires previous formatting of the import data to comply with the format of the respective import template.

Pretrained large language models and other NLP tools can support the importing by entity-recognition and mapping of values to ontology terms. Tools such as ChatGPT can be used to support the semantification of a natural language statement against a storage model with a set of specified subject and object-positions, each characterized with a thematic label. When specifying appropriate ontologies for each position, the tool can identify matching values for each position based on the information from the natural language statement. The result can be returned to a user for verification using the dynamic labels. If necessary, the user can manually adjust the result and then confirm it.

By using the storage models of statement KGBBs as a sort of **lingua universalis**, the KGBB-Framework allows the specification of **operational data and metadata schema crosswalks**. If you have data and metadata stored in different databases that use different schemata and formats, and you want to export them into again different schemata and formats, you only have to specify the imports to the KGBB storage models of respective statement types and for them respective access templates for the export. The KGBB storage model thus functions like a broker between all the different schemata and formats: one only has to specify the mappings to and from the broker, but not between all possible combinations of schemata and formats, thus substantially reducing the mapping burden.

**Display templates—human-readable data views**

Using the structure from the storage model that the statement KGBB class provides, the KGBB-Engine generates a data structure from the data graph of a given statement unit. For being able to display the content of the semantic unit in a human-actionable way, the frontend needs some **display template** that specifies how to translate the data structure into a human-readable statement. Display templates organize information from the data structure together with additional labels and headlines to be presented in the UI (see also [Compound display templates](#) further below).

If, for instance, a weight measurement with 95% confidence interval were stored in a corresponding statement unit, the respective storage model would specify three literal-object-positions and one resource-object-position, i.e., a main value, an upper and a lower bound value, and a unit. The form-based display template could specify that this information should be presented in the frontend as 'weight (95% conf. interval): [literal-object_1:literal] ([literal-object_2:literal]-[literal-object_3:literal]) [resource-object_4:resource-label]'. Another example is a travels-statement, where the respective storage model specifies one required resource-object-position (i.e., destiny location) and three optional-object-positions (i.e., departure location, transportation, and date-time). The corresponding display template would present a travels-statement in the frontend as '*PERSON* travels by *TRANSPORTATION* from *DEPARTURE_LOCATION* to *DESTINATION_LOCATION* on the *DATETIME*'. In other words, the subject-position and the different object-positions (i.e., the syntactic positions with their associated semantic roles) map to corresponding variables within a string to form a human-readable statement (see Fig. 14). We call such textual display templates **dynamic labels [E3.;E3.1]**.

In addition to such textual display templates, it is also possible to specify graphical display templates for a **mind-map-like representation of a statement unit** using **dynamic mind-map patterns [E3.;E3.1]**. Dynamic mind-map patterns use a label for the predicate that underlies the corresponding type of statement, and if there is more than one object-position also labels for relating the various objects to the predicate. As a result, the statement can be visualized as a mind-map like graph, where each subject and object is represented as a node with the label of the corresponding resource from the underlying data graph of the statement unit (see Fig. 14). Such graphical representations of statement units can also be combined to form a mind-map of larger contexts and interrelationships that connect the dynamic mind-map patterns of several statement units, for instance to graphically display compound units. Mind-map-like representations of complex interrelationships between various entities are easier to comprehend than form-based representations, and thus increase the human-actionability of the knowledge graph.

**Dynamic Label (Textual Display)**

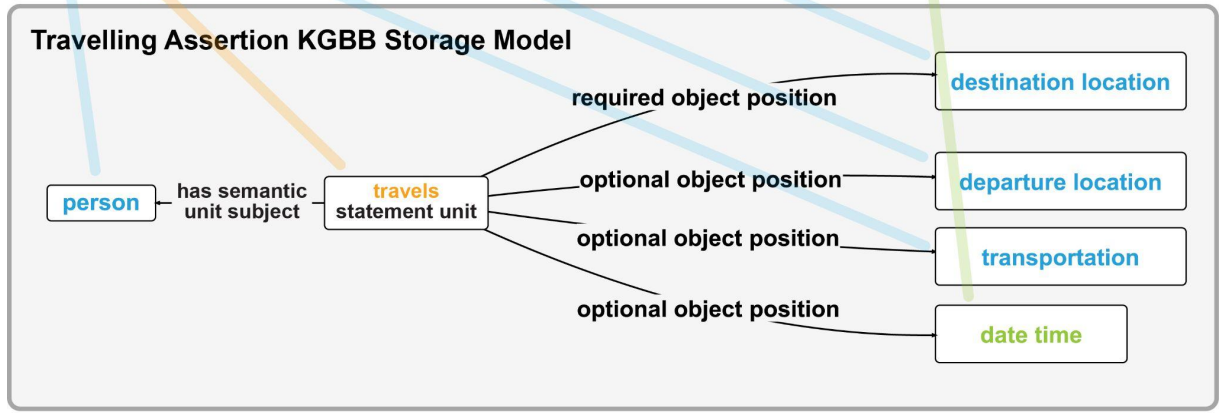

**Dynamic Mind-Map Pattern (Graphical Display)**

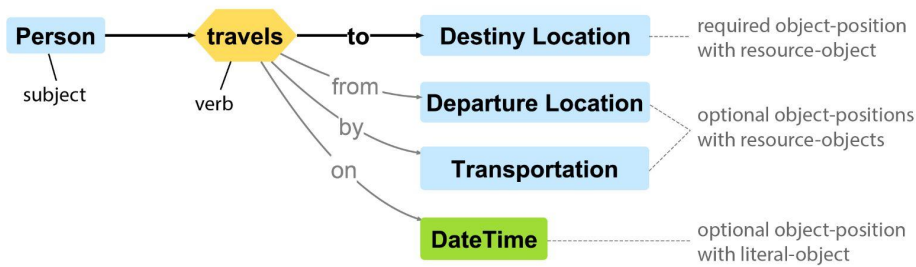

**Figure 14: Textual and graphical displays of a statement unit based on its storage model. Middle:** The subject-related storage model of a travelling statement unit, provided by its statement KGBB class. **Top:** A textual display of the travelling statement unit, based on the dynamic label provided by its statement KGBB. See how the subject-position and the object-positions from the storage model map to variable-positions (i.e., positions with associated semantic roles) in the dynamic label template (arrows). **Bottom:** A graphical display of the travelling statement unit, based on the dynamic mind-map pattern provided by its statement KGBB. The mapping of the subject-position and the object-positions from the storage model to nodes in the dynamic mind-map pattern are comparable to the mappings for the dynamic label template (for clarity of presentation, arrows here not shown).

The dynamic label templates can also account for whether it is a **lexical**, **assertional**, **contingent**, **prototypical**, or **universal statement unit**. In the case of a has-part statement, the assertional dynamic label template could take the form 'This *SUBJECT* has part this *PART*', the contingent template the form 'A *SUBJECT* can have part some *PART*', the prototypical template the form 'A *SUBJECT* typically has part some *PART*', and the universal template the form 'Every *SUBJECT* necessarily has part some *PART*'.

The dynamic label of a statement unit can also be utilized as the label for its statement unit resource when it is displayed in the graph for users to explore the graph **[E3.2]**. This way, users could access the information contained in statement units on the statement unit level of representational granularity, without having to access the actual triple level of the graph that is often not so easily comprehensible for a human reader.

## Compound KGBB

Compound KGBBs work differently than statement KGBBs, as they organize and manage the data from statement units and other compound units but do not have to provide storage models for their data graphs, since their data graphs are retrieved through merging the data graphs of their associated semantic units. Consequently, their storage model only covers the semantic-units graph (cf. Figure 8) and thus the specification of the information related to the compound unit resource itself.

**Compound-unit-specific extension to the general semantic-units graph storage model**
Due to the fact that a compound unit has two or more semantic units associated with it and that two compound units can be linked to each other via a linking statement unit that shares its subject resource with the subject resource of one compound unit and its object resource as the subject resource with the other, the general semantic-units graph storage model inherited from *KGBB* needs to be extended to include the following additional slots (range):

11. **HasAssociatedSemanticUnit** (*UPRI of semantic unit*), optional: if the semantic unit has another semantic unit associated with it, the property **hasAssociatedSemanticUnit** specifies the UPRI of that semantic unit. Only compound units have other semantic units associated.
12. **HasLinkedSemanticUnit** (*UPRI of semantic unit*), optional: if the semantic unit is a compound unit *A* that is connected to a compound unit *B* via a statement unit, where compound unit *A* and statement unit share the same subject resource and the object resource of the statement unit is the subject resource of the linked compound unit *B*, then this link between compound unit *A* to unit *B* is made explicit via the property **hasLinkedSemanticUnit**.

Many compound KGBB classes additionally add further constraints on their subject. A material entity item KGBB class, for example, would constrain its subject to be an instance of a corresponding ontology class, such as 'material entity' (BFO:0000040). Such constraints should ideally be specifiable in a LinkML template.

In addition to the metadata that is directly stored with each compound unit, the KGBB-Engine dynamically fetches the contributor, last-update, and imported-from provenance metadata for the compound unit as well as copyright licenses, access-restrictions, and specifications of the logical framework from all its associated semantic units and makes the information available to the presentation-layer.

**Compound KGBBs define the proposition-space of a knowledge graph**
Compound KGBBs not only facilitate structuring the knowledge graph into various types of compound units and thus collections of semantic units to form larger subgraphs that can be referenced by their own UPRIs. They also structure data input/import by restricting which statement units can be instantiated in particular contexts by connecting statement KGBBs and other compound KGBBs to it. For example, a process-item KGBB allows the use of an occurrent-part-of statement KGBB but a material-entity-item KGBB does not, because the occurrent-part-of relation is a relation that applies only between occurrents such as processes, and not between continuants such as material entities.

By linking to other KGBBs, **a set of compound KGBBs specifies the domain of discourse of a knowledge graph—it defines its proposition-space** and limits it to combinations of statements and contexts that are meaningful and that are within the scope of the respective knowledge graph. Compound KGBBs relate to other KGBBs so that data newly added through a compound KGBB can trigger other KGBBs to instantiate their corresponding semantic units. Resources can be shared across different KGBBs for subsequent re-use in newly added semantic units, and these newly added units are connected to the overall data graph via sharing resources.

By relating different KGBBs to one another in such a way, the structure of the knowledge graph can be specified in a highly modular and expandable way through specifying the relevant KGBBs to be used in a given context. Some of these relations are specified on the KGBB class level and apply to every KGBB instance of that class, but further relations can be added when describing the specific knowledge graph application with instances of KGBB classes using the KGBB-Editor. Moreover, since each statement KGBB that is related to compound KGBBs this way provides its own storage model, all propositions of the same type would still use the same storage model and all data and metadata in the respective knowledge graph would be truly FAIR **[F6.1/I7.1]**.

In other words, using an instance of an item group KGBB one can specify with the KGBB-Editor the type of item units that may be associated with it by linking to respective item KGBB instances. These possible interactions between different KGBBs are described in a **KGBB specification graph** that provides the specification for the respective FAIREr knowledge graph application. In this specification graph, one can specify for each of the item KGBB instances possible interactions with statement KGBB instances, and thus define the types of statement units that may be associated with this type of item unit. The specification graph also specifies which types of statement units are required and thus *must* be provided before being able to add the item unit to the knowledge graph and to this type of item group unit in the first place. It also specifies whether the item unit may contain only one or more statement units of that type (see [Specifying interactions across KGBB instances in a KGBB-driven knowledge graph application](#) for specifying relations between different KGBBs).

**Compound display templates**

A compound display template is a display template for a compound KGBB class and determines how the data from its compound unit instances should be presented as textual information in the UI **[E3.;E3.1]**. In general, every compound KGBB can have one or more such compound display templates specified in its class description. A compound display template determines the order in which the data of its associated semantic units are presented, including placeholder texts, tooltips, headlines, labels, etc. (Fig. 15). A compound display template can be specified for each compound KGBB class using a JavaScript template engine or the web template engine [Jinja2](#). Different templates can be defined for different contexts, enabling each knowledge graph application that uses the respective KGBB to define their own UI and present the data belonging to the semantic unit following their own UI look and feel. Moreover, it also enables differentiating UIs for different applications or different types of users.

In other words, compound display templates for compound KGBBs are always form-based templates. Compound units do not require mind-map-like display templates—they can be represented in the UI in a mind-map-like representation by merging the respective mind-map-like representations of their associated statement units.

Each newly defined KGBB class should come with at least one form-based display template that is ready-for-use and which could be changed and adapted, if required. Form-based display templates for compound units thereby embed the form-based display templates of their associated statement and compound units. In case of statement KGBBs, the form-based template should be complemented with a mind-map-like display template.

**Figure 15: Example of a UI page for editing a population item unit, utilizing information from the compound display template of the respective item unit KGBB class, embedding representations from its associated statement units based on their respective dynamic display templates.** The entire structure of this page as well as the labels shown are defined in the respective compound display templates. The page shows data about the population of Wuhan for a specific time period, provided by the data graphs of a material entity item unit and its associated statement units. The entire structure of this page as well as the labels shown are defined in the respective display templates, with the compound display template of the material entity item KGBB embedding the dynamic display templates from its associated statement KGBBs. The figure shows the item unit's representation in the dark-blue-bordered box, and the representation of the contents of its associated statement units in light-blue-bordered boxes. Light grey labels indicate the sources for various types of information within the different boxes.

## Question KGBB and KGBB-Query-Builder

A new question unit can be created by using the storage model of an existing statement KGBB and allowing named-individual, some-instance, every-instance, and class resources at the respective subject and object positions and then classifying it as a question unit. Based on the storage model used, the KGBB-Engine can transfer this question into a query that would retrieve a Boolean *true/false* answer if the subject and object-positions contained fully specified inputs (i.e., named-individual resources) or a list of statement units as the answer that match the statement unit description provided by the question unit if one or more of these positions contained underspecified input (i.e., some-instance, every-instance, or class resources). The KGBB-Engine derives a display template for this question unit based on the dynamic label template of its corresponding statement KGBB. So, if the statement KGBB's dynamic label template took the form '*PERSON* travels by *TRANSPORTATION* from *DEPARTURE_LOCATION* to *DESTINATION_LOCATION* on the *DATETIME*', the question unit's display template would take the form 'Did *PERSON* travel by *TRANSPORTATION* from *DEPARTURE_LOCATION* to *DESTINATION_LOCATION* on the *DATETIME*?' and the respective question unit could be represented in any graph visualization on the statement unit level of representational granularity using a corresponding dynamic label as a node or a dynamic mind-map as a graph pattern.

The same idea provides the foundation for the query interface of a KGBB-Frontend: users can create their own queries using the input forms for creating statement units using named-individual, some-instance, every-instance, class resources, and datatype as inputs for subject and object-positions without them having to use any graph query languages **[E3.4]**. The KGBB-Query-Builder translates the question units into actual graph queries. Here, the generic structure of the storage model of statement KGBBs substantially supports the development of procedures for this translation step.

Question units thus not only allow documenting questions in the graph, but also indirectly document the corresponding queries. This way, competency questions for a FAIREr knowledge graph can be documented in a readily actionable way within the knowledge graph itself.

The KGBB-Query-Builder can translate a **statement question unit** following a general query pattern, using the subject-position and the various object-positions as variables, and combine different statement question units to form a **compound question unit** using the Boolean operators AND and OR.

A statement question unit is stored the same way as a statement unit by applying the storage template of its corresponding statement KGBB, but classifying it as an instance of *statement question unit* and relating it to its corresponding statement KGBB using the property *based on statement KGBB*. A compound question unit is stored similar to a compound unit by indicating which statement question units it has associated and by relating them using the Boolean operators in its data graph.

# Specifying a KGBB-driven knowledge graph application by specifying interactions across KGBB instances

**Connecting data graphs through sharing resources across semantic units**
All triples belonging to the data graph of a statement unit form a connected graph, but how are these data graphs connected across several statement units in semantically meaningful ways? This can be achieved by specifying the possible interactions between respective KGBBs. Each KGBB is a class. Instances of these classes can be used to describe their possible interactions using the KGBB-Editor. Every interaction between two KGBB instances results in two corresponding semantic unit instance data graphs being connected. Five types of connections can be distinguished:

1. **Subject-subject connection**: Statement units A and B can be connected by sharing the same resource in their subject-position (Fig. 16A).
2. **Subject-object connection**: Statement units A and B can be connected by sharing the same resource in one of the object-positions of statement unit A and the subject-position of statement unit B (Fig. 16B).
3. **Subject-semanticUnit connection**: A statement unit and a semantic unit can be connected by having the resource of the semantic unit in the subject-position of the statement unit (Fig. 16C).
4. **Object-semanticUnit connection**: A statement unit and a semantic unit can be connected by having the resource of the semantic unit in one of the object-positions of the statement unit.
5. **Object-object connection**: Statement units A and B can be connected by sharing the same resource in one of their object-positions.

This list of connection types is exhaustive—no other type of connection exists. While the latter two connection types depend on the possibility to reuse existing resources in an object-position, the possibility of which depends on the input-constraints of the data graph storage models of the respective statement KGBBs, the former three must be defined as possible interactions between particular KGBBs.

## A) Association

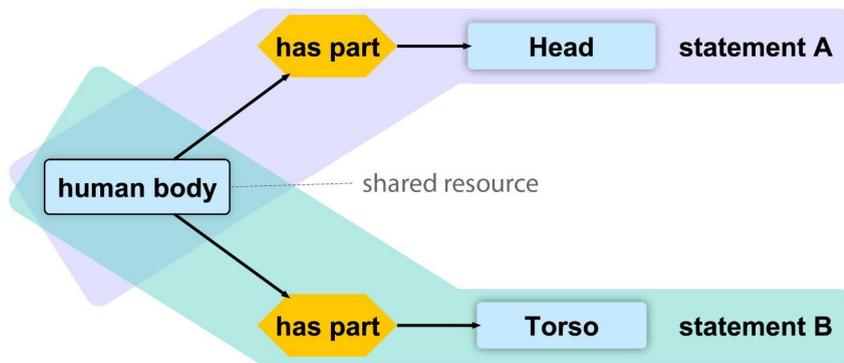

## B) Link

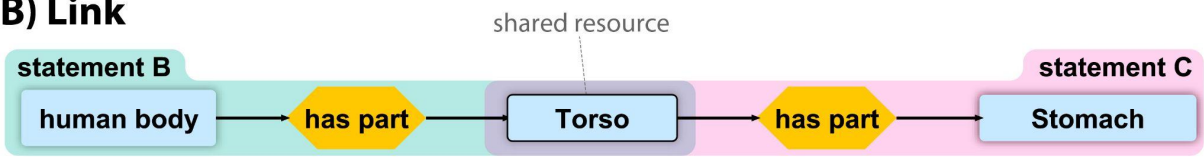

## C) Reference

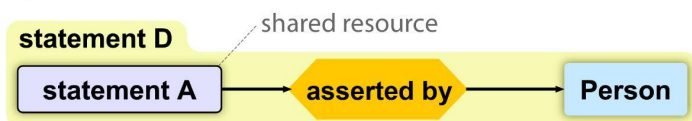

**Figure 16: Three ways to connect statement units in a graph by specifying possible interactions between KGBBs. A) Via association.** A resource can be shared by two or more statement units as their respective subject. This type of connection is described via an association node. **B) Via link**. A resource can be an object of one statement unit and the subject of another statement unit, thus connecting the two statements. This type of connection is described via a link node. **C) Via reference**. The resource of a semantic unit can be the subject resource of another statement unit, thus connecting them. This type of connection is described via a reference node.

A KGBB-driven FAIRer knowledge graph application can be specified by relating instances of KGBB classes to each other via specific types of **interaction-nodes** that define the possible interactions between the KGBBs. Interactions that result in subject-subject connections are specified via **KGBB association nodes**, whereas those that result in subject-object connections are specified via **KGBB link nodes**, and those that result in subject-semanticUnit connections via **KGBB reference nodes** (Fig. 16).

**Example for specifying a knowledge graph application by relating KGBB instances to one another**

Some propositions logically depend on others, such as a weight measurement on the existence of a 'material entity' (BFO:0000040) with the quality 'weight' (PATO:0000128). A quantitative-weight-measurement proposition thus depends on a corresponding qualitative-weight-quality proposition and requires the latter to be specified first. This dependency can be modeled in a KGBB-driven FAIREr knowledge graph application using an instance of a quality statement KGBB and an instance of a weight measurement statement KGBB, and relate each of them via a combination of a link and an association node to an instance of a measurement compound KGBB that models, among other measurements, also weight measurements.

An **association node** instance relates an instance of the measurement compound KGBB as the source KGBB to an instance of the quality statement KGBB as the corresponding target KGBB. Consequently, all measurement compound units will share their subject-resources with the subject resources of the corresponding quality statement units, with the latter having a dynamic label of the form '*OBJECT* has *QUALITY*'. The possibility of connecting the quality statement unit with the weight measurement statement unit by reusing the object-resource of the *QUALITY*-position of the former as the subject-resource of the latter is then specified via a **link node** between an instance of the quality statement KGBB as the link's source KGBB and an instance of the weight measurement KGBB as its target KGBB.

When the quality statement KGBB instantiates a quality statement unit that has '<u>weight</u>' (PATO:0000128) as its resource-object, the KGBB-Engine triggers the measurement compound KGBB to become active and provide the respective input-form for a user to provide the input required for creating the corresponding weight measurement statement unit based on the information provided via the link node and the weight measurement KGBB, with the dynamic label of the measurement statement unit taking the form '*SUBJECT[WEIGHT]* of *VALUE UNIT*'. The weight measurement KGBB then takes the resource-object of the quality statement unit as its subject and builds the data graph together with the node for the weight measurement statement unit.

If the process of measuring the mass of a given object that underlies a weight measurement statement must be described in detail to provide it as relevant metadata for the measurement compound unit, one would need to specify a corresponding process compound KGBB. This KGBB would allow, for example, to reference the measurement protocols that have been followed during the weight measurement process, but also which devices were used in which settings and who performed the measurement where and when. The process compound KGBB would involve the interaction of several instances of various statement KGBBs, all specified via respective association nodes. Now, to indicate that the process compound unit provides metadata for the measurement compound unit, the two compound unit resources must be part of a corresponding statement unit in which one of them takes the subject-position and the other one the object-position. Whereas the possibility to use a compound unit resource in the object-position is specified as an input-constraint in the storage model of respective statement KGBBs, using a compound resource in the subject-position of a statement unit must be specified via a **reference node**.

We thus distinguish three different types of KGBB nodes to specify three types of connections that may occur between the data graphs of semantic units that may be defined as interactions between KGBBs: association nodes, link nodes, and reference nodes—the other two possible types of connections (i.e., *object-semanticUnit* and *object-object*) are specified via input-constraints in respective data graph storage models.

These three types of KGBB nodes together with the KGBBs is all it takes for specifying all possible interactions between the KGBBs of a KGBB-driven FAIREr knowledge graph application. Consequently, a KGBB-driven knowledge graph application can be fully specified as a KGBB specification graph that relates KGBB instances to each other via a set of KGBB node instances. In the following, we describe the three types of KGBB nodes in more detail.

**KGBB association node**

Relating two KGBB instances to each other via a *'KGBB association'* node specifies which and how many target semantic units of a given type may be associated with a source's semantic unit, following which constraints. All target semantic units associated this way relate to the source's semantic unit via the property *hasAssociatedSemanticUnit*. The source KGBB is always a compound KGBB, whereas the target KGBB can be either a statement or a compound KGBB. Moreover, relating a source KGBB to a target KGBB using a *'KGBB association'* node implicitly specifies that the **subject resource of the source semantic unit is also the subject resource of the target semantic unit**. The specification of a *'KGBB association'* node includes the following properties (ranges) (Fig. 17):

1. **MinCount** (*positive integer range, including '0'*): specifies the number of semantic units of the target KGBB that are required for the source KGBB to exist. In other words, if the association node is set to a number higher than '0', then the specified number of target's semantic units must be created along with the source's semantic unit. This is specified by relating the *'KGBB association'* node to a positive integer value via the property *minCount*, which relates to SHACL's SH:minCount. A value of '0' indicates that no such requirement exists.

2. **MaxCount** (*positive integer range, including '0'*): specifies the number of semantic units the target KGBB may associate to a single semantic unit of the source KGBB. This is specified by relating the *'KGBB association'* node to a positive integer range via the property *maxCount*, which relates to SHACL's SH:maxCount. A value of '0' indicates no quantity limitation[11].

3. **CarryOverSubjectRangeConstraintTo** (*UPRI of an object-position class of the target KGBB*), optional: indicates that the constraint that is specified in the source KGBB for the subject of the source's semantic unit must be applied to the here specified object-position of the target KGBB. This is documented at the *'KGBB association'* node by relating it to the UPRI of the corresponding object-position class via the property *carryOverSubjectRangeConstraintTo*. A given association node can have more than one such carry over specification. When the association node is instantiated, the constraint will be tracked with a constraint-node that is linked to the target KGBB's semantic unit resource (see Statement-unit-specific extension to the general semantic-units graph storage model).

A compound KGBB can relate to various other compound and statement KGBBs through *'KGBB association'* nodes, and each respective node indicates whether (and how many of) the target semantic unit is required and whether only one or multiple semantic units of this type may be associated with this compound unit. For example, a user-driven knowledge graph application may have profile item units for its users, with each profile item unit requiring a user email statement unit. Only when a user provides their email address, the corresponding profile item will be created.

In the case of the weight measurement example from above, the measurement compound KGBB relates to the quality statement KGBB via a *'KGBB association'* node. This node is further specified to be *minCount* '1', indicating that a measurement compound KGBB must have at least one quality statement unit associated. The node is further specified to have *maxCount* '1', indicating that exactly one quality statement unit may be associated with the measurement compound unit—no more and no less.

---

[11] Target KGBBs that manage statement units that model predicates with the **logical property of being functional** can only have a maxCount of '1'.

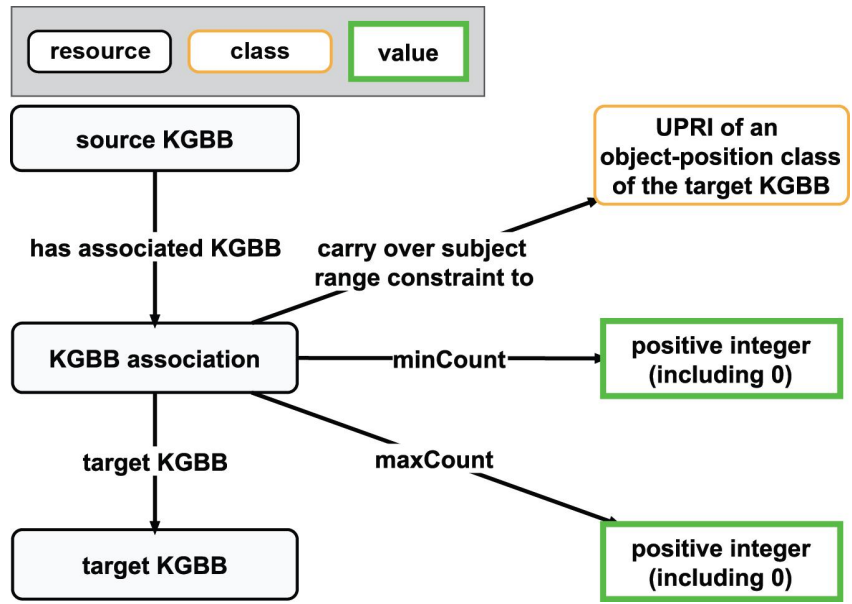

**Figure 17: Specification for a KGBB association node.** The KGBB association node links a source KGBB to a target KGBB via the properties *hasAssociatedKGBB* and *targetKGBB*. The source is a compound KGBB and the target can, in principle, be any type of KGBB. The node itself specifies, via the property *minCount*, how many instances of the type of semantic unit managed by the target KGBB are required, and thus must be specified by a user to add the source's compound unit to the knowledge graph, with '0' indicating no such requirements. The property *maxCount* specifies, how many instances of the type of semantic unit managed by the target KGBB may be added to the source's compound unit, with '0' indicating no quantity limitation. The property *carryOverSubjectRangeConstraintTo* specifies the UPRI of the object-position in the target's storage model, to which the same constraints must be applied as specified in the source's storage model for the subject-position. For every association node holds that the subject resource of the source compound unit is also the subject resource of the target semantic unit.

In summary, via association nodes, specific types of statement units can be related to a specific type of compound unit for organizing and structuring the data graph into subgraphs at various levels of representational granularity, and therewith also determine, which information belongs to a given type of compound unit and will thus be presented together in the UI (see also (14)).

**KGBB link node**

A *'KGBB link'* node is always required when an object resource of a given statement unit should be used as the subject resource of another semantic unit. The linking KGBB is always a statement KGBB, whereas the target KGBB can be either a statement or a compound KGBB. A KGBB link specifies that a linking statement unit, to which a target semantic unit gets linked, relates to this target semantic unit via the property *objectDescribedBySemanticUnit*. The specification of a *'KGBB link'* node that relates a linking KGBB to a target KGBB includes the *minCount* and the *maxCount* properties that are also used for specifying a *'KGBB association'* node. However, instead of the *carryOverSubjectRangeConstraintTo* property, it requires the *useAsSubject* and the *ifObject* properties (Fig. 18):

1. **UseAsSubject** (*UPRI of an object-position class of the linking KGBB*): indicates that the specified object-position class of the linking KGBB should be used by the target KGBB as the subject for its semantic unit. This is specified by relating the *'KGBB link'* node to the UPRI of the corresponding object-position class via the property *useAsSubject*.
2. **IfObject** (*URI of an ontology resource*), optional: specifies a specific ontology resource, of which the resource from the object-position specified via *useAsSubject* must be an instance, a some-instance, or an every-instance resource of, or it must be an instance of one of its subclasses. Only if the resource meets this constraint, the target KGBB is allowed to create its semantic unit. This is specified by relating the *'KGBB link'* node to the corresponding ontology class URI via the property *ifObject*. With this information, the KGBB-Engine can decide, based on the type of resource of the respective object-position of a given source semantic unit, which target KGBB should be activated. As a consequence, the same source KGBB can link to multiple target KGBBs, with each link being mediated by its own link node.

Whereas association nodes specify what can be associated to a given type of compound unit, thus limiting that compound unit's proposition-space, the statement unit of a linking KGBB can be understood to be an **opening proposition** that enables a user to add a new semantic unit to an already existing compound unit via this opening proposition. In other words, link nodes define how a graph may grow through adding new compound units to already existing compound units.

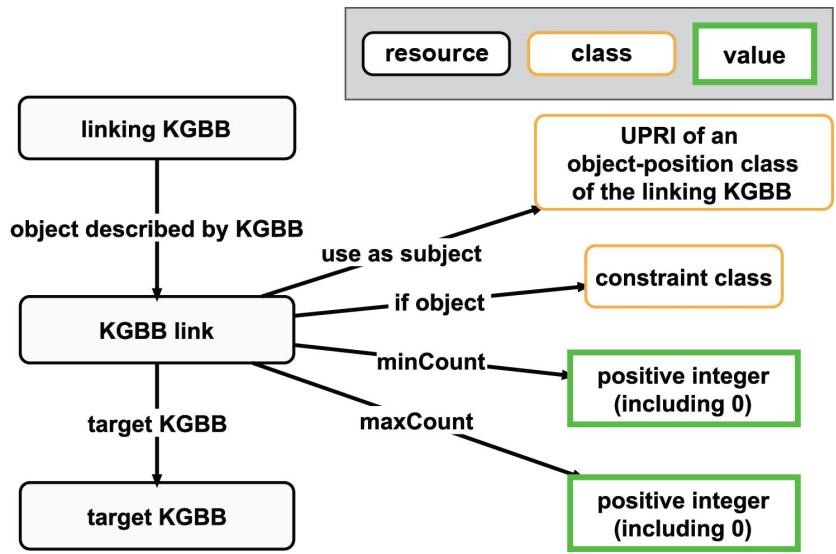

**Figure 18: Specification for a KGBB link node.** The KGBB link node links a linking KGBB to a target KGBB via the properties *objectDescribedByKGBB* and *targetKGBB*. The linking KGBB is always a statement KGBB, whereas the target KGBB can be any type of KGBB. The node itself specifies, via the property *minCount*, how many instances of the type of semantic unit managed by the target KGBB are required, and thus must be specified by a user to add the linking KGBB's semantic unit to the knowledge graph, with '0' indicating no such requirements. The property *maxCount* specifies, how many instances of the type of semantic unit managed by the target KGBB may be added to the linking semantic unit, with '0' indicating no quantity limitation. The property *useAsSubject* specifies an object-position (via the UPRI of the object-position class) from the storage model of the linking KGBB, indicating that an instance of that class must be reused as the subject resource in the target semantic unit. Finally, *ifObject* specifies a constraint that the resource that must function as the target's subject must meet. Only if the resource meets this constraint, the target KGBB is allowed to use it as its subject and create the target semantic unit.

In the case of the weight measurement example from above, the quality statement KGBB relates to the weight measurement statement KGBB via a *'KGBB link'* node. This node is further specified to be *minCount* '0', indicating that a quality statement unit does not have to have a weight measurement statement unit linked to it. With *useAsSubject* and the UPRI of the *QUALITY*-position, the node is specified to use the resource in the *QUALITY*-position of the quality statement unit as the subject for the weight measurement statement unit. With *ifObject* and the value 'weight' (PATO:0000128), the node is also specified to only trigger the weight measurement statement KGBB if the QUALITY-position has 'weight' (PATO:0000128) as its resource. And finally, the node is specified to have *maxCount* '0', indicating that more than one weight measurement statement unit may be linked to the quality statement unit.

**KGBB reference node**

A *'KGBB reference'* node is always required when the UPRI of a given semantic unit should be used as the subject of another semantic unit. The source KGBB can be any type of KGBB, whereas the target KGBB can only be a statement KGBB. A KGBB reference node specifies that a source semantic unit, to which a target statement unit gets linked, relates to this target statement unit via the property *refersToStatementUnit*. The specification of a *'KGBB reference'* node that relates a source KGBB to a target KGBB only requires the *minCount* and the *maxCount* properties that are also used for specifying a *'KGBB association'* node (Fig. 19):

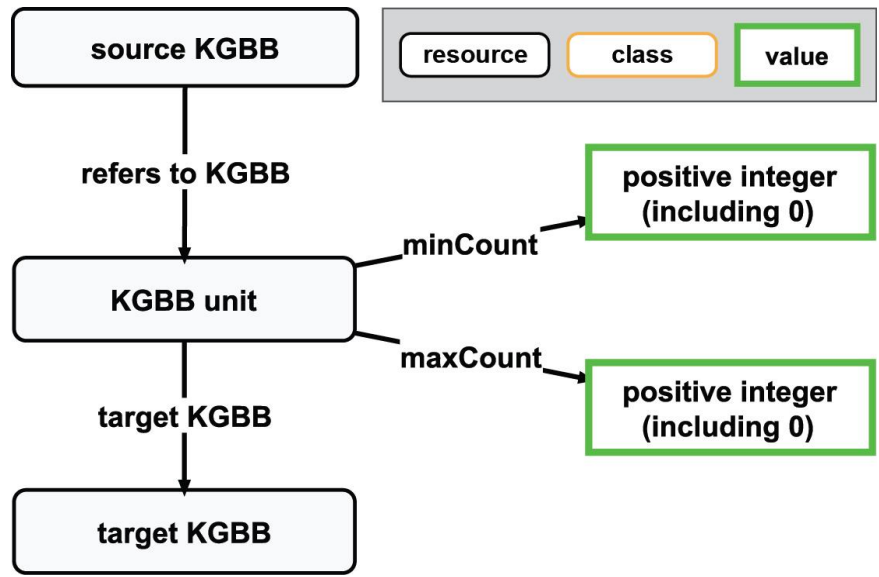

**Figure 19: Specification for a KGBB reference node.** The KGBB reference node links a source KGBB to a target KGBB via the properties *refersToKGBB* and *targetKGBB*. The source KGBB can be any type of KGBB, whereas the target KGBB is always a statement KGBB. The node itself specifies, via the property *minCount*, how many instances of the type of semantic unit managed by the source KGBB are required, and thus must be specified by a user to add the target KGBB's semantic unit to the knowledge graph, with '0' indicating no such requirements. The property *maxCount* specifies, how many instances of the type of semantic unit managed by the target KGBB may be added to the source semantic unit, with '0' indicating no quantity limitation. For every reference node holds that the UPRI of the source semantic unit is the subject resource of the target statement unit.

Reference nodes are required for making statements about statements. Whenever one wants to refer to a semantic unit as a whole, to make statements about it in a statement unit, a reference node is involved relating the respective statement KGBB with the KGBB of the referenced semantic unit. For instance, when the description of the phenotype of a particular multicellular organism is covered by a corresponding description item group unit and this item group unit should be related to the multicellular organism it describes, represented in the graph by an instance of multicellular organism (UBERON:0000468), via an is-about statement unit, it would be managed by the KGBB-Engine using the information in the respective reference node. The interaction required between the item group KGBB and the is-about statement KGBB cannot be described using the association or the link node, because none of them allows using the resource of an item group unit as the subject of an is-about statement unit.

**Combining KGBB association and KGBB link nodes to form KGBB relationship loops**

Sometimes, the addition of a statement unit to an already existing compound unit should result in the automatic creation of another compound unit of the same type. For instance, when adding a has-part statement to an organism item unit, indicating that the organism has a head as its part, it should automatically add a corresponding head item unit, with both item units being of the type material entity item unit. This can be specified on the level of KGBB relationship chains that form loops, starting at the material entity item KGBB that has a *'KGBB association'* node specified whose target has-part statement KGBB links via a *'KGBB link'* node back to the same material entity item KGBB (Fig. 20). The target has-part statement KGBB is a linking KGBB, and its instantiation automatically instantiates the related source material entity item KGBB. This results in material entity item units that have associated has-part statement units, with their PART-objects being further described by another material entity item unit. The semantic relation between these two material entity item units is documented in the graph by the property *hasLinkedSemanticUnit*, connecting the former item unit to the latter.

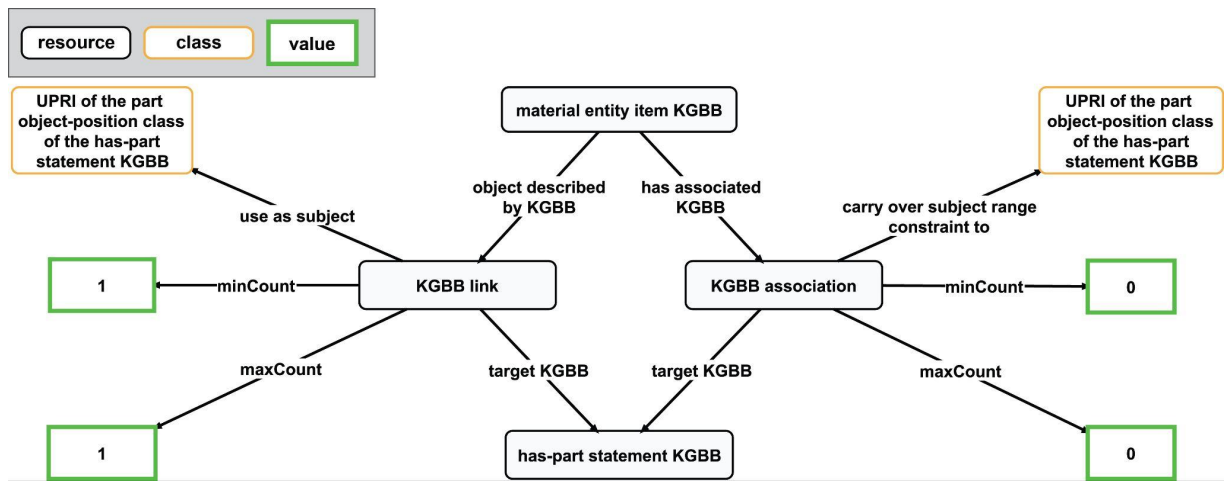

**Figure 20: Specification for a KGBB relationship loop.** The material entity item KGBB is connected back to itself via a has-part statement KGBB using a combination of association and link node. The association node specifies that the constraints that apply to the subject resource of the material entity item unit must also be applied to any added part in associated has-part statement units (*carryOverSubjectRangeConstraintTo* UPRI), that the associated has-part statement unit does not have to be created for creating the material entity item unit (*minCount* 0), and that multiple has-part statement units can be associated (*maxCount* 0) with the material entity item unit. The target of the association node is a has-part statement KGBB, which in turn is linked back to the material entity item KGBB via a link node. The link node specifies that the object resource from the part object-position of the has-part statement unit must be used as the subject of the material entity item unit (*useAsSubject* UPRI), that the linked material entity item unit must be created for creating the has-part statement unit (*minCount* 1; which will result in the automatic creation of, e.g., a *head Y item unit* when creating the has-part statement 'organism X has-part head Y'), and that only one material entity item unit can be linked from this has-part statement unit (*maxCount* 1). The *ifObject* property is not specified for this link node, since the has-part statement KGBB is only linking to one KGBB.

Thus, when providing input to the has-part statement KGBB, a new material entity item unit is automatically instantiated based on the input. In other words, when for the item unit that describes *organism X* a user specifies that *organism X* has part *head Y*, the respective has-part statement KGBB will instantiate the has-part statement unit and at the same time also the item unit describing *head Y*. As a result, the *organism X* item unit is related to the *head Y* item unit via *hasLinkedSemanticUnit*. In this newly created item unit, the user can then specify the parts of *head Y*, with each specified part instantiating a corresponding has-part statement unit and its linked material entity item unit. This way, several linked item units can be created that result in the creation of an item group unit that describes the partonomy of *organism X*.

KGBB relationship loops can be used in a knowledge graph to enable users to provide information on the level of granularity they want by first creating a partonomy of interrelated item units and providing additional information to an item unit of the level of granularity of their choosing.

The use of such loops is not restricted to parthood relations, but applies to any partial order relation such as class-subclass subsumptions, derives from, or before-after relations.

The combination of an association node and a link node can also be used to connect different types of compound KGBBs to each other via some statement KGBB. In this case, however, the target KGBB of the linking node is not the source KGBB of the association node and no loop results.

**The KGBB specification graph: Specifying a KGBB-driven FAIREr knowledge graph application**

Once all the KGBBs are defined for the different types of semantic units that a FAIREr knowledge graph application wants to cover, the application can be fully specified by defining the possible interactions between the different KGBBs by relating instances of them to each other via association, link, and reference nodes as described above. The specification itself **takes the form of a KGBB specification graph that comprehensively characterizes a particular KGBB-driven FAIREr knowledge graph application by defining its data- and proposition-space—the space that its data and metadata may populate**.

With the KGBB-Editor, the KGBB Framework supports the specification of KGBB-driven knowledge graph applications. The KGBB-Editor will allow domain experts to describe their own knowledge graph application in a FAIREr way. The Editor will have access to all KGBB classes from the KGBB-Repository, from which a domain expert may create the KGBB instances they need for their FAIREr knowledge graph. Different instances of the same KGBB class can be connected to instances of several KGBBs in varying ways using the different interaction nodes discussed above, thereby **specifying their particular usage depending on a particular context**. One could, for instance, use the same has-part statement KGBB for describing the partonomy of parts of a material entity but also for describing the partonomy of the different steps of a method. In the former case, one would use an instance of *'<u>KGBB association</u>'* to relate an instance of material entity item KGBB to an instance of the has-part statement KGBB, specifying in the association node that only material entities are allowed as subjects, whereas in the latter case one would use another instance of *'<u>KGBB association</u>'* and relate an instance of method item KGBB to another instance of the has-part statement KGBB, this time specifying in the association node that only methods are allowed as subjects.

All compound KGBBs are described using association, link, and reference nodes and can be stored in the repository as compound KGBB classes that can be reused by any other domain expert for their knowledge graph application. With the property <u>*description*</u> (DCTERMS:description), each KGBB class provides a human-readable description of the type of data it manages. The property <u>*label*</u> (RDFS:label) provides a human-readable label, and the property *<u>manages</u>* relates the KGBB instance to the class of statement units it is managing. The property *<u>hasStorageModelSpecification</u>* relates the KGBB instance to a corresponding LinkML specification. The various object-position classes that the storage model refers to are related to the KGBB instance via the property *<u>hasObjectPositionClass</u>*. Additional properties, i.e., *<u>hasAccessTemplate</u>*, *<u>hasImportTemplate</u>*, *<u>hasDynamicLabelTemplate</u>*, and *<u>hasMind-MapGraphTemplate</u>*, relate to one or more access, import, and display templates.

Finally, since the constraints specified in the storage models of a KGBB class may refer to particular ontology classes for restricting the set of UPRIs that are allowed in a specific object- or subject-position, some KGBBs only work in combination with particular ontologies. The property ***useWithOntology*** specifies these ontologies[12].

All KGBB instances belonging to a given KGBB-driven knowledge graph application are related to that application via the property ***employsKGBB*** and the application's UPRI. Specific KGBBs can be indicated as starting points for adding semantic units to a knowledge graph application by linking the application's UPRI to them via the property ***dataEntryStartingPoints***, which is especially useful for any crowdsourced knowledge graph application.

In summary: every KGBB-driven knowledge graph application can be described and specified as a connection of interrelated KGBB instances using the nodes and properties mentioned above. If this were provided in RDF/OWL and the respective KGBB classes were also specified in RDF/OWL with all properties and KGBB-specific terms defined in this ontology as well, **specifications of KGBB-driven FAIREr knowledge graph applications could be represented themselves in a semantically transparent and truly FAIR way as KGBB specification graphs, resulting in FAIREr knowledge graph software tools**.

## KGBB-Engine

We want to develop a whole set of tools for creating and managing FAIREr knowledge graphs. This tool-set will provide a **development framework for FAIREr knowledge graph applications**, i.e., the **KGBB Framework**, that will guarantee that each semantic unit created by a KGBB-driven knowledge graph is a FAIR Digital Object, with data and metadata that are not only machine-actionable but also fully human-actionable and therefore human-friendly, thus meeting the requirements of **cognitive interoperability** of human readers (14). The KGBB Framework will support researchers and domain experts in setting up their own FAIREr knowledge graph applications.

Each KGBB is a standalone knowledge graph processing module that can be specified independent of other KGBBs and be published as an OWL file, which can be read by the KGBB-Engine and which can be made openly accessible and reusable via a KGBB-repository. This represents the first step towards our goal. The modular architecture allows researchers to reuse KGBBs that others have specified.

The KGBB-Engine takes in a central function in the KGBB Framework and is thus one of its key components. It brings the information contained in the various KGBB classes and in the KGBB specification graph of a knowledge graph application into action, turning KGBBs into small knowledge-processing modules that manage a FAIREr knowledge graph application. The KGBB-Engine must provide the following functions (Fig. 21):

1. Through its KGBB API, it **loads and processes the information from the KGBB specification graph together with the corresponding KGBB class specifications with their storage, display, access, and import templates** from the KGBB ontology. The storage templates in particular, which are specified as YAML files following the notation of [LinkML](LinkML) and which can be translated into [SHACL](SHACL) shapes, help to hide the complexity of the underlying data model from the end user while supporting expert users and developers to interact with the graph. The storage templates can be used for query generation and data visualizations, with their

---

[12] For concrete knowledge graph applications, it may be beneficial to build their own ontologies that grow with their contents, and refer to them instead of referring to external ontologies. In that case, the specification of the KGBB must indicate that.

   reusability ensuring data consistency across all KGBB-driven knowledge graph applications that share the same KGBBs.
2. With its build in **KGBB-Query-Builder**, the KGBB-Engine supports an easy and intuitively usable query interface for the **KGBB-Frontend** that utilizes the CRUD queries derived from the storage models of each KGBB employed in the FAIREr knowledge graph application. Users can combine read queries of different KGBBs by forming unions or intersections of their individual query results and thus create new and more complex queries. Moreover, by utilizing the input-forms provided by each KGBB, users can narrow-in and thus refine the individual queries by adding query parameters such as value-restrictions or class-resource-restrictions for the ranges of each object- and subject-position of a particular type of semantic unit. These queries can even be documented as **statement and compound question units** in the graph, and users can explore them alongside the other semantic units.
3. The KGBB-Engine processes all **CRUD operations on all semantic units in the graph** through its CRUD API. With the support of its KGBB-Query-Builder, it derives the CRUD queries from the information provided by the KGBB storage models. This API allows the use of different database technologies in the backend, including RDF tuple stores, property graphs such as [Neo4j](), other NoSQL databases like key-value databases, [Wikibase](), and relational databases. This way, it **decouples the persistence-layer from the application-layer** and even allows mitigating an existing knowledge graph from one back-end technology to another or the use of different back-end technologies in a federated knowledge graph.
4. When a new statement unit is added to the knowledge graph, the KGBB-Engine decides on the basis of the unit's subject resource whether the statement unit is an assertional, a contingent, a prototypical, or a universal statement unit and **classifies** it accordingly and uses the corresponding display templates when displaying their contents in the UI. It also classifies any some-instance or every-instance identification unit that specifies a cardinality restriction as an instance of *cardinality restriction unit*, and any negated semantic unit as an instance of *negation unit* (for a detailed discussion of these categories of semantic units, see (23)).
5. Whenever a new named-individual, some-instance, every-instance, or class resource is added to the knowledge graph as a consequence of being the subject or the resource-object of a semantic unit that is newly added to the knowledge graph, the KGBB-Engine automatically adds a corresponding identification unit with this resource as its subject to the knowledge graph. Identification units carry all information related to a particular resource such as whether it is a named-individual, a some-instance, or an every-instance resource and also its class affiliation and in case of named-individuals its label (see discussion of identification units in (23)).
6. The KGBB-Engine maps data from the data graph of a semantic unit, stored according to the unit's KGBB storage model, onto the models specified by the KGBB's access and display templates for accessing and displaying the data. This can be utilized in various ways using the Access API: (i) the same data could for instance be visualized in different ways as graphs in the UI, following different graph patterns from **access templates**; (ii) based on the form-based and mind-map-like **display templates** provided by the KGBBs, data can be communicated with the presentation-layer in the UI, with the templates filtering the complex data structure for the information relevant to a human user, using **dynamic labels and dynamic mind-map patterns** for presenting statement, compound, and question unit nodes, therewith **decoupling human-readable data display from machine-actionable data storage**;

and (iii) the data can be exported in different formats, following different standards that are based on the different access templates that have been specified in each KGBB class (**schema/metadata crosswalks**). Moreover, the KGBB-Engine can also provide the data from the data graphs of one or more particular semantic units in the form of JSON, RDF, or as CSV that can be read by R for example. The engine can also provide the underlying data structure from the storage models as Java or Python data classes. Thus, data do not necessarily have to be organized in the low level of abstraction that triples provide. As a consequence, the data managed by the KGBB application becomes **separated from the application and can be readily used in other frameworks**.

7. Through its **Access API**, the KGBB-Engine enables data retrieval through a REST API call with a set of key-value parameters, instead of having to send a SPARQL or Cypher query. It can be used by external applications to access the contents of the FAIREr knowledge graph.
8. Through its **Import API**, the KGBB-Engine enables importing legacy data that is structured in accordance with one of the import templates provided by the KGBB classes.
9. All APIs in the KGBB Framework will follow the open API initiative such as **SmartAPI** which aims at maximizing the FAIRness of web-based APIs. In combination with the description of each KGBB-driven knowledge graph application in the KGBB specification graph as interconnected instances of KGBB classes as described above, **the KGBB Framework provides a FAIREr knowledge graph framework**.

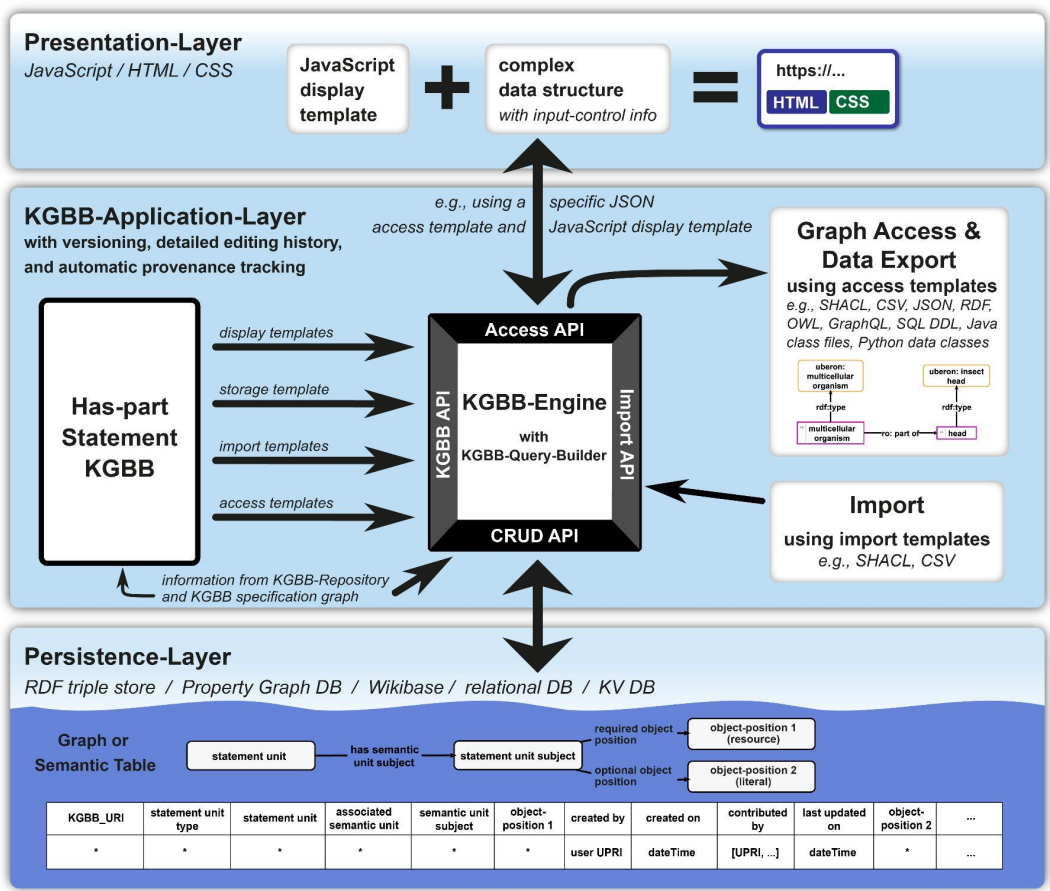

**Figure 21: Schematic representation of the interaction between the Knowledge Graph Building Block Application Engine (KGBB-Engine), a given Has-part Statement Knowledge Graph Building Block, the presentation-layer, and the persistence-layer.**

**Versioning and tracking editing history**

It could be important to be able to track the editing history on the level of particular semantic units–especially, if the knowledge graph were rapidly evolving or if its contents were the result of some collaborative editing effort or even of crowdsourcing, with every user being able to edit any semantic unit in the graph, including semantic units that other users have created. And if the knowledge graph allowed its users to cite one or more of its particular semantic units, thus turning the knowledge graph into a valuable resource for scientific communication, it would also require some versioning mechanism that provides a citable resource for sustaining the cited contents. With such a mechanism, the knowledge graph can continuously evolve through user input while still being citable. The versioning mechanism of the KGBB-Engine supports this and follows two basic ideas:

**1) Soft-delete: no input can be deleted or changed after it has been added to the graph.** Every editing step results in adding new triples to the data graph of the respective semantic unit, with the information provided being linked to a corresponding object-position instance. In other words, object-position instances belonging to a particular object-position class are not updated but, with each editing step, new instances are added to the graph. In order to identify newly added object-position instances, the Boolean property *currentVersion* is used and set to '*true*'. If a user, for instance, updated the VALUE-position of a weight measurement statement unit, the VALUE-position instance that so far contained the latest added value and thus the one that has the value '*true*' for property *currentVersion*, would be set to '*false*', and a new VALUE-position instance would be added with the *currentVersion* value '*true*' and its information would be displayed when a user navigates to the statement unit.

Each object-position instance added this way instantiates the corresponding object-position class, so that the corresponding KGBB can identify all instances of the same object-position of a given semantic unit. All previously added object-position instances of the same object-position class are still in the data graph of the corresponding statement unit, and their information could be accessed anytime and sorted by their creation date if needed, thus providing a **detailed editing history** for each object-position of a statement unit, but also for the statement unit as a whole, and even for any semantic unit it is associated with. So, in case of the weight measurement example from above, information about who entered when which weight measurement value could be presented in the UI. Each object-position possesses the respective information. This information is valuable in the context of managing collaborative efforts of editing contents of a knowledge graph, and also for crowdsourced FAIREr knowledge graphs to identify who added/edited what and when.

The same approach is applied not only to individual object-positions, but to particular semantic units as well. The respective semantic unit resource also specifies through the property *currentVersion* its status. In case a user wants to delete a semantic unit, instead of actually **deleting** the respective unit and maybe even all of its associated units, *currentVersion* is set to '*false*' for this unit and possibly its associated units. By default, the application does not display information about resources that have *currentVersion* set to '*false*', and in the UI it looks like the information has been deleted, although it is still available in the knowledge graph and all corresponding provenance metadata can still be accessed **[A2.]**.

**2) A user can create a version of any semantic unit and that version cannot be edited anymore and is thus persistent through time.** Users can create versions of particular semantic units. Each such version is represented by its own node in the graph. It possesses its own UPRI, which could take the

form of a DOI. Consequently, such versions can be referenced and cited in publications. The version node also tracks metadata through respective properties, like any other semantic unit. Each version could, additionally, also have a **content identifier (CID)** that uses cryptographic hashing for **decentralized content addressing**.

Several versions can exist for a given semantic unit. The node representing the newest version is linked to its semantic unit node via the property *hasCurrentVersion* (PAV:hasCurrentVersion), whereas all other version nodes are linked to it via the property *hasVersion* (PAV:hasVersion). The version nodes of a semantic unit are linked to each other through a chain of *previousVersion* (PAV:previousVersion) properties, starting from the newest version to its respective previous versions.

All object-position instances that belong to the versioned semantic unit will also be updated to include the UPRI of the versioned semantic unit as a value for their respective *versionID* property. With this information, it is always possible to access the data belonging to the respective version by querying for all object-position instances with the UPRI as a value for this property. This property can be applied several times to the same object-position instance to track more than one version UPRI. This is necessary, because not all object-positions of a given semantic unit may have been edited between two versions, and thus a particular object-position instance may belong to more than one version.

Both aspects, the soft-delete and the versioning, have been demonstrated in a small Python-based prototype of a FAIR scholarly knowledge graph application that employs semantic units and KGBBs and that uses Neo4j as the persistence-layer technology (62). It includes versioning of semantic units and automatic tracking of their editing histories and their provenance. The prototype is available via https://github.com/LarsVogt/Knowledge-Graph-Building-Blocks.

**Storing and identifying statement units**

As already mentioned, the KGBB-Engine is planned to allow storing semantic units in five different types of persistence-layer technologies. Storing in an **RDF** tuple store follows the general schema of **nanopublications** (36–38,63) as described in (23) and requires the tuple store to allow for Named Graphs, with the Named Graph that contains the data graph of a statement unit having the same UPRI as the statement unit itself so that the resource stands for both the statement unit and the statement unit's data graph.

For NoSQL databases such as **labeled property graphs** (e.g., Neo4j), each relation in the knowledge graph specifies to which semantic units it belongs via respective **property-value pairs**, with some values being arrays of UPRIs of different semantic units because a given relation can be associated with more than one semantic unit. The property *statementUnitURI* identifies to which statement unit a given relation belongs, whereas *compoundUnitURI* identifies the compound units, and *listUnitURI* the list units. The properties *versionID* and *datasetUnitID* specify to which versioned semantic units and which dataset units they belong. This notation allows for identifying all triples belonging to a given semantic unit by a straightforward and simple Cypher query. In the case of a statement unit, this query could look like this:

'MATCH (n {current_version:"true"}) WHERE ("x" IN n.statementUnitURI) RETURN n',

with *x* being the UPRI of the particular statement unit. This applies to the other categories of (versioned) semantic units correspondingly.

When storing the data associated with a semantic unit in a **relational database**, several tables need to be created for each KGBB instance, based on the various parts of its overall storage model. The **semantic header table** stores the data from a unit's semantic-units graph and only differs in the presence of some of its properties between compound and statement units. Every newly added semantic unit adds another row to this table. The table consists of the same slots and values as its corresponding semantic-units graph storage model specifies (cf. [The general semantic-units graph storage model](#) and [Statement-unit-specific extension to the general semantic-units graph storage model](#)).

For storing data belonging to a statement unit, additional tables are required that capture the data from the unit's data graph. The **subject-related data graph table** stores the data from the subject-related data graph storage model of a statement unit (cf. [Subject-related data graph storage model](#)). In a given row of the table, the properties *requiredObjectPosition* and *optionalObjectPosition* can be present several times, depending on the number of object-positions in the corresponding storage model of the respective KGBB. Statement units have, in addition, for each object-position in the storage model a separate additional **object-related data graph table**, consisting of the same slots and values as defined in their object-related data graph storage model (cf. [Object-related data graph storage models](#)).

The combination of a semantic header table and a set of subject- and object-related tables constitutes the representation of the data of a semantic unit in a **semantic table** that can be stored and managed in a relational database. Semantic tables are semantically isomorph to graph-based representations, such as in RDF with Named Graphs or as in a labeled property graph—no information gets lost. In other words, one can move data back and forth between these different backend technologies without loss of information.

We plan to extend the storage options for KGBB-driven knowledge graph applications to other NoSQL databases such as **key-value databases**, but we still have to evaluate their principle suitability. The same applies also to **[Wikibase](#)** as another storage option.

**A first KGBB-Engine prototype**

In a master thesis project, we evaluated the capabilities of the KGBB Framework by developing a basic prototype of the KGBB-Engine that processes a defined set of KGBB specifications to create a KGBB-driven application. So far, it processes all types of n-ary statement units, item units, and typed statement units whose storage models are documented in YAML, following the LinkML notation schema. Moreover, it manages to read semantic specifications of KGBB-driven knowledge graph applications, using instances of KGBB classes and the association and the link nodes as described above, and creates a corresponding knowledge graph application. The prototype can also make use of multiple database technologies using the KGBB-CRUD API and LinkML storage templates, providing both support for RDF triple stores via SPARQL-query generation and support for a Neo4j database via Cypher-query generation. Due to its modular approach, it is also possible to support other storage technologies, i.e., SQL-Databases or key-value stores, in the future (https://gitlab.com/TIBHannover/orkg/semantic-building-blocks). The KGBB-Engine has been developed following the hexagonal architecture approach (53,54), which supports the development of new features in a collaborative way by integrating these features via their own APIs and adapters to the stack of already existing features.

## KGBB-Functions

**KGBB-Functions** can communicate with the KGBB-Engine via its Access API and thus retrieve data from the graph and send data back to the graph in the form of JSON, RDF, or as CSV, or they can utilize the Java and Python data classes generated by the engine to better interact with the underlying data structure. This reduces the barrier for software developers to contribute new KGBB-Functions to the KGBB Framework. KGBB-Functions can be stored in the KGBB-Repository and shared for reuse.

An example for such a function would be *units-conversion*, a KGBB-Function that can be associated with any measurement statement KGBB to convert the values of measurement statement units across various measurement units, such as from kilometers to miles. Another example would be *proportion-calculation* for automatically calculating for instance the density value for a given object based on its mass and volume. The result of the calculation could be fed back to the KGBB-Engine through its Import API and a respective density statement unit could be added automatically to the graph or made available as additional dynamic in-memory information. As a result, the three corresponding statement KGBBs for mass, volume, and density measurement statement units can be specified in the *proportion-calculation* KGBB-Function to form a functional module. Other types of dynamic information such as rankings, deductions, inductions, and basically anything based on reasoning or rule-based inferences, can be provided through KGBB-Functions.

But also tools for visualizing data from semantic units could be implemented in KGBB-Functions, such as interactive data graphics using, for example, [Vega-Lite](Vega-Lite) (64), or a tool for managing geospatial data using GeoSPARQL, OpenStreetMap, and other map-based visualization functions.

By developing KGBB-Functions against the different APIs of the KGBB-Engine (in particular the Access REST API), KGBB-Functions can be added to any knowledge graph application as **functional building blocks that are associated with particular Knowledge Graph Building Blocks, forming interactive units**.

Some KGBB-Functions will be essential and integral parts of the KGBB-Engine such as user access control, knowledge graph management operations, data validation services, or convenience services like searches for a specific type of label (rdfs:label, skos:prefLabel, kgbb:dynamicLabel), others will contribute to a layer of optional services such as graph visualization services, data analytics tools, or commenting functionalities.

## KGBB-Editor

Another step towards our goal of an openly and freely usable KGBB Framework would be the development of an intuitively usable **low-code KGBB-Editor** that enables **researchers and domain experts with no experience in Semantics, graph modelling, and programming** to specify their own KGBBs and their own KGBB-driven FAIREr knowledge graph application (*user story 9*) by specifying it in the form of a KGBB specification graph.

When having to specify a new KGBB class, users of the KGBB-Editor would first select a suitable general KGBB class from the taxonomy of KGBB classes and add further constraints and parameters to it, resulting in a more specific KGBB. Alternatively, when having to specify a new statement KGBB, the KGBB-Editor would guide the user through the following list of questions, the answers of which enable the KGBB-Editor to create the new statement KGBB class and its associated statement unit class, **without the user having to make any semantic modelling decisions**:

1. **What is the predicate or verb of the statement?**
   Is part of the label of the statement unit class.
2. **Provide a description or characterization of the type of statement you want to add to the knowledge graph. What kind of propositions will it cover?**
   Provides a human-readable definition of the newly created statement unit class. It could also take the form of a definition of the statement's predicate.
3. **Provide some example statements.**
4. **Specify the number of object-positions the statement should cover.**
5. **For the subject-position and each object-position, provide a short and meaningful label.**
   Each label takes the function of a **thematic label** that characterizes the **semantic role** that is associated with each position (cf. (48)). It is used for various purposes, including as placeholder labels for input fields and as placeholder labels for variables for the textual and graphical display templates. It also provides the label for the corresponding object-position class.
6. **Specify, which of these object-positions are required (i.e., arguments) to form a semantically meaningful statement with the subject and the predicate, with all other object-positions being optional (i.e., adjuncts).**
   This specifies the object-positions that are linked in the subject-related storage model of the statement KGBB using the property *requiredObjectPosition*, with all other object-positions being linked using the *optionalObjectPosition* property. In the UI of the Editor, this could be done using checkboxes.
7. **For each object-position, provide a short description of the type of objects the object-position covers and give some typical examples.**
   This provides the human-readable definition for the corresponding object-position class. This text can also be used as tooltip for the corresponding input field for adding or editing respective statement units.
8. **For each object-position, decide whether the object must be represented in the form of a resource or a literal. In the case of a resource, choose an ontology class that best specifies the kind of object that is allowed for the object-position, and in the case of a literal, choose the datatype and, if required, define datatype specific constraints.**
   This provides input-constraints for each object-position, which are documented in the object-related data graph storage model of the statement KGBB.
9. **Decide whether any logical properties apply to the statement's predicate (e.g., transitivity), and if they apply, which object-position is affected by them.**
   The KGBB-Editor adds this information to the respective object-position class via the corresponding Boolean annotation property (e.g., *transitive*). This information is optional.
10. **Write a human-readable statement using the thematic labels for the subject and the various object-positions and the predicate.**
    Provides the dynamic label pattern for this statement unit.

Based on the generic structure of the storage model of statement KGBBs, the KGBB-Editor will then process the information and automatically specify the storage model, create the respective KGBB class, the statement unit class, all object-position classes, and the display template for the dynamic label. The object-position classes hereby take a function similar to the syntactic positions and their associated semantic roles in ProbBank (50). Now, the newly created KGBB is already ready to be used.

The KGBB-Editor will use the thematic labels of the subject-position and the different object-position classes in any subsequent steps in which a user has to refer to the storage model, for instance, when specifying an access template, where resources have to be mapped from the storage model to the access model (cf. Fig. 7, 9, 12, 13, 14).

## Discussion

One of the main points of criticism against the KGBB Framework is the need to specify a statement KGBB for each type of proposition required in a knowledge graph. This limitation results from the fact that in a KGBB-driven knowledge graph application, the statement KGBBs define the proposition-space.

While this criticism is valid, we want to reply that truly FAIR data and metadata nevertheless requires that all statements in a knowledge graph can be identified individually and that each statement references the graph pattern on which it has been modeled (schematic interoperability [F6.1/I7.1]; *user story 2*). Having to model each type of statement used in a knowledge graph is not unique to the KGBB-based approach, but applies to FAIR knowledge graphs in general. The difference to other knowledge graph frameworks is that the KGBB-Editor will substantially support the creation of semantic models for new types of statements and will not require experience in Semantics, RDF, OWL, or any graph query language. Moreover, once specified and made available in the KGBB-Repository, a statement KGBB can be reused by any other party, reducing the effort needed for defining any KGBB-driven knowledge graph application.

## KGBBs, the EOSC Interoperability Framework, and cognitive interoperability

The [European Open Science Cloud](#) (EOSC) is an environment for hosting and processing research data to support EU science. In the context of the FAIR Guiding Principles, EOSC created an EOSC Interoperability Framework that differentiates between technical, semantic, organizational, and legal interoperability (10). We are convinced that the KGBB Framework can help improve the first two types of interoperability in particular.

Institutions that operate their own KGBB-driven knowledge graph applications can share their data and even built a federated virtual knowledge graph with their partners, while data stewardship remains in their own hands, thus ensuring full control over their data's ethical, privacy, or legal aspects (following Barend Mons' *data visiting as opposed to data sharing* (9)). Moreover, by decoupling the application-layer from the persistence-layer and allowing mitigating an existing knowledge graph to various different storage technologies, by providing the possibility to specify additional access models and enabling accessing the data of the knowledge graph in various formats (including Java and Python data classes), by allowing the documentation of data and metadata at various levels of granularity, by including the KGBB specification graph in the knowledge graph itself and thus providing semantic transparency regarding the application itself, and by making KGBBs, KGBB-Functions, the KGBB-Engine, the KGBB-Editor, and the KGBB-Frontend openly and freely available, the KGBB Framework meets the recommendations of the EOSC Interoperability Framework for **technical interoperability** (10).

On the other hand, by using ontology terms and other controlled vocabularies that provide UPRIs and publicly-available definitions for their class terms, by tracking and documenting metadata and allowing to extend the documented metadata to include also disciplinary metadata and by providing a way to specify metadata crosswalks, by decoupling human-readable data display from machine-actionable data storage, with the former focusing on providing human-readable data views and the latter involving the consistent application of KGBB-specific storage models that guarantee the semantic interoperability of data and metadata, by providing the possibility to apply community-specific standards for accessing data and metadata, and by documenting for every data graph of every semantic unit which storage model has been used for modelling its data, KGBB-driven knowledge graph applications also meet the recommendations for **semantic interoperability** (10).

In addition to these two interoperability layers, we argue that the KGBB Framework also supports increasing the cognitive interoperability of data and metadata by adding more structure to the graph and organizing it into various layers of semantic units that belong to different levels of representational granularity, by providing human readers intuitive tools for exploring the graph that leverage these added layers and levels of granularity as well as different granularity trees and reference frames, reducing the graph's complexity to what is momentarily of interest to a human reader and organizing it in ways that enables them to filter out information that is only relevant to machines, and enabling them to zoom in and out across the different levels of representational granularity. Increasing the **human explorability of data and metadata** thus supports how the graph's contents can be communicated with human readers—something increasingly important with an increasing size of the graph—thereby guaranteeing not only the machine-actionability of data and metadata but also their human-actionability and thus their human-friendliness, therewith meeting the recommendations for **cognitive interoperability** (for a detailed discussion see (14)).

The KGBB Framework also meets the requirement of cognitive interoperability with respect to developers and operators of knowledge graph applications by providing easy-to-use tools that take away some complexity layers and requirements such as having to be proficient in query languages, shape languages, or RDF and OWL. The KGBB-Editor will automatically specify the storage model based on the specification of the required and optional inputs for a given type of statement unit and the KGBB-Query-Builder will derive generic CRUD queries based on this storage model—no need for developers and operators to define storage models and write the queries themselves. Moreover, with its modular architecture, KGBBs once defined and KGBB-Functions once developed can be reused and connected in a straightforward way using the low-code KGBB-Editor to specify one's own knowledge graph application in a KGBB specification graph. Since every KGBB-driven knowledge graph application is documented by such a KGBB specification graph that represents the different KGBBs employed in the graph and their connections in a semantically transparent and FAIR way, understanding the general architecture of the underlying data model of a knowledge graph is also supported and contributes to the overall cognitive interoperability provided by the KGBB Framework.

## Revisiting the user stories

The trade-off between human-actionability and machine-actionability of data and metadata does not exist in the KGBB Framework because KGBB-driven knowledge graph applications **decouple data display from data storage**. Each KGBB provides form-based textual and mind-map-like graphical display templates for one or more views on the data and metadata of the corresponding type of semantic unit. In other words, each KGBB provides information for the presentation-layer on how to display which information from which nodes of the respective unit's data graph, even allowing displaying the same data and metadata in different ways using different predefined display templates. This provides the conceptual basis for specifying context-dependent data and metadata representations. The storage model thereby allows the storing of information in a machine-actionable format with all the information that machines need made explicit, while the display templates define how to reduce the information from the data graph to what is actually relevant to a human reader and ways to present this information in a UI. The concept of semantic units (23) implemented in the KGBB Framework thus provides a new conceptual framework for developing adequate UIs that increase the **cognitive interoperability** of data and metadata, turning a FAIR knowledge graph into a FAIREr knowledge graph, i.e., a knowledge graph that is easier to explore for humans (14). Thus, **using the KGBB Framework, Tom can produce large, complex graphs of FAIR data that are still intuitively explorable (*user story 1*)**.

The use of KGBBs and the possibility to connect different KGBB instances to one another, with each KGBB class providing display and access templates and a storage model with input-constraints, allows developers of knowledge graph applications to specify their application on the abstract level of KGBB instances and their possible interactions. The resulting KGBB specification graph specifies all interaction that the application allows between its KGBBs. When adding data to such a knowledge graph application, users only see input forms that guide them when adding contents to the graph–they do not have to make semantic and data modelling choices themselves. Since each semantic unit of the same type is stored using the same KGBB-specific storage model and documented in the graph as a FAIR Digital Object, all data contributed through KGBBs will be fully interoperable and FAIR. Therefore, **as users of a KGBB-driven crowdsourced knowledge graph, Catherine and Felix can add FAIR data without having to make any semantic modelling decisions themselves (*user story 2*)**.

By organizing the data graph layer of a knowledge graph into different types of semantic units, by representing these semantic units with their own UPRI, by specifying their affiliation to a semantic unit class, and by associating a corresponding KGBB to each semantic unit class that provides a storage model from which the KGBB-Engine automatically derives CRUD queries, users of FAIREr knowledge graph applications can employ queries from different KGBBs and combine them like query-building-blocks in a modular way supported by the **KGBB-Frontend** and the **KGBB-Query-Builder**, forming unions or intersections of individual query results, and they can even save and document them in the form of **question units**. The input-forms provided by each KGBB can be used for specifying the query parameters for more detailed queries so that users can intuitively create queries without having to be experts in any graph query language. Since the KGBB-Engine derives the CRUD queries automatically from the storage model specification provided by each statement KGBB, researchers with no programming experience and no knowledge about graph query languages like Cypher or SPARQL will be able to use the KGBB-Query-Builder to combine queries provided by the engine to create new queries. As a result, **for a KGBB-driven knowledge graph application, Sara and Bob do not necessarily have to know graph query languages for querying the knowledge graph because they can use the KGBB-Query-Builder of the KGBB-Engine to intuitively create their own queries (*user story 3*)**.

The KGBB-Engine automatically organizes a knowledge graph into various **semantically meaningful subgraphs** at multiple levels of representational granularity, identifying different granularity trees and frames of reference in the graph, and assigns a UPRI to each such subgraph. Thus, the resulting semantic units are **identifiable** and can be **referenced** in triple statements, with the consequence that making statements about statements becomes straightforward. Automatically structuring and organizing the data graph into different semantic units provides an efficient way of organizing a knowledge graph to allow users to intuitively make statements about statements, and also provides a clear and straightforward implementation schema that is beneficial for developing relevant search schemata. As a consequence, **by employing the KGBB Framework, Karl can operate a KGBB-driven knowledge graph application that provides his users intuitive ways for making statements about statements (*user story 4*)**.

The KGBB Framework **decouples data access from data storage** by mapping the data from the storage model to one of the defined access models for each statement KGBB. Users can access the data and metadata belonging to a given semantic unit according to any of its specified access templates. Each KGBB allows the specification of one or more such access model. Additional access models can be added anytime using the low-code KGBB-Editor. If several access models are defined, the KGBB-Engine even enables **schema crosswalks** from one access model to another. Thus, the KGBB Framework allows the inclusion of newly emerging standards for already existing KGBB-driven knowledge graphs and therefore allows adapting to an ever-evolving research landscape. Moreover, it is possible to include different community-specific data models for the same type of data. The same applies to all provenance and other types of metadata, thus also enabling comprehensive **metadata crosswalks**. When accessing or exporting the data and metadata, users only have to choose which model to use from the set of specified access models (or access model families). Therefore, **by specifying different access models for KGBBs, Dan and Anna can contribute their measurement data as FAIR data while still being able to access them in a format or model that is compliant with their community-specific domain standards (*user story 5*)**.

The KGBB Framework **decouples the storage model from the storage technology**. The KGBB-Engine provides an API with which the knowledge graph can be stored in and accessed from various database technologies, allowing a knowledge graph application to change its database technology at any time. Moreover, the organization of the overall data graph into semantic units significantly supports the querying-properties of the data graph, since queries can be systematically split into different steps. This gives the storage model more flexibility with regard to querying. Because more detailed models can be provided through the specification of various access models, the storage model can be optimized for efficiently storing data and metadata and thus for scalability and does not have to take into account any domain-specific standards and requirements. Moreover, if, for instance, a given knowledge graph were stored in a graph database that at some point does not scale well anymore, the corresponding data and metadata could always be moved to a relational database, stored as semantic tables and thus in a tabular format due to the flat graph pattern of the generic structure of the storage models. All semantics are preserved. If an RDF or Neo4j representation of the data is required, the table can be accessed as an RDF or a labeled property graph employing respective access templates. Therefore, **with a KGBB-driven knowledge graph application, John could always change the technology used in the persistence layer if the current technology ran into scaling issues (*user story 6*)**.

The organization of a knowledge graph into semantic units allows assigning KGBBs and their subject-position and object-positions to specific rows and columns of tabular datasets. Deep learning text-mining algorithms can be used for identifying values in the cells and mapping them to datatypes and ontology terms. As a result, the import and step-wise semantification of legacy data from a tabular format could be significantly supported and accelerated. Moreover, if the knowledge graph already possesses a lot of data, this data can be utilized as training data for deep learning algorithms. Since the data is partitioned and associated with KGBBs and their storage model parameters, adequate training datasets for KGBB-specific AI-based NLP tasks could be easily identified. Therefore, **KGBB-driven knowledge graph applications could possibly support Mila in importing semantically unstructured (legacy) data from a tabular format (e.g., CSV, Excel sheets) into a FAIR knowledge graph (*user story 7*)**.

By organizing a knowledge graph into different semantic units and classifying each statement unit into either a universal, a contingent, a prototypical, an assertional, or a lexical statement unit and by having resources to represent named-individuals, some-instance, every-instance, and class entities, KGBB-driven knowledge graphs provide a conceptual framework that facilitates the formal distinction between universal, contingent, prototypical, assertional, and lexical statements in a knowledge graph. Moreover, the use of the four types of resource entities together with semantic units allows a notation for negation and cardinality restrictions that does not require the use of any blank nodes and is also easier to comprehend than the alternative OWL based notations (see (23)). Thus, **by using a KGBB-driven knowledge graph application, Nick can document data from clinical studies in a FAIREr knowledge graph and differentiate between universal, prototypical, and assertional statements and can intuitively document absences, negations in general, and cardinality restrictions (*user story 8*)**.

By providing a generic storage model that can be applied to any proposition and thus any type of statement unit, the KGBB-Editor allows domain experts without any experience in Semantics to specify new statement KGBBs and with them new statement unit classes without having to think about their underlying data model–the domain experts only have to answer a list of questions. Thus, **by using the low-code KGBB-Editor, Paula could specify the statement KGBBs she needs for documenting the data from her project, without having to think about how to semantically model them (*user story 9*)**.

## Related and future work

To our knowledge, the concept of semantic units or something similar has not been proposed before. Since KGBBs conceptually depend on and relate to semantic units, there are no tools or frameworks that we can compare the KGBB Framework to, but we can compare some of its aspects and features to other existing frameworks.

Most similar in scope are probably systems implementing the **ontology-based data access (OBDA) paradigm**. It has become a popular paradigm for accessing data stored in legacy sources using Semantic Web technologies (65), by accessing the data through a conceptual layer that provides a query vocabulary over all the different relational legacy data sources connected to the system. The conceptual layer is expressed as an RDF(S) or OWL ontology and is connected to the underlying relational databases using R2RML mappings. This allows querying the relational databases using SPARQL.

The commercial enterprise knowledge graph platform **[Stardog](#)** follows the OBDA paradigm and enables enterprises to access and unify all their unstructured and semi-structured data sources stored in a variety of proprietary database systems of different types, resulting in a virtually connected enterprise knowledge graph. Stardog claims to help to unify all the different data silos that may exist in a company, leave the data where they are, but nevertheless leverage them in a federated approach via their query engine. Stardog's data virtualization engine accesses the siloed data sources to enrich the knowledge graph with more context. Their semantic graph, in turn, adds a metadata-driven abstraction layer over the siloed data sources that allows querying them, that infuses meaning into them, and brings back a consistent and unified data representation. It also allows the application of alternative schemata to support the particular needs and requirements of different stakeholders. Their inference engine uses machine learning and reasoning and creates new connections and complements existing data with new nodes and relationships.

[Ontop](#) is an open-source framework that also implements the OBDA paradigm by translating the SPARQL queries into SQL queries, which turns a set of relational databases into a **virtual knowledge graph**, with the contents being enriched by a domain ontology (66).

[Metaphactory](#) is an enterprise platform for building FAIR knowledge graph applications. It comes with several built-in functionalities that provide services for different categories of users, ranging from domain experts with no background in semantics or computer science, to expert-users and data scientists, and to data stewards and application developers. It provides a web-component based UI that uses a customizable templating mechanism that allows specifying template pages associated with resources of the same type, enabling expert-users to create their own UIs. This is comparable to the KGBB-specific form-based display templates that can also be configured by expert-users. Metaphactory also provides authoring-control based on knowledge graph patterns comparable to the controls provided by the storage templates of KGBBs. Moreover, like the KGBB Framework, metaphactory enables rapid building of knowledge graph applications with reusable components and supports collaborative knowledge modelling and knowledge generation. By enabling the integration of ontology-based data access engines such as Ontop or engines such as Stardog that additionally integrate a triple store, metaphactory can virtually integrate native RDF triple stores with relational databases that are accessible via SPARQL (67). Although following open standards, however, metaphactory is a commercial platform that is not open source.

Although some overlap exists, metaphactory's, Stardog's, and Ontop's overall scope and objective differs considerably from that of the KGBB Framework. All three systems do not support semantic units or any comparable approach for improving the explorability and with it the cognitive interoperability of knowledge graphs. They focus on integrating already existing non-semantic data sources into a virtual knowledge graph, whereas the KGBB Framework focuses on adding new data into a newly built FAIREr knowledge graph landscape.

[Apache Stanbol](#) is an open-source modular software stack with a reusable set of components for semantic content management. It provides a service for content enhancement that adds semantic information to content, reasoning services for retrieving additional semantic information based on the semantic information retrieved via the content enhancement, and services to define and manipulate data models (e.g., ontologies) used for storing the semantic information. [Semantic MediaWiki](#) is a free, open-source extension to [MediaWiki](#) that can turn a wiki into a flexible knowledge management system by storing and querying data within the wiki's pages. Neither Apache Standbol nor MediaWiki support semantic units or any comparable concept, and both do not create knowledge graphs, but instead enrich non-semantified contents with semantics.

[SHACL](#) and [ShEx](#) are shape constraint languages for describing RDF graph structures (i.e., shapes) that identify predicates and their associated cardinalities and datatypes. Shapes can be used for communicating data structures, creating, integrating, or validating graphs, and generating UI forms and code.

[Reasonable Ontology Templates](#) (OTTR) takes up the idea of shapes and uses them as building blocks for knowledge bases. The templates provide an abstraction level that is better suited for managing a knowledge base than the low-level RDF triples or OWL axioms. The templates are stored in a template library which supports reuse and therewith uniform modelling across different knowledge bases. Since templates can refer to other templates, OTTR's modularity is comparable to the KGBB Framework. Moreover, with its clear separation of templates and their instantiations in the form of data, OTTR clearly separates between knowledge base design and knowledge base contents. This is also comparable to the KGBB Framework. Tools exist for mapping CSV or relational data to templates.

Comparable to OTTR is LinkML. [LinkML](#) is a general purpose, platform-agnostic, object-oriented data modelling language and framework that aims to bring Semantic Web standards to the masses and that makes it easy to store, validate, and distribute FAIR data (68). It can be used with knowledge graph applications for schematizing a variety of data types, ranging from simple flat checklist standards to complex interrelated normalized data that utilizes inheritance. LinkML fits nicely into frameworks common to most developers and database engineers, such as JSON files, relational databases, document stores, and Python object models, while providing a solid semantic foundation by mapping all elements to RDF URIs. With LinkML you can model data schemata and data dictionaries by authoring YAML files using the LinkML notation.

LinkML also comes with various tools. Generators provide automatic translation from the schema YAML to various formats, including JSON-schema, JSON-LD/RDF, SPARQL, OWL, SQL DDL, ShEx, GraphQL, Python data classes, Markdown, and UML diagrams. Loaders and dumpers convert instances of a schema between these formats. Generators thus allow integration with tools offered in other technical stacks. The SPARQL generator allows generating a set of SPARQL queries from a schema and the Excel Spreadsheet generator to create a spreadsheet representation of a schema. LinkML currently supports four different strategies for data validation: 1) validation via Python object instantiation; 2) validation through JSON-Schema; 3) validation of triples in a triple store or RDF file via generation of SPARQL constraints; or 4) validation of RDF via generation of ShEx or SHACL.

Generators also exist for object models particular to specific languages, such as Python, Pydantic, Java, or Typescript. The Java Generator produces Java class files from LinkML models and provides optional support for user-supplied jinja2 templates. Moreover, generators exist for specific database frameworks, including SQL DDL and SQL Alchemy. In other words, LinkML supports both complex interlinked normalized relational data and flat denormalized data, such as from spreadsheets and CSVs.

LinkML provides the notation that we plan to use for documenting and the tool set for processing the various KGBB storage models.

We also follow with interest the ongoing [Abstract Wikipedia](#) project, especially the part relating to [Constructor Units](#) and the abstract content language. Constructor Units provide abstract representations of predicate statements and thus follow an idea that shares some similarities with our approach of the generic storage for statement KGBBs, with the differentiation of required and optional object-positions. Moreover, the possibility of super-constructors that contain constructors within them shares some similarities with semantic units, where compound units consist of a set of associated semantic units, some of which may be compound units themselves. Interesting to us is also the idea to verbalize a constructor unit in more than one sentence for improving its readability and to have several possible sentence-like realizations of it, e.g., in different languages, all being provided in rendering time.

# Conclusion

Today, with the pressure of having to rapidly produce new and FAIR data and metadata that meet the requirement of being integrable with existing data and metadata sources from various domains, knowledge graph management platforms must support rapid development of data management solutions that are suitable for everyday research, that demonstrate the added value of knowledge graph technologies, and that can also be implemented by smaller projects with a limited budget or even by individual researchers for their personal research knowledge graph (69).

Ideally, such a platform supports the needs of different categories of user groups: **Domain experts**, who are not interested in the internals of the application logic or the underlying data structure of the graph. They are not interested in the additional information that has to be added to make data and metadata machine-actionable, but want to access the graph in intuitive ways with the graph being reduced to the information they are currently interested in.

**Expert users and data scientists**, on the other hand, want to interact with the data, maybe analyze it using Jupyter Notebooks, R, or other data analysis tools, and require the data and metadata of the knowledge graph to be easily accessible in the formats they prefer and want the knowledge graph to provide convenient and efficient tools that support their data management operations.

And finally **data stewards, ontology engineers, and application engineers**, who require that the knowledge graph infrastructure can be easily incorporated into the overall software stack of their organizations while retaining full control over their data's ethical, privacy, or legal aspects (following Barend Mons' *data visiting as opposed to data sharing* (9)). They want the platform to make their lives easier when having to develop new targeted applications on top of the knowledge graph.

With its main idea—besides implementing semantic units in a knowledge management system—to clearly distinguish between an internal in-memory application data model, a data storage model, data display models, and data access/export models and to decouple the application data model from the rather technical aspect of storing data in a database and displaying data in the UI, the KGBB Framework supports all this. We argue that this decoupling is essential for being able to solve many of the immanent problems of knowledge management systems. In a KGBB-driven FAIREr knowledge graph application, based on the information provided by a set of interlinked KGBB instances, the KGBB-Engine manages the growing knowledge graph and organizes and structures it into multiple semantic units of different types, allows intuitively making statements about statements, provides users with input-forms and human-readable data views in the UI, enforces graph patterns for internal interoperability and machine-actionability of data and metadata, making them FAIREr, allows the specification of additional data schemata for access and export with the possibility to add further schemata for newly upcoming standards, and provides CRUD queries derived from the storage models of each KGBB that can be intuitively used, combined, and further specified in a query builder for searching and exploring the graph. Moreover, being able to specify a KGBB-driven FAIREr knowledge graph in a KGBB specification graph provides a self-describing knowledge graph application that meets the requirements of a FAIR software.

With the low-code KGBB-Editor and an openly accessible KGBB-Repository, the KGBB Framework provides a set of resources and tools that increase the overall interoperability of the knowledge graph's data and metadata by allowing other software applications to interoperate with the graph via its Access REST API that enables data retrieval through a REST API call with a set of key-value parameters, instead of having to send a SPARQL or Cypher query. Domain experts can define their own KGBB-driven FAIREr knowledge graph applications and developers can access and export their data as JSON, RDF, or CSV. Users of these FAIREr knowledge graphs can explore the graph's contents in intuitive ways, while machines can access all information they require.

The KGBB Framework will offer an open source knowledge graph management application platform that provides, with its central KGBB-Engine, a data management solution that is reusable in different domains. Due to its flexible data and metadata access options, it is also compatible with other applications and can easily be integrated into existing software stacks. Its genuine modular architecture supports flexibility, extensibility, and customizability, while requiring only a low initial investment to set up.

We think, the time has come that, in addition to focusing on the machine-actionability of data and metadata, we start to focus also on their human-actionability and thus their human-friendliness, turning FAIR knowledge graphs into FAIREr knowledge graphs that meet the requirements of cognitive interoperability. The KGBB Framework is one suggestion for how we could arrive there.

# Acknowledgements

We thank Philip Strömert, Roman Baum, Björn Quast, Peter Grobe, István Míko, and Kheir Eddine for discussing some of the presented ideas. We are solely responsible for all the arguments and statements in this paper. This work was supported by the ERC H2020 Project 'ScienceGraph' (819536). We are also grateful to the taxpayers of Germany.